\documentclass[onecolumn,notitlepage,9pt,nofootinbib]{revtex4-1}
\usepackage{graphicx}
\usepackage{float}
\usepackage[utf8]{inputenc}
\usepackage[english]{babel}
\usepackage[intlimits]{amsmath}
\usepackage{mathtools}
\usepackage{amssymb}
\usepackage{verbatim}
\usepackage{amsfonts}

\newcommand{\be}{\begin{eqnarray}}
\newcommand{\ee}{\end{eqnarray}}
\newcommand{\dif}{\mathrm{d}}
\newcommand{\defeq}{\stackrel{\text{def}}=}
\newcommand{\R}{\mathbb{R}}
\newcommand{\C}{\mathbb{C}}

\newcommand{\Z}{\mathbb{Z}}

 \newcommand{\im}{\mathbf{i}}
  
 \newcommand{\abs}[1]{\left\vert#1\right\vert}

\begin{document}
\title{One step replica symmetry breaking and extreme order statistics of logarithmic REMs}
\author{X. Cao, Y. V. Fyodorov and P. Le Doussal}
\address{LPTMS, CNRS (UMR 8626), Univ. Paris-Sud, Université Paris-Saclay, 91405 Orsay, France}
\address{King's College London, Department of Mathematics, London  WC2R 2LS, United Kingdom}
\address{CNRS-Laboratoire de Physique Théorique de l'Ecole Normale Supérieure, 24 rue Lhomond, 75231 Paris, Cedex, France}
\begin{abstract}
Building upon the one-step replica symmetry breaking formalism, {duly understood and ramified},
we show that the  sequence of ordered extreme
values of a general class of Euclidean-space logarithmically correlated random energy models (logREMs) behave in the thermodynamic limit as a randomly shifted decorated exponential Poisson point process. The
distribution of the random shift is determined solely by the large-distance ( ``infra-red'', IR) limit of the model, and is equal
to the free energy distribution at the critical temperature up to a translation. the decoration process
is determined solely by the small-distance (``ultraviolet'', UV) limit, in terms of the biased minimal process. Our
approach provides connections  of the replica framework to results in the probability literature and sheds further light on the freezing/duality conjecture which was the source of many previous results for log-REMs.  In this way we derive the general and explicit formulae for the joint probability density of depths of the first and second minima (as well its higher-order generalizations)
in terms of model-specific contributions from UV as well as IR limits.In particular, we show that the second min statistics is largely independent of details of UV data, whose influence is seen only through the mean value of the gap. For a given log-correlated field this parameter can be evaluated numerically, and we provide several numerical tests of our theory using
the circular model of $1/f$-noise.
\end{abstract}
\maketitle

\section{Introduction}\label{sec:intro}
Extreme value statistics of logarithmically correlated random energy models (logREMs) is a subject of active recent research by both physicists and mathematicians. LogREMs (and their ancestors, \textit{uncorrelated} REM {and simply-correlated GREM}) originated in physics \cite{derrida1980random,derrida1985generalization,derrida1986solution,derrida1988polymers} as simplified models of mean field spin glasses, and then were singled out as a particularly interesting case of the general problem, the thermodynamics of a particle in random potential \cite{carpentier2001glass,fyosom2007,fyodorov08rem}. When approached with the \textit{replica trick} of disordered statistical mechanics, these models are known to belong to the class that can be solved by \textit{one-step replica symmetry breaking} (1RSB) scheme.  Such solution is certainly much simpler than the full replica symmetry breaking scheme deemed necessary for description of the low-temperature phase of other classes of the mean field spin-glass models \cite{sk75spinglass,parisi80sk},  see \cite{Talagrandbook} for the modern rigorous approach, as well as of general manifolds in disordered landscapes \cite{mezard1991replica,ledoussal08cuspshock}. Yet one can argue that among all models of 1RSB class, logREMs are a special marginal case precisely at the boundary with the full RSB models\cite{fyosom2007,fyodorov08rem}, and that at the mean-field level they share such features of the full RSB as presence of  massless replicon modes in their fluctuation spectrum in the whole low-temperature phase, see e.g. Appendix D of \cite{FyoLec2010}.

 Another line of research initiated in \cite{derrida1988polymers} heavily relies on relations between LogREMs and traveling wave equations of Fisher-Kolmogorov-Petrovsky-Piscounov (FKPP) \cite{fisher1937wave,kolmo91kpp} type. Such a relation is exact for a particular instance of LogREM,
the  branching Brownian motion (bBm), and the deep mathematical analysis in the FKPP framework due to Bramson\cite{Bramson1983memoirs} provided  a paradigm
for understanding 
generic logREMs.  In physics literature these ideas were promoted to greater generality by  a renormalization group argument in \cite{carpentier2001glass}. More recently, in the mathematical literature adopting and generalizing Bramson's methods and ideas to other log-correlated processes and fields, most notably to the two-dimensional Gaussian Free Field (2DGFF),  helped to develop understanding of extreme and high values of those objects to a great depth , see \cite{BramsonZeitouni1,BramsonZeitouni2,ding2015convergence} and references later in this section. Besides providing mathematic tractability, the logREM-FKPP interplay looked upon from the physical viewpoint\cite{derrida1988polymers,carpentier2001glass}  revealed the fundamental \textit{freezing} phenomenon manifesting itself via the temperature-independence of extensive free energy in the whole low-temperature, 1RSB phase of logREMs. Basic thermodynamics implies then that the entropy vanishes, \textit{i.e.}, the Boltzmann-Gibbs measure is dominated by a few extremely low energies. The freezing phenomenon is the central feature of logREMs, with consequences in other areas of physics and mathematics. To that end one may mention the multi-fractal properties of electronic wave-functions in disordered gauge field \cite{chamon1996localization,castillo97dirac}, as well as  understanding of high values of multifractal patterns of more general nature \cite{rosso12counting,fyodorov2015high}. Most recently freezing was employed to conjecture the statistics of high values of characteristic polynomials of random matrices and eventually high values of the Riemann zeta-function in intervals along the critical line \cite{fyo12zeta,FyoKeat14,FyoSim15} which
 generated a new stream of activity \cite{ABB2015,arguin2015maxima,PaqZeit2016,ChMadNaj2016}.  As such, the whole subject of freezing and logREM extrema was under an intensive research in recent years and was addressed from several directions and viewpoints. Not attempting a comprehensive review, we very briefly survey this body of work below:

\begin{itemize}
\item[-] 
Building upon the picture of the freezing phenomenon  developed in \cite{carpentier2001glass} for \textit{Euclidean-space logREMs} (also known as logREMs with statistical translational invariance), the paper \cite{fyodorov08rem} proposed for the first time exact free-energy distribution for a particular regularized version of the 2DGFF sampled along a circle, which is also known as (log-)circular model of $1/f$-noise. The treatment was based on assuming the validity of the \textit{freezing scenario}  conjecturing that  in the thermodynamic limit the whole \textit{distribution} of the Gumbel-convoluted free-energy $F - G / \beta$, with $F$ being the free energy at temperature $1/\beta$, and $G$ being a standard Gumbel variable independent of $F$,  is {\it freezing}, that is becomes  temperature-independent below the transition temperature, modulo a translation. Further work \cite{fyodorov2009statistical} observed the co-existence of freezing with the \textit{duality-invariance} property, and extended the above freezing scenario to the \textit{freezing-duality conjecture} (FDC), which has been checked numerically for several observables in various logREMs \cite{fyodorov2009statistical,fyodorov2010freezing,fyodorov2015moments,cao15gff,cao16maxmin}. Although a relation between FDC and 1RSB was hinted briefly in \cite{fyodorov2010freezing}, up to now very little has been worked out explicitly in this direction. Intriguingly, recent mathematical work \cite{madaule2013glassy,subag2015freezing} provided a proof of the freezing of the distribution for $F - G / \beta$ without any reference to duality, though physical approach points towards an intimate relation between the two, at least for all the Gaussian models that are exactly solved (these models are sometimes referred to as \textit{integrable}, although quantum/classical integrable structures behind them are not clearly identified yet).
In fact, a lot of general and rigorous results  concerning the extreme values of 2d GFF and the associated Boltzmann-Gibbs-Liouville measure have been obtained in recent years in mathematical literature. Note that the 2d GFF is a random generalized function \cite{Sheffield07} so that studying its value distribution necessarily involves a regularization scheme. Among most popular rigorous schemes are the 'multiplicative chaos' framework \cite{kahane1985chaos}, see applications to LogREM-related models in \cite{duplantier2014critical,David2016lft}, and various variants of the discretization/lattice models \cite{ding2015convergence,daviaud2006extremes,ding2014extreme,arguin2014poisson,arguin2015poisson,biskup2016extreme,biskup2014conformal}. Although providing so far no explicit distribution of free-energy or the global minimum (see however \cite{ding2013exponential}), these studies corroborated the freezing picture of FDC.

\item[-] In a closely related development, an understanding of the \textit{full} process of minimal values of the branching Brownian motion (bBm) was initiated by \cite{brunet2011branching} and elaborated in full rigour in \cite{arguin2011genealogy,arguin2012poissonian,aidekon2013branching,arguin2013extremal}. The bBm is the prototype of \textit{hierarchical logREMs}. Such models are exactly described by FKPP equations, which describes only approximately the Euclidean-space ones discussed in the precedent paragraph. It is now known that in the thermodynamic limit, the minima process tends to a \textit{randomly shifted, decorated Gumbel Poisson point process} (SDPPP) \cite{subag2015freezing}. With respect to the Gumbel Poisson point process which describes the minima of the uncorrelated REM, the novel ingredient of SDPPP, \textit{i.e.}, the decoration process, describes the internal structure of blocks of extreme values which share a near ancestor, and thus are highly correlated. Such a picture is expected to apply also in the context of non-hierarchical logREMs, in particular those related to 2d GFF and $1/f$-noise, in which the blocks are those of extreme values given by nearby points. However, until recently \cite{biskup2016full}, this expected picture has not been established in    worked out in sufficient quantitative details which can be compared to numerical simulations of particular instances of logREMs. Also, despite the insightful discussions made in \cite{subag2015freezing}, we feel that the relation between SDPPP, freezing and 1RSB needs to be spelled out in a more explicit and comprehensive manner in the context of GFF-type logREMs. In particular, we feel necessary to start bridging the remaining gaps between physicists' and mathematicians' description of this common object of interest.
\end{itemize}
The goal of our present work is to provide such a detailed account at the physical level of rigour by employing the standard and powerful, albeit heuristic framework and language of the replica approach to a general class of logREMs. In doing so we will assume a few conditions, similar to \cite{ding2015convergence}, which allow us to define what physicists usually call the \textit{ultraviolet} (UV, short-distance)  and \textit{infrared} (IR, long-distance) data of a given logREM model in the thermodynamic limit. Under these assumptions, using 1RSB scheme, we are able to show that the ordered sequence of minimal values forms a SDPPP. More precisely, the\textit{ random shift} is equal to the free energy at critical temperature (modulo a translation), which depends only on the IR data. The free energy distribution at the critical temperature is shown to determine also the distribution in the whole low temperature phase (and thus the minimum) by the freezing scenario, of which we spell out the relation with the duality invariance property in the 1RSB framework. The\textit{ decorating process} depends, on the other hand, solely on the UV data, and is equal to the process of \textit{biased minima}. {In this way we derive the general and explicit formulae for the joint
probability density of depths of the first and second minima (as well its higher-order generalizations)
in terms of model-specific contributions from UV as well as IR limits. In particular, we show that the second min statistics is largely independent of details of UV data, whose influence is seen only through the mean value of the gap in a universal way.}

The paper is organised as follows. In Sect. \ref{sec:synopsis}, after defining the class of logREMs that we will consider and { specifying} their IR and UV data, we will state our results restricted to first and second minima for clarity, and then eventually describe the whole sequence of minimal values. In Sect. \ref{sec:num} we test some of our {analytical predictions numerically} using the circular $1/f$-noise model.  Sect. \ref{sec:minimum}  { we provide basic information} on the 1RSB framework. After describing in general the 1RSB approach for logREM in Sect. \ref{sec:1rsb}, we apply the method to retrieve the known freezing result of the free energy/minimum of the potential (Sect. \ref{sec:1rsbmin}), and elaborate on the 1RSB-FDC relation in Sect. \ref{sec:FDC}. After this warm up, we spell out the derivation for the second minima (eq. \eqref{eq:main}) in Sect. \ref{sec:derivation}. For a reader { interested only in getting flavour of the method rather than in full technical detail}   Sections \ref{sec:1rsbmin} and \ref{sec:derivation} (without sub-subsections) constitute a summary of the derivation of \eqref{eq:main}.

Sect. \ref{sec:fullminima} generalizes the above results to the whole sequence of minima values, where the structure of the decorated Poisson point process becomes evident.  Sect. \ref{sec:derivationgeneral} gives the derivation of \eqref{eq:maingeneral0}, which generalises \eqref{eq:main}, and discusses its consequences, including the joint distribution of all order statistics, which coincides with SDPPP (this will be shown in Appendix \ref{sec:dppp}). Sect. \ref{sec:sdppp} is devoted to the calculation of some marginal distributions (\textit{e.g.}, that of $k$-th minimum) of SDPPP. Sect. \ref{sec:uvtechnical} concentrates on the UV sector; we shall first study a variant of the uncorrelated REM that mimics the UV structure of logREMs, and confirms the 1RSB treatment of the UV sector by a replica-free calculation; this will be followed by a discussion on the nature of the decoration process.

A few appendices are devoted to more technical topics. Appendix \ref{sec:frsb} gives a full replica symmetry breaking analysis of logREM and its multi-scale generalisations. Appendix \ref{sec:EA} computes by 1RSB the Edwards-Anderson order parameter (defined in \cite{cao16maxmin}) for a general class of \textit{spherical logREMs}. Appendix \ref{sec:dppp} recalls the definition of SDPPP and calculates its joint minima value distribution, of which Appendix \ref{sec:sdppp} computes some marginal distributions. Motivated by recent work \cite{biskup2016full}, Appendix \ref{sec:minposition} takes the minima positions into the picture. Finally, Appendix \ref{sec:REMFyo} provides a derivation of a technical result on uncorrelated REM which is used in Sect. \ref{sec:gREM}.

\section{Synopsis}\label{sec:synopsis}
In this section, after defining the UV and IR data of logREMs in Sect. \ref{sec:modeldef}, we give the main results in Sect. \ref{sec:mainresult} restricting ourselves initially to the value distribution of the first and second minima which will be eventually the scope of direct numerical checks (in Sect. \ref{sec:num}). That particular case captures already all the most essential structures of the problem. Finally, in Sect. \ref{sec:fullsummary} we provide our results for the general case as well.

\subsection{Log-REMs and their UV and IR limits}   \label{sec:modeldef}
In this paper we are going to consider a general log-REM associated with a real random Gaussian potential $V_{j,M}$, { with $j=1,..M,$  indexing $M$ sites of a} lattice (which we assume to be either $d$-dimensional or a hierarchical tree). The covariance $\overline{V_{i,M}V_{j,M}}^c$ will be assumed to decay logarithmically with the distance between sites, \textit{e.g.} in the case of $d$-dimensional lattice we may informally assume $\overline{V(\mathbf{x})V(\mathbf{y})}^c \sim -2d\ln{\left(|\mathbf{x}-\mathbf{y}|/M\right)}$, a more detailed description following later on. The property of logarithmic decay can be conveniently characterized quantitatively as follows\begin{equation}
\forall i \,,\,  \frac{1}{\ln M} \ln \abs{\{j: \overline{V_{i,M} V_{j,M}}^c > 2q \ln M \}}  \rightarrow 1 - q  \,,\, M\rightarrow\infty \,,\, 0 \leq q \leq 1 \,.
\label{eq:logremdef}\end{equation}
where $\abs{A}$ is the size of the set $A$. Throughout this work, $\overline{[\dots]}$ means averaging over disorder, \textit{i.e.}, over the $(V_{j,M})_{j=1}^M$ in the above case, unless otherwise stated. In particular, we require the (log)REM variance
\begin{equation} \overline{V_{i,M}^2}^c \sim 2\ln M + O(1) \,,\, M\rightarrow\infty\,, \label{eq:variance} \end{equation}
a key assumption to be supplemented below in \eqref{eq:intra}. Note that this definition applies to both hierarchical logREMs and logREMs in Euclidean space, \textit{i.e.} with Statistical Translational Invariance.

The partition function of the logREM is defined as usual
\begin{equation}
\mathcal{Z} \defeq \sum_{j=1}^M \exp(- \beta V_{j,M}) \,, \label{eq:defdiscretepartition}
\end{equation}
where $\beta$ is the inverse temperature. The definition \eqref{eq:logremdef} has the advantage of fixing the critical temperature
\begin{equation} \beta_c = 1  \,,\label{eq:betac} \end{equation}
regardless of the geometric nature of the model. In fact eq. \eqref{eq:betac} holds for both uncorrelated REM and logREM that satisfy \eqref{eq:variance}, and  a way to see this for uncorrelated REM is as follows \cite{derrida1980random}. By \eqref{eq:variance}, one computes the microscopic ensemble entropy, defined as the log of the number of states up to a given energy: $S(E = -2q \ln M) = \ln M (1 - q)$ for $q\in [0,1]$ and $S = 0$ for $q > 1$, and notices the non-analyticity at $q = 1$ where $\beta_c = \left.\frac{\partial S}{ \partial E}\right\vert_{q = 1-\epsilon} = 1$. For correlated REMs, strictly speaking, the existence of one (and only one) transition is a result of RSB analysis, see Sect. \ref{sec:1rsb} and \ref{sec:frsb}.

\begin{figure}
	\includegraphics[scale=.3]{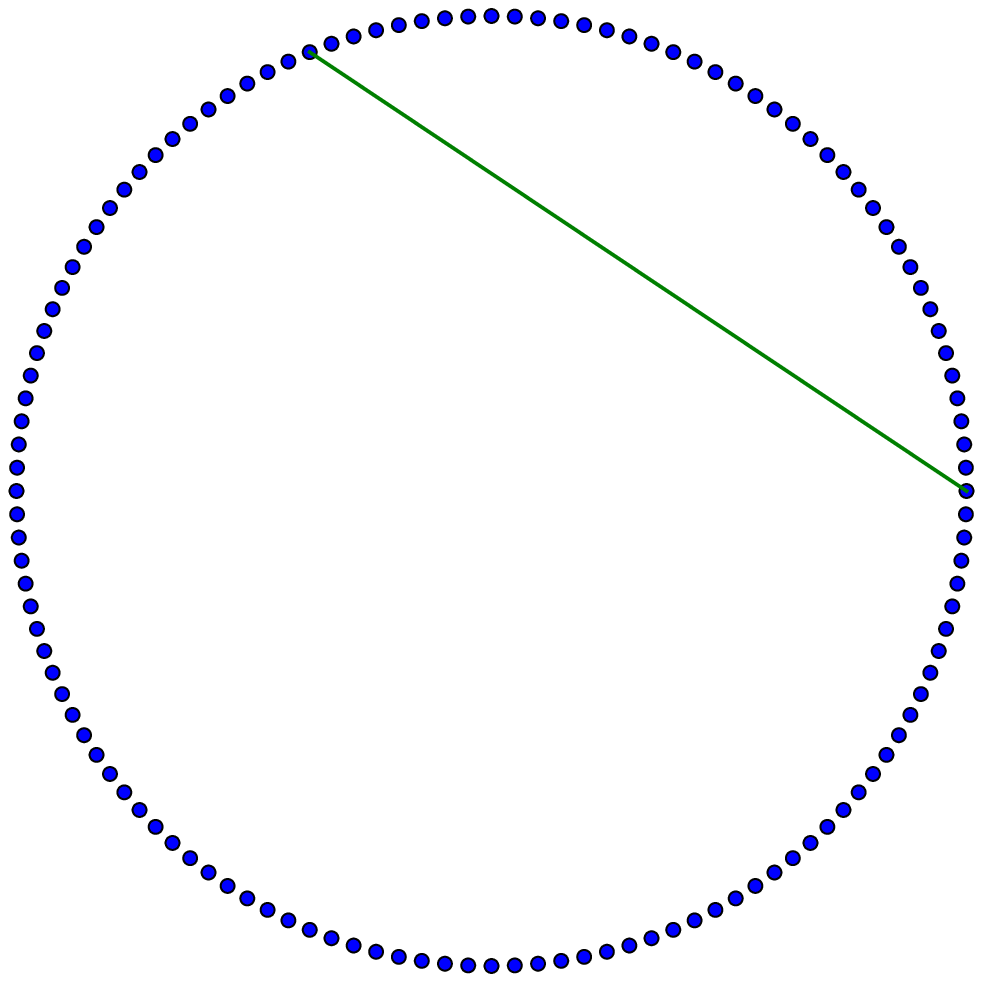}	\includegraphics[scale=.3]{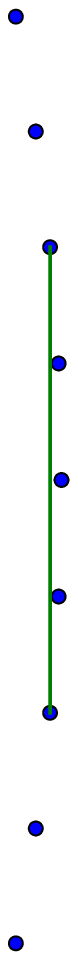}
	\caption{An illustration of the IR (left) and UV (right) covariances in a circle model.} \label{fig:uvir}
\end{figure}

The general criterion \eqref{eq:logremdef} specifies only the covariance in the scaling regime $0 < q < 1$. Although this is enough to guarantee the super-universal properties of logREM predicted in \cite{carpentier2001glass}, to fully define the model and calculate the distribution of minima values one needs to specify the model behaviour at the system-size scale and the lattice-spacing scale, in terms of IR and UV limit data, respectively, which we define below. An illustration is given in Fig. \ref{fig:uvir}.

\subsubsection{IR data}\label{sec:irdata}
The \textit{IR data} consist first of a $d$-dimensional manifold $X$ discretized by a family of point meshes $\xi_{i,M} \in X$ that tend (as $M\rightarrow\infty$) to a point density measure:
\begin{align}
 & \mu(\xi) \defeq \exp(- l(\xi)) \abs{\dif^d \xi} \,,\, \text{such that} \label{eq:muxi} \\
 & \frac{1}{M^n} \sum_{i_1, \dots, i_n} f(\xi_{i_1,M}, \dots, \xi_{i_n, M}) \stackrel{M\rightarrow\infty}\longrightarrow \int_{X} \prod_{a=1}^n \dif \mu(\xi_a) f(\xi_1, \dots, \xi_n) \,. \label{eq:irmeasure}
\end{align}
The other part of the IR data is a continuous mean value (sometimes called the background charge) $c(\xi)$ and a continuous { covariance kernel} $C(\xi, \eta), \xi,\eta \in X$ satisfying:
\begin{align}
&  \xi_{j,M} \stackrel{M\rightarrow\infty}\longrightarrow \xi \Rightarrow \overline{V_{j,M}}  \stackrel{M\rightarrow\infty}\longrightarrow c(\xi) \\
&  (\xi_{j,M}, \xi_{k,M}) \stackrel{M\rightarrow\infty}\longrightarrow (\xi, \eta) \,,\, \xi \neq \eta    \Rightarrow  \overline{V_{j,M} V_{k,M}}^c  \stackrel{M\rightarrow\infty}\longrightarrow C(\xi, \eta) \label{eq:irdata}
\end{align}
To ensure matching with \eqref{eq:variance}, {the kernel} $C$ should be logarithmically diverging when $\xi \rightarrow \eta$: more precisely, if $X$ is $d$-dimensional,
\begin{equation} C(\xi,\eta) \sim -2 d \ln \abs{\xi - \eta} \,. \label{eq:limitIRcov} \end{equation}

The IR data allow to define the continuous replica partition function
\begin{equation}
\overline{Z_b^n} \defeq \int \prod_{a=1}^n  \exp(-b c(\xi_a) - (1 + b^2) l(\xi_a)) \abs{\dif^d \xi_a}  \prod_{a<a'} \exp(b^2 C(\xi_a, \xi_{a'})) \,.  \label{eq:contZ}
\end{equation} 
The reason for which we write $b$ instead of $\beta$ in \eqref{eq:contZ} will be clear later on. {The replicated partition function} is usually given a Coulomb-gas integral, except that the (identical) charges attract each other. This causes the integral to diverge, generically for $n b^2 \geq 1$. Nevertheless we shall assume that by analytical continuation, \eqref{eq:contZ} can be defined for generic $n, b \in\C$ (this has been carried out for several concrete models starting from \cite{fyodorov08rem,fyodorov2009statistical},  see \cite{ostrovsky2009mellin,ostrovsky2013selberg,ostrovsky2013theory,ostrovsky2016barnes,Ostrovsky2016A,ostrovsky2016riemann} for the relevant rigorous mathematical developments). Note that here the $\overline{[\dots]}$ is rather a metaphor since we do not define the continuum random partition function $Z$ at the first place; heuristically it could be defined as
 $$ Z_b \stackrel{\text{def}}= \int \exp(-l(\xi))  \abs{\dif^d \xi} \exp(-b (v(\xi) + c(\xi))) \,,\, \overline{v(\xi)} = 0 \,,\, \overline{v(\xi)^2} = -2 l(\xi)\,,\, \overline{v(\xi)v(\eta)} \stackrel{\eta \neq \xi}= C(\xi, \eta)\,.$$
Then \eqref{eq:contZ} can be ``derived'' by Wick theorem. Here $v(\xi)$ can be seen as a regularized Gaussian Free Field with zero mean. It does not make probabilistic sense directly, since \textit{e.g.}, its variance is negative, however a rigorous construction of the integrand of the RHS is possible in  the multiplicative chaos framework {developed originally in \cite{kahane1985chaos}, see \cite{RhodesVargas2014} for recent developments}. We will return to the meaning of $Z_b$ later in \ref{sec:minposition}.

When approaching the critical temperature $b \rightarrow 1_-$, it is now known \cite{duplantier2014critical} that the well-behaved object is not the partition function $Z_b$ but its derivative $(\partial_b Z_b)\vert_{b=1}$. This echoes the fact that, in all known exactly solved cases \cite{fyodorov08rem,fyodorov2009statistical,fyodorov2010freezing,cao15gff,cao16maxmin,fyodorov2015moments}, the analytically continued continuum integral $\overline{Z^{n}_b}$ has a zero of order $n$ at $b = \beta_c = 1$. Therefore we define
\begin{equation}
 \overline{\widetilde{Z}^n} \stackrel{\text{def}}= \lim_{b \rightarrow 1-}\frac{\overline{Z_b^n}}{(1-b)^n}  \,. \label{eq:partialZcont}
\end{equation}
 The same object (modulo a constant) was called $\overline{z^n}$ \textit{op. cit.}. We shall assume that \eqref{eq:partialZcont} exists for generic $n\in \C$.

\subsubsection{UV data}
The \textit{UV data} describe the limit of covariance between points $i$ and $j$ separated by a fixed distance \textit{in units of lattice spacing} in the $M\rightarrow\infty$ limit. From the IR point of view, $\xi_{i,M}$ and $\xi_{j,M}$ tend to the same point $\xi \in X$ in this limit, and $\overline{V_{i,M}V_{j,M}} \approx C(\xi, \xi)$ has a divergence. However, if we zoom in to the lattice spacing scale, the points $\xi_{i,M}$ close to $\xi$ will form a $d$-dimensional hyper-cubic lattice $\xi_{i,M} \rightarrow \xi + a \mathbf{x}_{i}, \mathbf{x}_{i} \in \Z^{d}$ (as $M\rightarrow\infty$), where the grid side $a \propto (M \exp(-l(\xi)))^{-1/d}$ because of \eqref{eq:muxi}. Thus we expect from \eqref{eq:limitIRcov} that $C(\xi_{i,M}, \xi_{j,M}) \approx -2d \ln \abs{ \xi_{i,M} -\xi_{j,M}} \approx 2 (\ln M  - l(\xi))$; note that this agrees with, and supplements, eq. \eqref{eq:variance}. Now the UV data are defined as the limit of all further corrections, which we express as a symmetric matrix $(f_{ij})$:
\begin{equation}
\overline{V_{i,M} V_{j,M}}^c - 2 (\ln M - l(\xi)) \stackrel{M\rightarrow\infty}\longrightarrow  f_{ij} \,. \label{eq:intra}
\end{equation}
We assume that the limit exists, does not depend on $\xi$ (homogeneity of UV data), and locally translation invariant in the sense that $f_{ij} = f(\mathbf{x}_{i} -  \mathbf{x}_{j})$. This implies in particular that diagonal elements $f_{ii}$ are { all equal}.

The notion of UV limit data will play a crucial rôle in the correct calculation of the higher order statistics because different minima can be at the distance of order of lattice spacing from each other, and probe the covariance matrix at that scale. Note that  only the IR data are needed for calculating the minimum value distribution. Obtained by different limit procedures, the UV and IR data are independent of each other, in the sense that one can modify the IR data {for a given model} while keeping its UV data intact, and vice versa. Let us the latter point by reviewing some concrete $1$-d models.

\subsubsection{Example: Circular model}\label{sec:circularmodel}
The \textit{circular model} of $1 / f$-noise \cite{rosso12counting,fyodorov08rem} is a Gaussian log-REM defined by the following mean and covariance matrix:
\begin{equation}
\overline{V_{i,M}} = 0 \,,\, \overline{V_{i,M} V_{j,M}} = \sum_{k=1}^{M/2} 2\lambda_k \cos  \frac{2\pi k (i-j)}{M} \,,\, \lambda_k =  k^{-1} \,. \label{eq:circledef}
\end{equation}
Its \textit{IR data} are the following, as one can easily check by \eqref{eq:irdata}:
\begin{equation}
\xi_{j,M} = \exp\frac{2\pi j}{M} \in X = \{ \xi \, s.t. \,\abs{\xi} = 1\} \Rightarrow C(\xi, \eta) = -2 \ln \abs{\xi-\eta} \,,\, c(\xi) = 0 \,,\, l(\xi) = \ln(2\pi) \,, \dif \mu(\xi) = \frac{\dif \xi}{2\pi \im \xi} \,. \label{eq:circleIR}
\end{equation}
The points $\xi_{j,M}$ provide thus uniform meshes of the unit circle, whereas $C$ is the Green function of the planar 2d Gaussian free field. The {associated continuum Coulomb} integral \eqref{eq:contZ} is the Dyson integral \cite{dyson1962statistical1}, which has a closed form expression that can be analytically continued:
\begin{equation}  \overline{Z_b^{n}} =  \frac{1}{(2\pi)^{n b^2}}\int_0^{2\pi} \prod_{a < a'} \abs{e^{\im \theta_a} - e^{\im \theta_{a'}}}^{-2\beta^2} \prod_{a=1}^{n} \frac{\dif \theta_a}{2\pi}  =  \frac{1}{(2\pi)^{n b^2}} \frac{\Gamma(1 - nb^2)}{\Gamma(1 - b^2)^{n}} \Rightarrow \overline{\widetilde{Z}^{-t}} = (4\pi)^t \Gamma(1 + t) \,. \label{eq:dyson}\end{equation}
In particular, \eqref{eq:partialZcont} is satisfied. Its \textit{UV data} can be calculated from \eqref{eq:intra} using $l(\xi) = \ln(2\pi)$ from \eqref{eq:circleIR}
\begin{equation}
f_{ij} = \lim_{M\rightarrow\infty}\left( -2 \ln M + \ln(2 \pi)  + \sum_{k=1}^{M/2} \frac{2}{k} \cos  \frac{2\pi k (i-j)}{M} \right) = \begin{dcases} -2 (\ln\abs{i-j} -\text{Ci}(\abs{i-j} \pi )) & i \neq j \,, \\
   2 (\gamma_{E} + \ln \pi )   & i = j \,, \end{dcases}\label{eq:fijlimitFB}
\end{equation}
where $\text{Ci}(x) = - \int_{x}^{\infty} \cos t / t \dif t$ is the cosine integral. As $\abs{i-j} \rightarrow\infty$, $f_{ij} \approx -2 \ln \abs{i-j}$ grows logarithmically, which is a general feature of UV asymptotics for logREMs.

A few variants of the circular model have been studied. One example is a generalization based on 2D field in a disc with Dirichlet boundary conditions \cite{fyodorov2009statistical}, which modifies the covariance of \eqref{eq:circledef} {by replacing $\lambda_k=1/k$ with  $\lambda_k = (1 - q^k) / k,$} where $q \in (0,1)$. One can check that the UV data \eqref{eq:fijlimitFB} remain unchanged, while the continuous covariance matrix of IR data is modified to $C(\xi, \eta) = -2\ln\abs{\xi - \eta} + 2 \ln\abs{\xi - q \eta}$ whereas $l(\xi)$ and $c(\xi)$ are not changed. For these models the analytical continuation of the continuum integral \eqref{eq:contZ} is known \cite{cao15gff} only as an infinite sum, not as a closed-form formula similar to \eqref{eq:dyson}.

One can also change solely the UV data. A such example is the ``long range'' circular model \cite{fyodorov2009statistical}, defined as $\lambda_k = 2\pi/\sqrt{2(1 - \cos(2\pi k / M))}$. This implies $f_{ij} = \int_0^1 \dif x  \frac{2\pi (1 - \cos(2\pi x \abs{i-j}))}{\sqrt{2(1 - \cos(2\pi x))}}$, different from \eqref{eq:fijlimitFB}. However, the IR limit remains the same, mainly because $\lim_{M\rightarrow \infty} \lambda_k = k^{-1}$ as in \eqref{eq:circledef}.

\subsubsection{Example: Interval model}\label{sec:interval}
An example of less trivial $l(\xi)$ and $c(\xi)$ is the \textit{interval model} \cite{fyodorov2009statistical,fyodorov2015moments}. Its general discrete-version definition depends on four parameters $c_0, c_1, d_0, d_1$ \footnote{they should be sufficiently close to $0$ for the results of this work applies; in other parameter regimes, there can be a pinned phase, not considered in this work}. One starts by {defining} the non-uniform mesh points whose density follows a Beta distribution:
\begin{equation} \int_0^{\xi(j,M)} \xi^{d_0} (1-\xi)^{d_1} \dif \xi  = \frac{j}{M} \int_0^1 \xi^{d_0} (1-\xi)^{d_1} \dif \xi \,,  \label{eq:intervalmesh}\end{equation}
in terms of which the mean and covariance of $V_{j,M}$ are given by
\begin{equation}\overline{V_{j,M}} = c_0 \ln \xi(j,M) + c_1 \ln (1 - \xi(j,M)) \,,\quad \overline{V_{j,M} V_{k,M}}^{c} =
\begin{dcases}  - 2 \ln \abs{\xi(j,M) - \xi(k,M)} & j \neq k \,, \\
 2 \ln \left(M  \xi^{d_0} (1-\xi)^{d_1}\right) + f_0 & j = k \,.  \end{dcases}   \,. \label{eq:intervaldef} \end{equation}
Here $f_0$ is a constant, which can be fixed arbitrarily, as long as the covariance matrix above is semi positive definite for all $M$.
The IR data of the model do not depend on $f_0$, and are given by $c(\xi) = c_0 \ln \xi + c_1 \ln (1 - \xi)$ and $l(\xi) = -d_0 \ln\xi - d_1 \ln(1 - \xi)$, and its continuum limit representation is given by the Selberg integral \cite{forrester2010log}:
\begin{equation}
\overline{Z^n_b} = \int_0^1 \prod_{a=1}^n \left(\xi_a^{b e_0} (1 - \xi_a)^{b e_1} \dif \xi_a \right) \prod_{a < a'} \abs{\xi_a - \xi_{a'}}^{-2b^2}  \,,
e_i = - c_i + (b + b^{-1}) d_i \,,\, i = 0, 1\,.
\end{equation}
The continuation of its exact solution (due to Selberg) to complex values of $n$ is discussed in \cite{fyodorov2009statistical} and  \cite{ostrovsky2009mellin,ostrovsky2013selberg,ostrovsky2013theory,ostrovsky2016barnes,Ostrovsky2016A,ostrovsky2016riemann}. In particular, one knows that it has a $\Gamma(1 - b^2)^{-n}$ factor, like \eqref{eq:dyson}, so that \eqref{eq:partialZcont} makes sense.

On the other hand, the UV data depends only on $f_0$ in \eqref{eq:intervaldef}, and is given by $f_{ii} = f_0$ and $f_{ij} = -2\ln\abs{i-j}$ for $i\neq j$.

\subsection{First and second minima}\label{sec:mainresult}
\subsubsection{Observables}
From now on, we will omit the system-size label, \textit{i.e.}, ${}_{\dots,M}$ will be dropped in the subscripts to ease the notations, so $V_i = V_{i,M}$ denotes the values of some logREM with well-defined IR and UV data. We consider the following observables related to the first and second minima:
\begin{align}
&G_\beta(y) \defeq \overline{\prod_i\theta_\beta(V_i - y)} \,,\quad \theta_\beta(x) = \exp (-\exp(-\beta x)) \stackrel{\beta\rightarrow\infty}\longrightarrow \begin{dcases} 1 & x > 0 \\ 0 & x < 0 \end{dcases} \,.\label{eq:defG}\\
&H_\beta(y_0, y)  \defeq \sum_j \overline{ (1 -  \theta_\beta(V_j - y_0)) \prod_{i\neq j} \theta_\beta(V_i - y) } \,,\quad y_0 < y \label{eq:defH}
\end{align}
The first observable is the generating function of the partition function (\textit{i.e.}, the Laplace transform of its probability density) commonly used in the disordered system context; see \eqref{eq:Ggen} and \eqref{eq:laplace}. In particular, it decreases from $1$ to $0$ as $y$ goes from $-\infty$ to $\infty$. $H_\beta$ can be also interpreted in a similar way, see \eqref{eq:HGgen}. They have the following zero-temperature limits:
\begin{align}
&G_\beta(y) \stackrel{\beta\rightarrow\infty}\longrightarrow \mathbb{P}(V_{\min} > y) \,,\quad -\partial_y G_\infty(y) = \overline{\delta(V_{\min} - y)} \label{eq:Ginfty}  \\
&H_\beta(y_0, y) \stackrel{\beta\rightarrow\infty}\longrightarrow \mathbb{P}(V_{\min} < y_0, V_{\min,1} > y ) \,,\quad - \partial_{y,y_0} H_\infty = \overline{\delta(V_{\min} - y_0) \delta(V_{\min,1} - y)} \,, \label{eq:Hinfty}
\end{align}
where here and below we denote $\partial_{y_1,\dots,y_p} \defeq \partial_{y_1} \dots \partial_{y_p}$.
Namely, $1 - G_\infty$ is the cumulative density function of the minimum value, and derivatives of $H_\infty(y_0, y)$ provide the joint probability distribution of the first and second minima values. Note that if we had not summed over $j$ in \eqref{eq:defH}, the zero temperature limit \eqref{eq:Hinfty} would contain in addition the minimum position.

\subsubsection{Main result}
The main result is encoded in the equation which holds in the $M \rightarrow \infty$ limit for all $\beta$ :
\begin{equation}
H_\beta(y_0, y) = -\frac{1}{\min(1,\beta)} \partial_{y} G_\beta(y) K_\beta(y - y_0) \,, \label{eq:main}
\end{equation}
where $K_{\beta}(\Delta)$ is the UV correction factor definition of which will be given below. In spite of the product form, we do not have an obvious interpretation of \eqref{eq:main} in terms of statistical independence; in particular, the minimum and the first gap $V_{\min,1}- V_{\min}$ are not uncorrelated, see \eqref{eq:pdfmin1} below. Nevertheless, the two terms of the RHS have contrasted nature, as we detail below:
\begin{itemize}
\item[-] The shape of $G_\beta(y)$ (\textit{i.e.} modulo a translation) depends only on the IR data \eqref{eq:contZ} and manifests freezing:
\begin{align}
&\int  -\partial_y G_\beta(y) e^{t y} \dif y \,\cdot\, \exp(-t F(M,\beta,f_{ij})) \stackrel{M\rightarrow\infty}\longrightarrow
\begin{dcases} \overline{Z_\beta^{-t/\beta}}\Gamma(1 + t/\beta) & \beta < 1 \, \\ \overline{\widetilde{Z}^{-t}} \Gamma(1 + t)  &  \beta \geq 1 \end{dcases} \,, \label{eq:Gfreezing} \\
& F(M,\beta,f_{ij}) = \begin{dcases} -(\beta + \beta^{-1}) \ln M  - \beta f_{ii}/2 &  \beta  < 1 \,, \\
                                     -2 \ln M + \frac{1}{2} \ln\ln M + a_0    & \beta = 1 \,,\\
                                     -2 \ln M + \frac{3}{2} \ln\ln M + a_1  & \beta > 1 \,.
\end{dcases} \label{eq:FM}
\end{align}
In \eqref{eq:FM}, $F(M, \beta, f_{ij})$ contains the $M$ dependence of the free energy, and a $M$-independent, UV-data-dependent correction. The latter can be exactly calculated in the $\beta < 1$ phase (recall that $f_{ii}$ is independent of $i$), see the beginning of Sect. \ref{sec:FDC}, eq. \eqref{eq:etfhighT}. In the other phases this correction remains an unknown constant ($a_0, a_1$ above); this is related to unsolved issues when $Z$ in \eqref{eq:contZ} is replaced by the renormalized value
$\widetilde{Z}$ in \eqref{eq:partialZcont}. As a consequence of \eqref{eq:Gfreezing}, the distribution of the minimum
(modulo a UV dependent translation) is determined by the critical partition function $\overline{\widetilde{Z}^{-t}}$ of eq. \eqref{eq:partialZcont}.
\item[-] The \textit{UV correction factor} $K_\beta(\Delta)$ depends only on the UV data $f_{ij}$.  In particular, distribution of the gap between the first and second minima only depends on UV data $(f_{ij})$ defined in \eqref{eq:intra}. To describe how the latter determine $K_\beta(\Delta)$, we define the {\it local} logREM, which is a sequence of Gaussian variables $u_i$ with zero mean and the following covariance:
 \begin{equation}
 \overline{u_i} \defeq 0 \,,\quad \overline{u_i u_j}^c  \defeq C_N + f_{ij} \, . \label{eq:localrem}
 \end{equation}
 Here the over-line $\overline{[\dots]}$ means averaging over the \textit{local} disorder $(u_i)_{i=1}^N$, with $C_N$ being a large positive number chosen to make the covariance matrix positive-definite (whose precise value turns out to be irrelevant, see below). Then $K_\beta(\Delta)$ is given in terms of the partition functions of the local logREM:
\begin{align}
& K_\beta(\Delta) =  \lim_{N\rightarrow\infty} \sum_{j=1}^N \frac{ \overline{w_{\beta}^{m}(j,\Delta)} - \overline{w_{\beta}^{m}(j,\infty)}}{\overline{z_\beta^{m}}} \, , \quad m = \min (1, 1/\beta)  \label{eq:Koffd} \\
& w_\beta(j,\Delta)  \defeq  \exp(-\beta (u_j + \Delta)) +  \sum_{i \neq j} \exp(-\beta u_i) \,, \quad
z_\beta = \sum_{i=1}^N \exp(-\beta u_i) \label{eq:smallw}
\end{align}
The numerical convergence as $N\rightarrow \infty$ is expected to be much faster than in the $M \to \infty$ limit in the originally defined logREM, eq \eqref{eq:logremdef}; the reason is that the observable $K_\beta(\Delta)$ captures essentially microscopic information, even when $N\rightarrow\infty$. We will discuss this in more detail and generality in Sect. \ref{sec:biased}.

\end{itemize}

\subsubsection{High temperature phase}
In the high-temperature $\beta < 1$ phase, where $m = 1$, \eqref{eq:Koffd} and \eqref{eq:smallw} become easy to evaluate:
\begin{equation}
 K_{\beta<1}(\Delta) = \lim_{N\rightarrow\infty} \sum_{j=1}^N \frac{ \overline{e^{-\beta(u_j - \Delta)}} +  \sum_{i\neq j} \overline{e^{-\beta u_i}} - \sum_{i\neq j} \overline{e^{-\beta u_i}}}{\sum_{i=1}^N \overline{e^{-\beta u_i}}}
 =  e^{-\beta \Delta} \,. \label{eq:KhighT}
\end{equation}
Therefore \eqref{eq:main} simplifies to
\begin{equation}
H_{\beta < 1}(y_0, y) = - \beta^{-1}\partial_{y} G_\beta(y) e^{\beta(y - y_0)} \,. \label{eq:H1highT}
\end{equation}
The equation is {(super)} universal since $K_{\beta<1}$, \eqref{eq:KhighT}, depends on neither UV nor IR data. At the same time, in the low-temperature $\beta > 1$ phase $K_\beta$ will have in general a non-trivial temperature dependence, in contrast to $G_{\beta > 1}$, which manifests freezing there. The \textit{non-freezing} of $K_\beta$ (and its higher order statistics extensions, see \eqref{eq:defKgeneral}), is the essentially new feature that prevents us from extending naïvely freezing to conclude ``$H_{\beta>1} = H_{\beta=1}$'', which as we find is \textit{false} in general. Nevertheless, the naïve freezing does hold for a modified version of $H_\beta$ (see \eqref{eq:Hfar}), whose zero temperature limit gives the statistics of the second minimum whose position is far away from that of the global minimum, see \eqref{eq:fargap} for precise definition.

\subsubsection{Zero-temperature limit: distribution of the first and second minima values}\label{sec:mainzeroT}
In the zero temperature $\beta\rightarrow\infty$ limit, \eqref{eq:main} and \eqref{eq:Hinfty} give the joint distribution of $V_{\min}$ and  $V_{\min,1}$ :
\begin{align}
&\overline{\delta(V_{\min,1} - y) \delta(V_{\min} - y_0)} = -G''_\infty(y) K'_\infty (y - y_0)- G'_\infty(y) K''_\infty (y - y_0)\,,\label{eq:pdfmin1}
\end{align}
where $K_\infty$ can be obtained by taking the $\beta\rightarrow\infty$ limit of $K_\beta$ defined in \eqref{eq:Koffd}. In that limit, the local partition functions \eqref{eq:smallw} are dominated by $u_{\min}$ and $u_{\min,1}$, the first and second minima of the local logREM, respectively. More precisely, we have $z_\beta \rightarrow e^{-\beta u_{\min}}$ for any $\delta \geq 0$ and can check that $w_\beta (j, \delta) \rightarrow e^{-\beta {u_{\min}}}$ if $u_j \neq u_{\min}$, and
$ w_\beta (j, \delta) \rightarrow e^{-\beta \min({u_{\min}+\delta,u_{\min,1}})}$ otherwise. Now inserting these into \eqref{eq:Koffd} with $\delta = \Delta$ and $\delta = +\infty$ and observing that only the case where $u_j = u_{\min}$ yields a non-vanishing contribution in \eqref{eq:Koffd} we obtain
\begin{equation}
K_\infty(\Delta) = \frac{\overline{\exp(-\min(u_{\min} + \Delta, u_{\min,1}))} -  \overline{\exp(-u_{\min,1})}}{\overline{\exp(-u_{\min})}} \,. \label{eq:Kinfty}
\end{equation}
Differentiating, we get
\begin{align}
 K'_\infty(\Delta) = - e^{-\Delta} \,\,\frac{\overline{\theta(u_{\min,1} - u_{\min} - \Delta) \exp(-u_{\min})}}{\overline{\exp(-u_{\min})}}\,.  \label{eq:dKinfty}
\end{align}
We would like to emphasize that the factors $\exp(-u_{\min})$ in the numerator and denominator do not cancel each other unless assuming independence of $u_{\min}$ and the local gap $u_{\min,1} - u_{\min}$, so $K_{\Delta}$ is not a function of {only the} local gap $u_{\min,1} - u_{\min}$. Nevertheless, \eqref{eq:dKinfty} implies that $-K'(\Delta)$ decreases from $-K'_\infty(0)= 1$ to $ K'_\infty(\Delta\rightarrow\infty) = 0$. This allows us to define a new pair of the {\it biased} extrema $v_{\min} < v_{\min,1}$ for which the following relation holds:
\begin{equation}
\overline{\theta(v_{\min,1} - v_{\min} - \Delta)} \, \defeq \,\, \frac{\overline{\theta(u_{\min,1} - u_{\min} - \Delta) \exp(-u_{\min})}}{\overline{\exp(-u_{\min})}} \,. \label{eq:gapbiased}
\end{equation}
The biased extrema are defined up to an overall translation, \textit{i.e.} only the gaps are meaningful observables and one may as well fix $v_{\min} = 0$. They are biased in the sense that events in which $u_{\min}$ is more negative have dominating (with weight $\exp(-u_{\min})$) contribution to the statistics of the gap $ u_{\min,1} - u_{\min}$.

Combining \eqref{eq:gapbiased} and \eqref{eq:pdfmin1} we see that the \textit{first gap} $g_1 =V_{\min,1} - V_{\min}$ satisfies
\begin{align}
&  \overline{\delta(g_1 - \Delta)} = K_\infty''(\Delta) = e^{-\Delta} \left(\overline{\theta(v_{\min,1} - v_{\min} - \Delta)} +
\overline{\delta(v_{\min,1}-v_{\min} - \Delta)}\right) \,, \label{eq:gapstat2} \\
& \overline{\theta(g_1 - \Delta)} =  -K_\infty'(\Delta) = e^{-\Delta} \, \overline{\theta(v_{\min,1} - v_{\min} - \Delta)} \, ,\,   \label{eq:gapstat1}  \\
& \overline{\theta(g_1 > \Delta) (g_1 - \Delta)} =  K_\infty(\Delta) =  \overline{\theta(v_{\min,1}-v_{\min,0} - \epsilon)\left(\exp(-\Delta) - \exp(v_{\min,1}-v_{\min,1}) \right)}  \,.    \label{eq:gapstat}
\end{align}
For the second minimum $V_{\min,1}$ we have
\begin{align}
&\overline{\delta(V_{\min,1} - y)} =  - G'_\infty(y) + g G''_\infty(y) \,\, \Leftrightarrow\,\,
\overline{\theta(V_{\min,1} - y)} =  G_\infty(y) - g G'_\infty(y)  \,,\quad g = \overline{g_1} \,. \label{eq:pdf2ndmin}
\end{align}
In other words, possible distributions of $V_{\min,1}$ lie in a one-parameter family determined solely by the IR data, with the
parameter being the mean value of the gap depending only on the UV data.

The above results are only a fragment of a larger picture, \textit{i.e.} that of the randomly shifted decorated Gumbel Poisson point process, with the decoration process being the \textit{biased} local minima process $(v_{\min}, v_{\min,1}, \dots)$. The best way to see how the latter emerges is to generalize the results to the full minima value process, which we are going to describe now.

\subsection{Higher order statistics}\label{sec:fullsummary}
Here we summarize the main results on the higher order statistics, referring to Sect. \ref{sec:fullminima}, Appendix \ref{sec:sdppp} and \ref{sec:minposition} for the detailed derivation and discussions, as well as for additional results. They are all obtained from the 1RSB calculation of the following observable that generalizes $H_\beta(y_0, y)$ of \eqref{eq:defH}:
 \begin{align}
H_\beta\left((j_0, y_0), \dots, (j_{k-1}, y_{k-1}),y \right) &\stackrel{\text{def}}=  \overline{\prod_{s=0}^{k-1} (1 -  \theta_\beta(V_{j_s} - y_s)) \prod_{i \neq j_1,..j_{k-1}}  \theta_\beta(V_{i} - y) } \label{eq:Hgeneral_def0}
 \end{align}
Here $y_0 < y_1  < \dots < y_{k-1} < y$ and $j_1, \dots j_{k-1}$ are all distinct; a position-value couple $(j_s, y_s)$ will be called a \textit{marker}. When the positions are fixed, the observable \eqref{eq:Hgeneral_def0} gives the joint distribution of the minima values and their positions; such calculation {however requires the knowledge of} new continuum integrals, see Appendix \ref{sec:minposition} for a more detailed discussion. If however we focus on the values of the minima only and sum over their positions, we will arrive to the following result generalizing \eqref{eq:main}:
\begin{align}
&\sum\limits^{*}_{j_0\dots j_{k-1}} H_{\beta}((j_0, y_0), \dots, (j_{k-1}, y_{k-1}), y) =  \sum_{\mathbb{P}(k)} (-1)^{p+k} \mathcal{D}_{\beta,p} G_{\infty}(y) \prod_{q=1}^p K_{\beta}(y-y_{q,0}, \dots, y-y_{q,k_q-1}) \,,\, \label{eq:maingeneral0}\\
&\mathcal{D}_{\beta, p}  = -b^{-1}\partial_y(1-b^{-1}\partial_y)\dots ((p-1)-b^{-1}\partial_y) \,,\, b = \min(1, \beta) \,.  \label{eq:Dp0}
\end{align}
Some explanations are in order. First, the equality of \eqref{eq:maingeneral0} holds in the $M\rightarrow\infty$ limit. On the LHS, the notation $\sum\limits^{*}_{j_0\dots j_{k-1}} $ means that the sum is over distinct positions $j_0,\dots, j_{k-1}$. On the RHS, the sum $\sum_{\mathbb{P}(k)}$ is over all the \textit{partitions} of the set of $k$ first minima $\{y_0, \dots, y_{k-1}\} = \bigsqcup_{q=1}^p \{ y_{q,0}  <  \dots < y_{q, k_q-1} \}$; here $1 \leq p \leq k$ is the number of parts and $k_1, \dots, k_p$, with $\sum_{q=1}^p k_q=k$, are the sizes of the parts (see Sect. \ref{sec:derivationgeneral} for detailed introduction to partitions). The term corresponding to a partition of $p$ parts contains a product of $p$ \textit{UV correction factors}; they are an infinite family of functions $\left\lbrace K_\beta(\Delta_1, \dots, \Delta_l)\right\rbrace_{l = 1, 2, \dots}$, defined \textit{independently of $k$ in \eqref{eq:maingeneral0}}; their explicit definition will be given in \eqref{eq:smallw_general}, here it suffices to mention that they depend only on the UV data $f_{ij}$ as well as on the temperature $\beta$, and reduce to \eqref{eq:Koffd} when $l = 1$.

In the $\beta\rightarrow\infty$ limit \eqref{eq:maingeneral0} implies the following joint distribution of the ordered minima $V_{\min} = V_{\min,0} < V_{\min,1} < V_{\min,2} < \dots$ of the logREM (\textit{i.e.}, $V_{\min,k}$ is the $(k+1)$-st minimum):
\begin{align}
&\overline{\theta(V_{\min,k} - y)\prod_{s=0}^{k-1}\delta(V_{\min,s} - y_s) } = \sum_{\mathbb{P}(k)} \mathcal{D}_p G_{\infty}(y) \prod_{q=1}^p \left[e^{y_{q,0}-y} D( y - y_{q,0},\dots,  y - y_{q,k_q-1})\right] \,, \label{eq:pdfvalue0} \\
&D(\Delta_{0},\dots, \Delta_{\ell-1}) \stackrel{\text{def}}= \,\overline{\theta(v_{\min,\ell}-v_{\min,\ell-1} -\Delta_{\ell-1}) \prod_{s=0}^{\ell-2} \delta(v_{\min,s} + \Delta_s - v_{\min,\ell-1} - \Delta_{\ell-1})}
\nonumber \\ \stackrel{\text{def}}= \, & \frac{\overline{\theta(u_{\min,\ell}- u_{\min,\ell-1} -\Delta_{\ell-1}) \prod_{s=0}^{\ell-2} \delta(u_{\min,s} + \Delta_s - u_{\min,\ell-1} - \Delta_{\ell-1})\exp(-u_{\min})}}{\overline{\exp(-u_{\min})}} \label{eq:decoration0}
\end{align}
This equation is equivalent to saying that all the minima $V_{\min}, V_{\min,1}, \dots$ are generated by a randomly shifted decorated Gumbel Poisson point process (SDPPP), such that the random shift has the same distribution (modulo a translation) as the free energy at critical temperature $\beta = 1$, and the decoration process is given by the biased minima $v_{\min}, v_{\min,1}, \dots$. Their statistics are characterized in terms of minima of the local logREM $u_{\min}, u_{\min,1}, \dots$ by \eqref{eq:decoration0}, which is a straightforward generalization of \eqref{eq:gapbiased}. SDPPPs have recently been shown to describe the minima of several models related to Branching Brownian motion \cite{aidekon2013branching,arguin2013extremal}, and are expected to describe minima of general logREMs. For logREMs generated by discrete GFF this was very recently shown rigorously in \cite{biskup2016full}. Our analysis provides a replica-approach derivation of this result in great generality, and provides {an alternative} characterization of the decoration process. Furthermore, the results concerning the first and second minima will be tested numerically to high precision in the next Section \ref{sec:num}.

In Appendix \ref{sec:sdppp}, we generalize \eqref{eq:pdf2ndmin} to the marginal distribution of all higher order minima in terms of the following generating function:
\begin{align}
\sum_{k=0}^{\infty} \overline{\theta(V_{\min,k} - y)} t^k = \frac{1}{1-t} \exp\left(- \sum_{k=1}^{\infty} t^k \overline{g_k} \partial_y\right)G_\infty(y) =   \frac{1}{1-t} G_{\infty}\left(y -\sum_{k=1}^{\infty} t^k \overline{g_k}\right), \quad g_k = V_{\min,k} - V_{\min,k-1}  \,. \label{eq:marginalgenerating}
\end{align}
See Appendix \ref{sec:kthmin}. Similarly, a generalization of \eqref{eq:gapstat} is obtained for the marginal distribution of all gaps $g_k$ in terms of the biased minima process
\begin{align}
&\sum_{k=1}^{\infty} \overline{(g_k - \epsilon)_+} t^k = -\ln \left( 1 - \sum_{j=1}^{\infty}  \frac{t^j}{j!}  \overline{\left( \exp(v_{\min} - v_{\min,j-1} - \epsilon) - \exp(v_{\min} - v_{\min,j}) \right)_+}  \right)   \,,\, x_+ \defeq \max(x,0) \,.
\label{eq:gapgenerating0}
\end{align}
See Appendix \ref{sec:gapstatistics} for more discussion.

{Note that our main formulae (40)-(46) are expected to be
valid for any SDPPP, so that any of its possible characterizations (e.g. like one used in \cite{subag2015freezing}) should necessarily imply them. Nevertheless, to the best of our knowledge such expressions did not seem to appear in the mathematical literature in an explicit form.}

\section{Numerical study}\label{sec:num}
\begin{figure}
	\includegraphics[scale=.4]{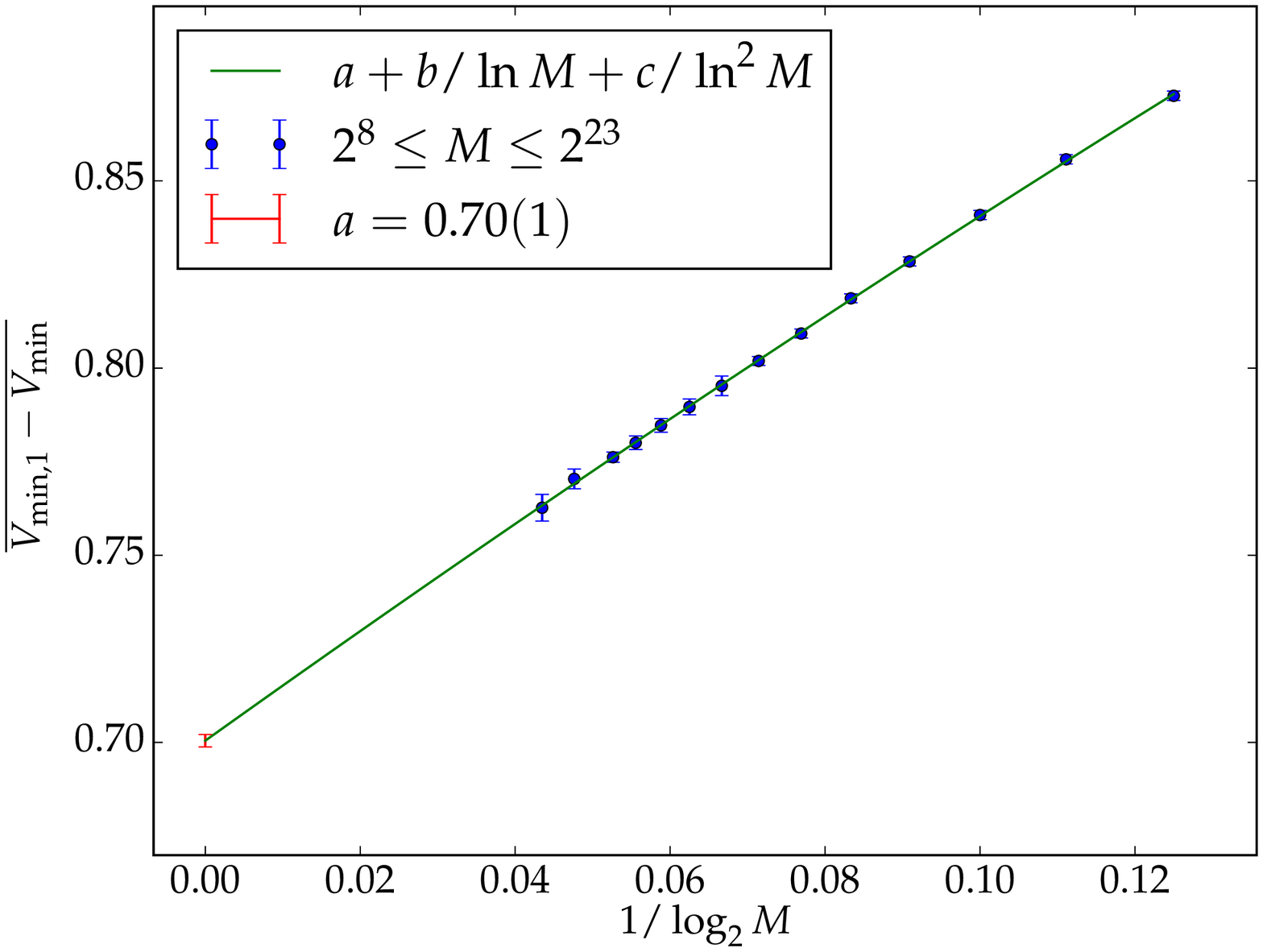}
	\includegraphics[scale=.4]{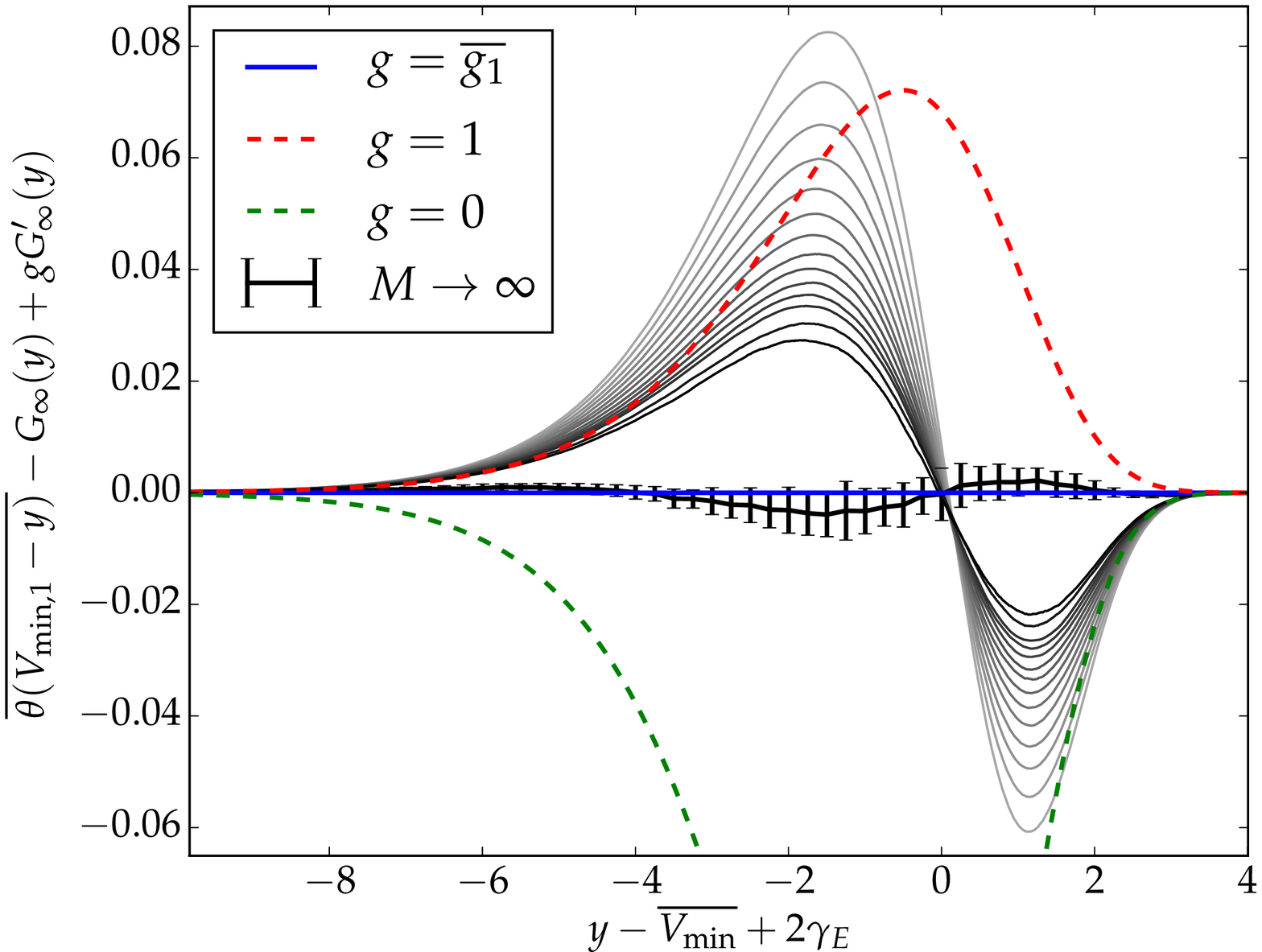}
	\caption{Left: the numerical measure of the mean of the first gap, as a function of the system size (points), is well described by a quadratic finite-size Ansatz $a + b / \ln M + c / \ln^2 M$. We use it to extract the $M \rightarrow \infty$ value $\overline{g_1} = a = 0.70(1)$. Right: The cumulative distribution function of the second minimum $V_{\min,1}$ of the circular $1/f$-noise model, with the theoretical prediction \eqref{eq:pdf2ndmin} subtracted, and the parameter $g = \overline{g_1}$ fed by the previous measurement. Grey curves are numerical data with system sizes $2^{8} \leq M \leq 2^{23}$, and the extrapolation to $M\rightarrow \infty$ (thick black curve with error bars) {is performed by applying the quadratic Ansatz pointwise.
 The error bars combine the error in the distribution with that in $\overline{g_1}$. For comparison we plot in dash lines \eqref{eq:pdf2ndmin} with other values of $g$.}}\label{fig:2ndmin}
\end{figure}

In this section, we test our predictions on the circular model defined in \eqref{eq:circledef}. Recall that its covariance matrix is cyclic: the Fourier mode $k$ has the eigenvalue ${\lambda_k} = \abs{k}^{-1}$ if $0 < \abs{k} \leq M/2$ while $\lambda_{k=0} = 0$. Therefore $\left(V_{j,M}\right)_{j=1}^{M}$ can be generated efficiently by the fast Fourier transform \cite{fyodorov2009statistical}.
{ Among all Euclidean-space logREMs this model is arguably the most amenable to both analytical and numerical analysis, and
is also singled out by its relevance in the random matrix context, see \cite{fyodorov2009statistical,fyodorov2015high}. However, despite intensive interest all previous studies focused only on its IR data, in particular on the distribution of the (shifted) global minimum derived originally in \cite{fyodorov08rem}:}

\begin{equation}
\overline{\theta(V_{\min} > y + y_M)} = G_{\beta > 1}(y + y_M) = 2e^{y/2} K_1(2e^{y/2}) \,.
\end{equation}
Here $K_n$ is the Bessel $K$-function and the unknown shift $y_M$ can be fixed by the average value: $y_M = \overline{V_{\min}} + 2 \gamma_E$ ($\gamma_E$ is the Euler's constant). {The expression \eqref{eq:pdf2ndmin} of the present paper allows us to predict the following explicit form of the distribution of the second deepest minimum:}
\begin{equation}
\overline{\theta(V_{\min,1} > y)} =  G_{\infty}(y) - g  G'_{\infty}(y) =
2e^{\tilde{y}/2} K_1(2e^{\tilde{y}/2}) + 2 g e^{\tilde{y}}K_0(2e^{\tilde{y}/2}) \,,\quad \tilde{y} = y - \overline{V_{\min}} - 2 \gamma_E \,,\, g = \overline{g_1} \,,\label{eq:prediction1}
\end{equation}
Let us test this prediction directly. For this we generate $S$ independent samples of the circular model of size $M = 2^8,  \dots, 2^{23}$, with $S = 10^7$ for $M \leq 2^{15}$ and $S \geq 10^6$ for $M \leq 2^{20}$ and $S \geq 10^{5}$ for even larger $M$. We first observe, in Fig. \ref{fig:2ndmin} (Left panel) that the mean value of the gap $ \overline{V_{\min,1}} - \overline{V_{\min}}$ has a strong $M$ dependence which should be taken into account to extract the thermodynamic limit value $\overline{g_1}$. A good numerical fit is provided by a quadratic function of $1/\ln M$ (the same Ansatz is known \cite{cao16maxmin} to work in other observables of this model). The result is $\overline{g_1} = 0.70(1)$, and is fed into the RHS of \eqref{eq:prediction1}. On the other hand, we {collect} the full distributions of the shifted second minima $V_{\min,1} -\overline{V_{\min}} + 2 \gamma_E$ where $\overline{V_{\min}}$ is the numerically measured mean value of the minimum of the same system size. The results, with the prediction \eqref{eq:prediction1} subtracted, and shown in Fig. \ref{fig:2ndmin} (Right). We observe again a slow convergence; nevertheless, the extrapolation to $M \rightarrow\infty$ with the same quadratic $1/\ln M$ Ansatz has an excellent agreement with the prediction \eqref{eq:prediction1}. To compare, we plotted also the RHS of \eqref{eq:prediction1}, with two other values $g = 0$ and $g = 1$; it is clear that both would give wrong predictions.

\begin{figure}
	\includegraphics[scale = .4]{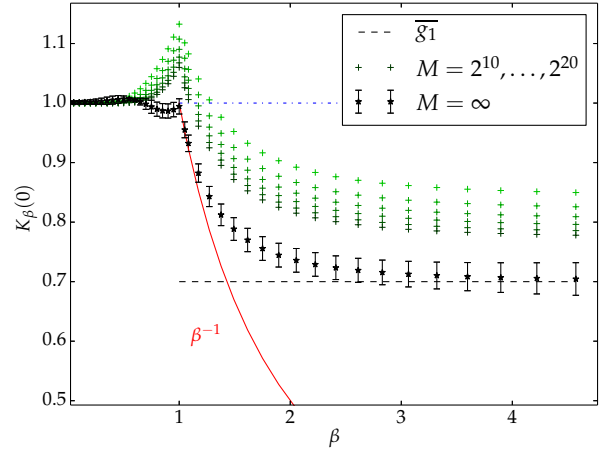}
	\caption{Numerical measurement of the right hand side of \eqref{eq:Kbzero} in the circular model, predicted to be equal to the UV factor $K_\beta(0)$. Finite-size system ($M = 2^{10}, \dots, 2^{20}$) raw data are plotted in green $+$'s, {with darker colour corresponding to larger system sizes}. The extrapolation to $M = +\infty$ value is performed by the $a + b /\ln M + c / \ln^2 M$ Ansatz. The naïve Ansatz in $\beta > 1$ phase ($K_\beta(0) = 1$),  {an (incorrect) “analytical continuation” $1/\beta$ and the numerical value of $\overline{g_1} = 0.7$ (Fig. \ref{fig:2ndmin}) are indicated by blue pointed, red solid, and black dashed lines, respectively.}  }\label{fig:kb0}
\end{figure}

Next, we focus on the UV sector of our theory, and study the \textit{temperature dependence} of the factor $K_\beta(\Delta)$. We focus on the cases $\Delta = 0$ for simplicity. To probe it numerically, we shall use the following equation, which we shall derive in Sect. \ref{sec:kbzeroder}:
\begin{equation}
 K_\beta(0) = \min(1,1/\beta) \sum_j (\overline{\ln \mathcal{Z}} -  \overline{\ln \mathcal{Z}_{\setminus j}}) \,,\quad \mathcal{Z} = \sum_k e^{-\beta V_k} \,,\, \mathcal{Z}_{\setminus j} = \sum_{k \neq j} e^{-\beta V_k} \,. \label{eq:Kbzero}
\end{equation}
Now the right hand side can be directly measured numerically. We do this for $40$ different values of $\beta = 0.05, \dots, \approxeq 4.5$, each in various system sizes $M = 2^{10}, \dots, 2^{20}$ in order to extrapolate to thermodynamic limit using a quadratic $1/\ln M$ Ansatz similarly to the procedure described above. The result is shown in Fig. \ref{fig:kb0}. In the $\beta < 1$ phase, we have the analytical prediction $K_\beta(0) = 1$, \eqref{eq:KhighT}. In that phase, away from $\beta = 1$, the numerics agrees perfectly with the prediction, with invisible finite-size effects. Finite-size effects become more pronounced as $\beta \gtrsim 0.8$, and have an intriguing non-monotone behaviour around $\beta = 1$. In the $\beta > 1$ phase the naïve freezing $K_{\beta>1}(0) = 1$ is clearly ruled out, and so is the ``analytical continuation'' $K_{\beta>1}(0) = 1/\beta$, which accounts for the explicit $\min(1,1/\beta)$ non-analyticity in \eqref{eq:Kbzero}. The $\beta$-dependence of $K_\beta(0)$ in the $\beta > 1$ phase is non-trivial and we do not have an educated guess to fit the data. To check consistency we look at the $\beta \rightarrow \infty$ limit; there, \eqref{eq:gapstat} predicts that $K_\infty(0) = \overline{g_1}$, the latter being the mean value of the first gap, which we measure numerically in Fig. \ref{fig:2ndmin} (Left panel) to be $ \overline{g_1} \approxeq 0.7$, and indicated in Fig. \ref{fig:kb0} as the black horizontal line. As expected, we observe that the latter coincides with the $\beta\rightarrow \infty$ asymptotic value of the extrapolated $K_\beta(0)$ data.

\begin{figure}
	\includegraphics[scale=.4]{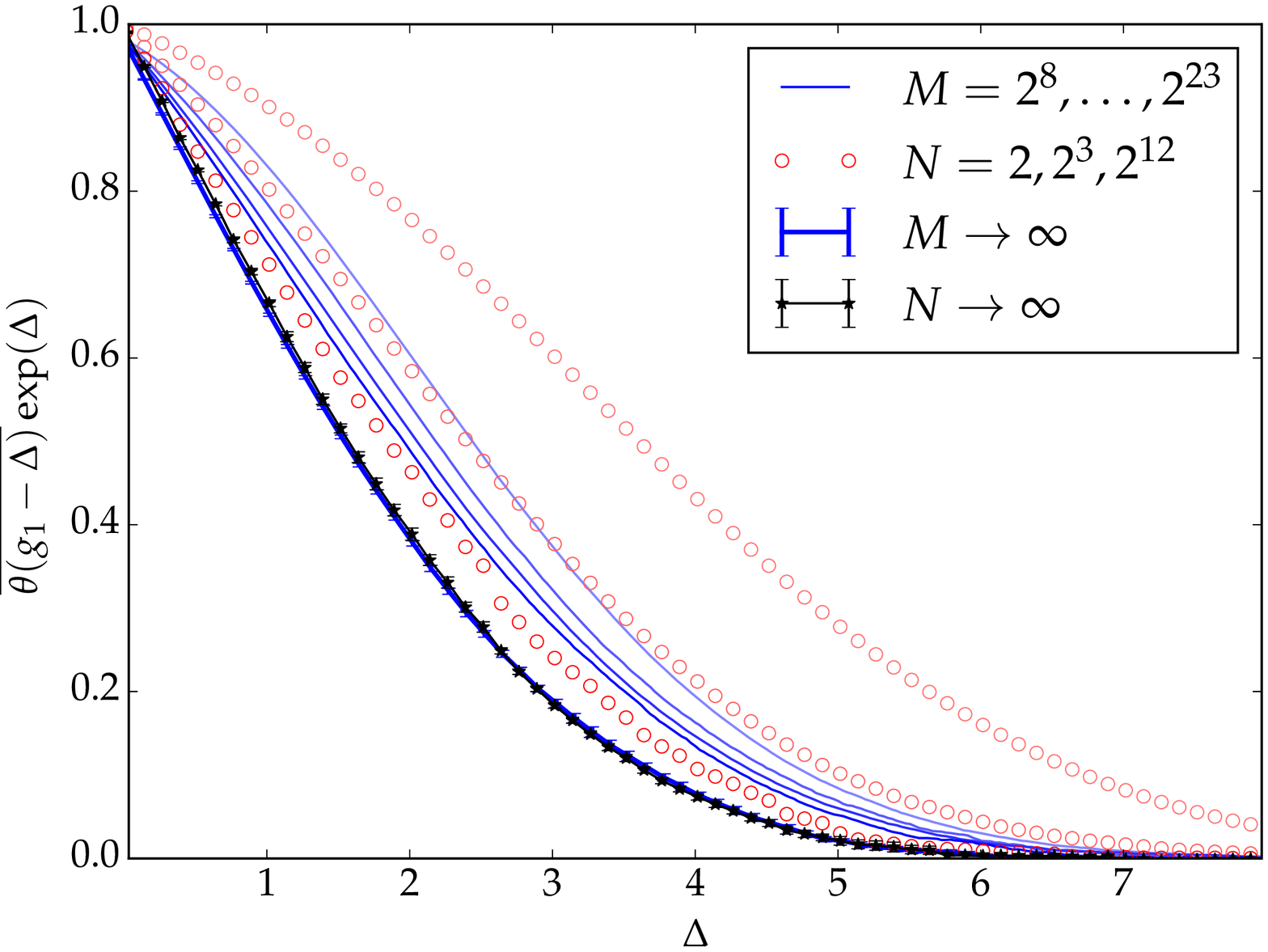}
	\includegraphics[scale=.4]{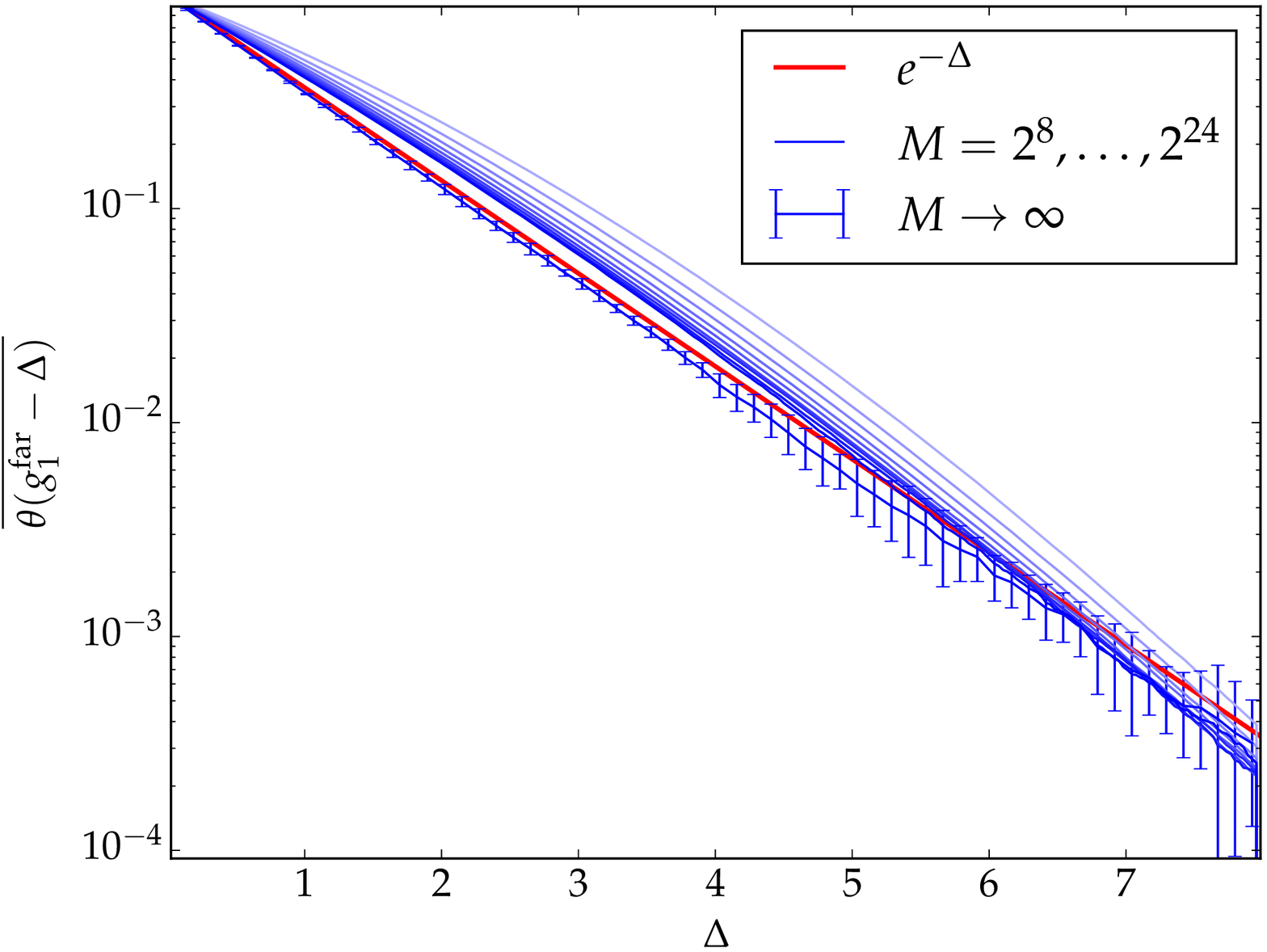}
	\caption{Left: Global \textit{vs} local first gap distribution in the circular $1/f$-noise model. The global gaps are measured directly from systems of sizes $M = 2^{8}, 2^{13}, 2^{18}, 2^{23}$, and extrapolated point wise by the $1/\ln M$-quadratic finite size Ansatz. The results (blue curves) are multiplied by $\exp(\Delta)$ (note that an exponential distribution would give a horizontal line at height $1$). The first gap of the biased process is measured by \eqref{eq:gapbiased}, in local logREMs of sizes $2\leq N \leq 2^{12}$ (red circles), and extrapolated using the same Ansatz to $N \rightarrow \infty$ (black circles). The good agreement between the latter curve (\textit{without} multiplying by $\exp{\Delta}$) and the $M \rightarrow\infty$ curve confirms the prediction \eqref{eq:gapstat1}. Right: the far-gap distribution $g_{1}^\text{far}$, defined in \eqref{eq:fargap}, with $N = \sqrt{M}$, compares well to the exponential, \eqref{eq:exponential} .}\label{fig:gapstats}
\end{figure}

In the $\beta > 1$ phase, $K_\beta(\Delta)$ is determined by the UV data of the model (given in \eqref{eq:fijlimitFB}) \textit{via} the local logREM, see eqs. \eqref{eq:localrem} and \eqref{eq:Koffd}. We now test these predictions in the zero temperature $\beta \rightarrow  \infty$ limit, where the gap distribution is related to the biased gap defined in \eqref{eq:gapbiased}, see eq. \eqref{eq:gapstat1}. Thanks to the translation-invariant structure of the covariance, the local logREM \eqref{eq:localrem} of the circular model can also be simulated efficiently using the fast Fourier transform. We do this for $N = 2^{1}, \dots, 2^{12}$, with $10^8$ independent samples for each size. We measure, using \eqref{eq:gapbiased}, the cumulative distribution function of the biased gap $\overline{\theta(v_{\min,1} - v_{\min} - \Delta)}$, and extrapolate to the $N\rightarrow\infty$ limit using the quadratic $1/\ln N$ Ansatz. We then compare this with $\overline{\theta(V_{\min,1} - V_{\min} - \Delta) e^{\Delta}}$  measured independently in {the original circular model }, and extrapolated to $M \rightarrow\infty$ using the same Ansatz in { $1/\log{M}$}. The result, Fig.\ref{fig:gapstats} (Left), shows an excellent agreement between the two limit distributions, validating the prediction \eqref{eq:gapstat1}. We remark that the finite size corrections of the local logREM are much smaller than those of the full model: even $N = 8$ provides an approximation as decent as $M = 256$. On the other hand, the measure of the biased minima \eqref{eq:gapbiased} is {sensitive to contributions of rare events,  and requires to collect better statistics to be reliably verified}.

{Despite being unable to deal} with non-trivial UV data analytically, we can get rid of them by defining the second minimum in a different way, as \begin{equation} V^{\text{far}}_{\min,1} \stackrel{\text{def}}= \, \min(V_{j,M}, \abs{j - j_{\min}} > N/2)  \,,\quad g_{1}^{\text{far}} \stackrel{\text{def}} = V^{\text{far}}_{\min,1} - V_{\min} \,, \label{eq:fargap} \end{equation}
where $j_{\min}$ the position of the minimum, whereas both $N\to \infty$ and $M\to \infty$ keeping $N\ll M$ in the thermodynamic limit; numerically, $N = \sqrt{M}$ is used. The idea is that when looking for the second minimum located far from the global minimum the UV data trivialize and one retrieves the exponential gap distribution in the thermodynamic limit:
\begin{equation}
\overline{\theta(g_{1}^{\text{far}} - \Delta)} \rightarrow \exp(-\Delta)\,, \quad M\rightarrow\infty\,, \label{eq:exponential}
\end{equation}
which is the case for the random energy model with uncorrelated potential. This prediction is well verified by the numerical data, which are shown in Fig. \ref{fig:gapstats} (Right). A derivation of this result within the 1RSB framework will be outlined in Sect. \ref{sec:remote2ndmin}.

\section{First and second minima}\label{sec:minimum}
In this section we will derive the main results concerning the first and second minima value distribution in Sect. \ref{sec:mainresult}. First we are going to give an outline of the essential ingredients of the 1RSB approach to logREM models, with calculation of the minimum value distribution presented as a good pedagogical example. Computations of the higher order statistics involve natural, albeit non-trivial, ramifications of that procedure.

As a starting point for our analysis we consider a finite temperature generating function \eqref{eq:defG} related to the global minimum value distribution:
\begin{equation}
G_\beta(y) = \overline{\exp(- e^{\beta y} \mathcal{Z} )} \quad
\Leftrightarrow \quad \int_\R  -\partial_y G_\beta(y) e^{t y} \dif y = \overline{\mathcal{Z}^{-t / \beta}} \Gamma(1 + t / \beta)  \,, \, \label{eq:laplace}
\end{equation}
where $\mathcal{Z}=\sum_{j=1}^M \exp(- \beta V_{j,M})$ is the standard partition function of the logREM as defined in \eqref{eq:defdiscretepartition}.  In the replica approach, it is also convenient to present \eqref{eq:laplace} as a \textit{formal} power series:
\begin{equation}
G_\beta(y) = \sum_{n=0}^{\infty} \frac{(-1)^n }{n!} e^{n\beta y} \overline{\mathcal{Z}^n} \,. \label{eq:Ggen}
\end{equation}
A direct consequence of the representation \eqref{eq:Ggen} which will be used later on is that differentiation on the LHS corresponds to multiplication on the RHS:
\begin{equation}
\partial_y G_\beta(y) = \sum_{n=0}^\infty \frac{(-1)^n }{n!} e^{n\beta y} n \beta  \overline{\mathcal{Z}^n} \,. \label{eq:laplaceder}
\end{equation}

\subsection{Overview of the 1RSB Ansatz}\label{sec:1rsb}
\begin{figure}
	\includegraphics[scale=.3]{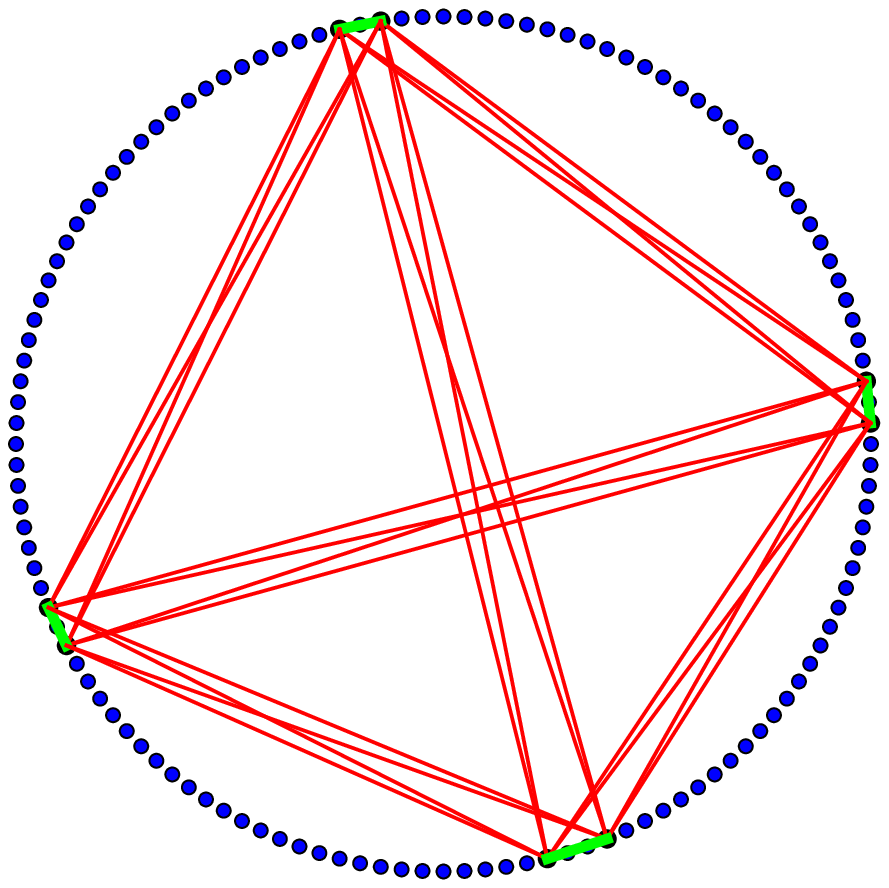}	\includegraphics[scale=.3]{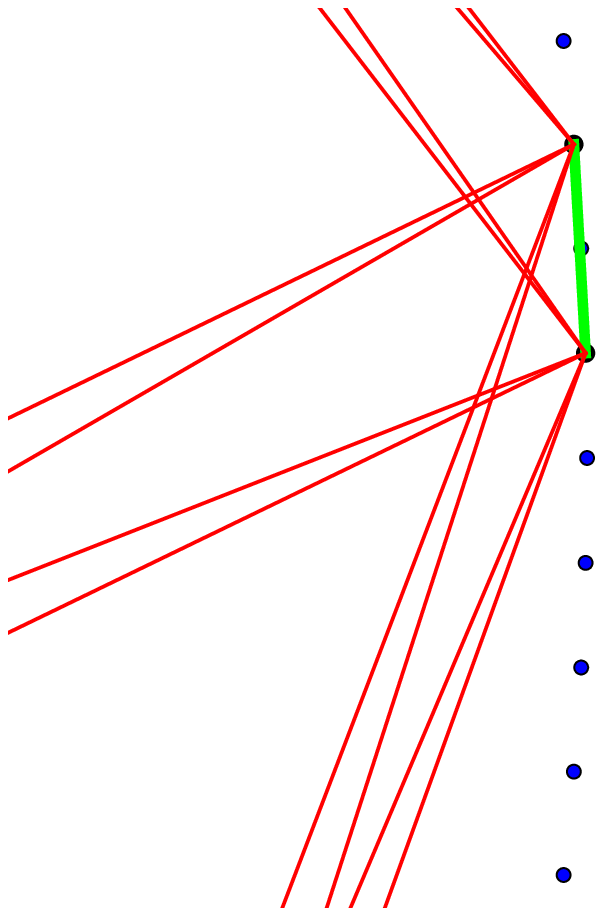}
	\caption{An illustration of the 1RSB Ansatz with $n = 8$ replicas and group size $m = 2$, and for the case of a circular model (compare with fig. \ref{fig:uvir}). The red (green) lines represent Wick contractions between replicas of different (same) group, called inter-group (intra-group, respectively) interactions in the text. The right figure is merely a zoom of the left one into the UV scale. }  \label{fig:1rsb}
\end{figure}
$\overline{\mathcal{Z}^n}$ is a sum over positions of $n$ replicas, which ``interact'' via disorder. Here, the interaction is pair-wise as a result of the Wick theorem:
\begin{align}
\overline{\mathcal{Z}^n} = \sum_{j_0, \dots, j_n = 1}^M \prod_{a=1}^{n} \exp(-\beta \overline{V_{j_a}}) \prod_{a,b=1}^n \exp(\beta^2 \overline{V_{j_a} V_{j_b}}^c / 2) \,.  \label{eq:Wick}
\end{align}
This is the prototype of replica averages to which replica symmetry breaking (RSB) Ansätze are applied. For logREMs (as well as uncorrelated REMs), it is known that a \textit{one-step} RSB (1RSB) Ansatz is correct. For hierarchical logREMs, this fact was first realized \cite{derrida1988polymers} using KPP equation. For Euclidean-space logREMs, 1RSB was argued to hold in \cite{carpentier2001glass} (Sect. III.E.2), by building upon a functional renormalisation group analysis, from which KPP-type equations emerge again, as flow equations. In Appendix \ref{sec:frsb}, we provide a \textit{full} replica symmetry breaking analysis of logREMs, which show  1RSB results from the general definition \eqref{eq:logremdef}. In the main text, we will take for granted the 1RSB Ansatz as the starting point of our derivations. The Ansatz claims the following (such a description of 1RSB, {which is a ramification of the standard one}, first appeared in \cite{fyodorov2010freezing}):

In the thermodynamic limit the replica sums such as \eqref{eq:Wick} are dominated by configurations such as depicted in Fig. \ref{fig:1rsb}. More precisely, the $n$ replicas form $n/m$ groups of equal size $m$. Within each individual group of size $m$ different replicas should be thought of as being within the mutual distances of the order of lattice spacing, which is the definition of the UV scale; therefore, their interaction should be calculated using \eqref{eq:intra}. On the other hand, distinct groups are separated typically by IR (system-size) scale, so one can treat the sum over group positions as continuum integral, and calculate inter-group interactions using IR data. Now, the group size $m$ is a non-analytical function of $\beta$:
\begin{equation}
m =  \min(1, \beta^{-1}) \,. \label{eq:moptimum}
\end{equation}
The unique transition at $\beta = 1$ \eqref{eq:betac} corresponds to the non-analyticity
of \eqref{eq:moptimum}. When $\beta > 1$ one has $m = 1/\beta < 1$ which is the standard, though odd, feature of the replica limit $n\to 0$ which forces one to consider all integer parameters as promoted to variables which can take in general any real values (strictly speaking, continuation to the complex plane is necessary, \textit{e.g.}, when performing inverse Laplace transform). We shall see how the above applies to the replica sums $\overline{\mathcal{Z}^n}$ in the next subsection.

We argue that the same method applies for partition sums {used to address} higher order statistics. Indeed, the object in question here is a discrete partition function of a particle in a potential $(U_j)$, obtained from that of a (log)REM by shifting a finite number of values:
\begin{equation}   \mathcal{W}((j_1, \Delta_1), \dots, (j_{k-1}, \Delta_{k-1})) \defeq \, \sum_{j=1}^M \exp(-\beta U_j)\,,\, U_j =  V_j + \sum_{i=0}^{k-1} \delta_{j,j_i} \Delta_i \,. \label{eq:Zshifted} \end{equation}
Here $\delta$ is the Kronecker delta, and $(j_1,\Delta_1), \dots, (j_{k-1}, \Delta_{k-1})$ are the positions and values of the shifts. The case $k = 0$ is the usual partition function $\mathcal{Z}$ \eqref{eq:Ggen};  we shall see in sect. \ref{sec:derivation} the partition function relevant for second minimum $\mathcal{W}(j,\Delta)$ \eqref{eq:HGgen} corresponds to $k = 1$; higher values of $k$ are needed for the full minima process, see \eqref{eq:bigWdef} below. RSB Ansätzse in general are assertions about which configurations of replica positions $j_1, \dots, j_n$ dominate $\overline{\mathcal{W}^n}$ in the thermodynamic $M \rightarrow \infty$ limit. As a consequence, the outcome of the approach depends only on the thermodynamics (that is, quantities proportional to $\ln M$) of the model, and is not altered by a finite ($O(1)$ as $M\rightarrow\infty$) energy shift performed in \eqref{eq:Zshifted}.



\subsection{1RSB for minimum}\label{sec:1rsbmin}
The general idea of the 1RSB calculation of $\overline{\mathcal{Z}^n}$ as described in Sect. \ref{sec:1rsb} has been briefly outlined
in \cite{fyodorov2010freezing}, and we describe it in detail below for the sake of clarity and completeness. For this, we divide the system into \textit{blocks} of $N$ sites, where $N$ is any intermediate scale such that $N\ll M$, whereas considering that both $N\to \infty$ and $M\to \infty$  in the thermodynamic limit. Each block will be labelled by their position in $\C$ (because the $\xi_{j,M}\in\C$ of a same block will converge to the same point); positions inside a block labelled $1, \dots, N$; in other words, we have a one-to-one correspondence $j(\xi, i)$ between the global address and the hierarchical one. Now, $n$ replicas form $n/m$ groups of size $m$; when $n$ is continued to complex values we will have $m = \max(1, 1/\beta)$ by eq. \eqref{eq:moptimum}. Let us index the groups by $g = 1, \dots, n/m$ and let $g(a), a = 1, \dots, n$ be the group that replica $a$ belongs to. Due to the permutation symmetry in replica space we need only consider, for instance, a particular $g(a) = \lceil a / m \rceil$ ($ \lceil x \rceil$ standing for the smallest integer larger or equal to $x$)  and multiply by the combinatorial factor counting the number of different groupings \footnote{Note that the $1 / \Gamma(1 + n/m)$ is included so the groups are treated as distinct.}
\begin{equation}
C_{n,m} = \frac{\Gamma(1 + n)}{\Gamma(1 + m)^{n/m} \Gamma(1 + n/m)} \,. \label{eq:Cnm}
\end{equation}
Each group is confined in a block with coordinates concentrated at UV distances around a macroscopic position $\xi_g$. Different groups are separated by a distance of system-size order so that the sum over block-positions can be approximated by a continuous integral:
\begin{equation} \sum_{\xi_1, \dots \xi_{n/m}} \dots \rightarrow \left(\frac{M}{N}\right)^{\frac{n}{m}} \int_{\xi_g \in X} \prod_{g} \left[ \abs{\dif^d \xi_g} \exp(-l(\xi_g)) \right] \dots \,. \label{eq:sumtoint} \end{equation}
Compared with \eqref{eq:irmeasure}, we have $M/N$ factors on the RHS because the LHS sums over block positions and there are $M/N$ blocks. The continuum limits also apply to mean values $\overline{V_j} \rightarrow c(\xi)$, as well as to inter-group Wick contractions $\overline{V_{j_a} V_{j_b}}^c \rightarrow C(\xi_{g(a)}, \xi_{g(b)})$ provided $g(a) \neq g(b)$. On the other hand, the \textit{intra}-group Wick contractions (depicted in Fig. \ref{fig:1rsb} by violet lines in the zoomed-in version) should be treated using the UV limit data \eqref{eq:intra}. We will organize them into intra-group interactions, see \eqref{eq:Exi} below; note however that the factors resulting from the shifts by $2l(\xi)$ in the LHS of \eqref{eq:intra} will be systematically combined with
other instances of $l(\xi)$ appearing in \eqref{eq:sumtoint}. Carrying out the above treatment,
we reduce \eqref{eq:Wick} in the $M \rightarrow \infty$ limit to the following expression:
\begin{align}
\overline{\mathcal{Z}^n} \, \rightarrow \,\, &  C_{n,m} \left(\frac{M}{N}\right)^{\frac{n}{m}}\int_{X} \prod_{g = 1}^{n/m} \left[ \abs{\dif^d \xi_g}  \exp(-m \beta c(\xi_g) - (1 + m^2 \beta^2)l(\xi_g)) \right] \prod_{g < g'} \exp(\beta^2 m^2 C(\xi_g, \xi_{g'})) \times \prod_{g=1}^{n/m} E(\xi_g) \,, \label{eq:Zn}  \\
E(\xi) &=  \sum_{i_1, \dots, i_m = 1}^N  \prod_{l,l' = 1}^m \exp(\beta^2 \overline{V_{j(\xi, i_l)}
	V_{j(\xi, i_{l'})}}^c / 2) =  \overline{z_\beta^m} \exp(\beta^2 m^2 (2\ln M -C_N) / 2) \,. \label{eq:Exi}
\end{align}
Here $E(\xi)$ is the intra-group interaction of the group at block $\xi$; we note that the bar in $\overline{z_\beta^m}$  denotes averaged moments of the local-logREM partition function $z_{\beta}$ defined in  \eqref{eq:smallw}.
Note that to alleviate formulae like \eqref{eq:Exi} and below we will tacitly assume that the limit $N\rightarrow\infty$ is taken
along with that of $M\rightarrow\infty$ in the order explained above. The expression \eqref{eq:Zn} can be rearranged in terms of the continuous partition sum defined in \eqref{eq:contZ} with parameters $b \to m \beta$ and $n \to n/m$
\begin{equation}
\overline{\mathcal{Z}^n} \rightarrow M^{n(1 + m^2\beta^2)/m} C_{n,m} \overline{Z_{\beta m}^{n / m}} \lim_{N\rightarrow\infty} \left( \overline{z_\beta^m} N^{-1}\exp(-\beta^2 m^2 C_N / 2) \right)^{n/m} \,. \label{eq:Zncont}
\end{equation}
Here $n$ should be understood as continued to a complex-valued variable, so that \eqref{eq:Zncont} yields the Laplace transform:
\begin{eqnarray}
&\overline{\mathcal{Z}^n} = \overline{\exp(t \mathcal{F})}\,,\quad  \mathcal{F} \stackrel{\text{def}}= -\beta^{-1}\ln \mathcal{Z} \,,\, t = - n\beta \,. \label{eq:etfdef} \\
\stackrel{\eqref{eq:Zncont}}{\Rightarrow}&\overline{\exp(t \mathcal{F})} =  \exp(-t\ln M (b + b^{-1})) C_{-\frac{t}{\beta},m} \overline{Z_{b}^{-\frac{t}{b}}} e^{t S_{[\text{UV}]}} \label{eq:Etf} \,,\quad b \defeq m \beta \,,\, \\
\text{where }& S_{[\text{UV}]} \defeq \, -b^{-1} \ln \left[ \lim_{N\rightarrow\infty}  \overline{z_\beta^m} N^{-1}\exp(-\beta^2 m^2 C_N / 2) \right]\,.
\end{eqnarray}
Eq. \eqref{eq:Etf} is the general expression for the free energy distribution (in terms of the Laplace transform). The free energy $\mathcal{F}$ has a universal extensive shift $-\ln M (b + b^{-1})$, an $O(1)$ fluctuations which depend on the IR data through the Coulomb gas integral  $\overline{Z_{b}^{- t/b}}$ and a shift $ S_{[\text{UV}]}$ which depends on $\beta$ and the UV data.

In the \textit{high temperature phase}, $\beta < 1$, by \eqref{eq:moptimum}, $m = 1$ and $b = \beta$, so $C_{n,m} = 1$ by \eqref{eq:Cnm} and $\overline{z_\beta^m} = N \exp(\beta^2 (C_N + f_{ii}) / 2)$ by \eqref{eq:smallw} and \eqref{eq:localrem}. Plugging this into \eqref{eq:Etf}, we have
\begin{equation}
\overline{\exp(t \mathcal{F})} = e^{t F(M,\beta,f_{ij})}  \overline{Z_{\beta}^{-t/\beta}} \,,\quad F(M,\beta,f_{ij}) = -\left(\beta + \beta^{-1}\right) \ln M - \beta f_{ii} / 2 \,, \label{eq:etfhighT}
\end{equation}
which is equivalent to the $\beta < 1$ case of \eqref{eq:Gfreezing} and \eqref{eq:FM} via the Laplace transform \eqref{eq:laplace}.

The $\beta > 1$ case is more involved and more interesting, because it is directly related to the freezing scenario; we will discuss it in Sect. \ref{sec:freezing}. Sect. \ref{sec:FDC} compares the current approach to the freezing and duality conjecture. The goal of those sections is to give some simple relations and arguments which, to our knowledge, have not been clearly spelt out, and which clarify the respective rôle of 1RSB and the duality invariance property in the freezing phenomenon.

\subsubsection{Freezing scenario}  \label{sec:freezing}
In this subsection, we relate the result \eqref{eq:Zncont} to the freezing scenario \eqref{eq:Gfreezing} in the $\beta > 1$ phase. Our arguments show quite generally how the 1RSB leads to the freezing scenario.

Throughout this subsection we shall consider the low temperature phase $\beta>1$. In that case \eqref{eq:moptimum} gives $m = 1/\beta$, and the continuum integral factor in \eqref{eq:Zncont}, $\overline{Z_{m\beta}^{n\beta}}$ is a Coulomb gas integral \eqref{eq:contZ}
 over $n\beta = -t$  charges (see \eqref{eq:etfdef}), and with the renormalised temperature $b = m\beta = 1$. Yet we recall from \eqref{eq:partialZcont} that $Z_b$ is ill-behaved at $b=1$ and we shall consider $\overline{\widetilde{Z}^{-t}}$ instead. To this end we let $b = 1 - \epsilon$ where $0<\epsilon$ is infinitesimal, so that  $\overline{Z_{b}^{-t}} = \overline{\widetilde{Z}^{-t}} (1 - b)^{-t}$ by \eqref{eq:partialZcont}, giving
\begin{equation} \overline{Z_{m\beta}^{n \beta}} \rightarrow \overline{\widetilde{Z}^{-t}} \epsilon^{-t} \,.   \label{eq:threehalf}\end{equation}
In the RHS, the factor $ \overline{\widetilde{Z}^{-t}}$ accumulates all the IR-data dependent contributions to the free energy. The factor $\epsilon^{-t} = \exp(t \ln (1/\epsilon))$ corresponds to a shift of $ -\ln \epsilon \rightarrow +\infty $ in the free energy.
For finite system sizes $M$ this divergence is regularized and as argued in \cite{fyodorov2015high} should give rise to the $\ln\ln M$ corrections in \eqref{eq:FM}.

Let us now discuss the problem of analytical continuation  of the combinatorial factor $C_{n,m}$
defined {\it a priori} only when $n,m$ and $n/m$ are all natural numbers. To choose the appropriate continuation
 we use the reflection formula for $\Gamma$-functions and, denoting  $s(x) = \sin\pi x$, transform \eqref{eq:Cnm} as follows:
\begin{align}
&C_{n,m} = \frac{\Gamma(1 + n)}{\Gamma(1 + m)^{\frac{n}{m}}\Gamma\left(1 + \frac{n}{m}\right)} = \frac{ \Gamma\left(1 - \frac{n}{m}\right)}{\Gamma(1 + m)^{\frac{n}{m}} \Gamma(1 - n)}
\frac{m s\left(\frac{n}{m}\right)}{s(n)} = \frac{\Gamma\left(1 - \frac{n}{m}\right)}{\Gamma(1+m)^{\frac{n}{m}} \Gamma(1 - n)} \label{eq:comb_cont}
\end{align}
In the last equality we discarded the fraction $\frac{m s\left(\frac{n}{m}\right)}{s(n)}$. This is justifiable because for a fixed $m$, $\lim_{n\rightarrow n_0} \frac{m s\left(\frac{n}{m}\right)}{s(n)} = 1$, provided $n_0 / m$ is an even integer or $m$ is an odd integer. We claim that the last form of \eqref{eq:comb_cont} is the correct continuation of the combinatorial factor. To see this, we plug \eqref{eq:comb_cont} back to \eqref{eq:Zncont} (and substitute $m = 1/\beta$ for the $\beta > 1$ phase), and separate terms of deterministic shift and fluctuation types:
\begin{align}
&\overline{\mathcal{Z}^n} = \overline{\widetilde{Z}^{n\beta}} \frac{\Gamma\left(1 - n\beta \right)}{\Gamma(1 - n)} \left( \mathbf{Z}_0 \right)^n \,,\quad
\mathbf{Z}_0 \defeq \, M^{2 \beta} \epsilon^{\beta} \Gamma(1 + \beta^{-1})^{\beta} N^{-\beta} \left(\overline{z_\beta^{1/\beta}}\right)^{\beta} \exp(-\beta C_N / 2)  \,, \label{eq:Znfreezing} \\
\Rightarrow \,\, &\overline{\exp(t\mathcal{F})} \Gamma(1 + t / \beta) \exp( t \beta^{-1}\ln \mathbf{Z}_0)  = \overline{\widetilde{Z}^{-t}} \Gamma\left(1 + t \right) \,,\quad \beta > 1 \,, \label{eq:freezing}
\end{align}
see also eqs \eqref{eq:threehalf} and \eqref{eq:etfdef} and \footnote{Note that this correct analytical continuation leads to the factor $\Gamma(1+t/\beta)$ in the l.h.s. of
\eqref{eq:freezing}, which is the Laplace transform of the PDF of $- G/\beta$, where $G$ is a standard Gumbel random variable.}
.  $\mathbf{Z}_0$ gathers all the terms contributing to the linear shift $-\beta^{-1} \ln \mathbf{Z}_0$ of $\mathcal{F}$. To justify the continuation \eqref{eq:comb_cont}, note that the RHS of eq. \eqref{eq:freezing} is positive and log-convex (in the interval containing $t = 0$ where it has no poles) if $\overline{\widetilde{Z}^t}$ is; had we chosen the naïve continuation of \eqref{eq:Cnm}, this would no longer be the case since we would have instead " $\overline{\exp(t \mathcal{F})} = \frac{\Gamma(1 - t \beta^{-1})}{\Gamma(1 - t)} \overline{\widetilde{Z}^{-t}}  e^{t c} \,\, ^{''}$ .

Now, it is easy to see that the prediction for the fluctuating part of the free energy $\mathcal{F}$ encoded in \eqref{eq:freezing}, which was obtained solely from arguments based on 1RSB and analytical continuations, is equivalent to the \textit{freezing scenario} \eqref{eq:Gfreezing}: Indeed, the RHS of \eqref{eq:Gfreezing} (for $\beta > 1$) and \eqref{eq:freezing} coincide. By \eqref{eq:laplace}, the LHS side of \eqref{eq:freezing} equals
$ e^{t \beta^{-1}\ln \mathbf{Z}_0 }\int_\R -\partial_y G_\beta(y) e^{ty} \dif y$, which is just \eqref{eq:Gfreezing}, provided we make the identification of $F(M,\beta>1,f_{ij})$ in \eqref{eq:FM} with the unknown and formally divergent constant in \eqref{eq:Znfreezing}:
\begin{equation} -\beta^{-1}\ln \mathbf{Z}_0 = F(M,\beta,f_{ij}) \,,\quad \beta > 1 \,. \label{eq:FM1}\end{equation}
Unfortunately, the present level of development  of the replica formalism is insufficient for determining this object to $O(1)$ precision in the $\beta > 1$ phase, because the relation between the formal divergence in \eqref{eq:Znfreezing} and the $\ln\ln M$ corrections of \eqref{eq:FM} is  not yet well understood with full quantitative control. More generally, size-dependent corrections of logREM are difficult to get access to analytically, especially using replicas; however, for simple versions of REM there is a recent progress using alternative approach \cite{derrida16kppfinitesize}.

\subsubsection{Freezing-duality conjecture}\label{sec:FDC}
This subsection discusses the freezing-duality conjecture (FDC), from a 1RSB viewpoint. We will show that duality invariance extends the optimization of the 1RSB parameter $m$ (which is the mechanism behind the formula \eqref{eq:moptimum}) to the level of fluctuations, and demonstrate that predictions of FDC so far are fully compatible with 1RSB. Our discussion will be quite general; for a non-trivial concrete example, we reproduce and substantially generalize the calculation of the Edwards-Anderson's order parameter of the circular model \cite{cao16maxmin} using 1RSB in Appendix \ref{sec:EA}.

Let us first recall the essence of FDC. To motivate it, consider the extensive free energy of logREMs, given by \eqref{eq:FM}. In the $\beta < 1$ phase, the free energy density $f(\beta) \defeq \, \lim_{M\rightarrow\infty}  F(M,\beta,f_{ij}) / \ln M =  -(\beta + \beta^{-1})$,  which shows that as a function of $\beta$, $f(\beta)$ depends on $\beta$ only through the combination $\beta + \beta^{-1}$.  This is usually called the duality invariance property; the name refers to the $\beta \rightarrow \beta^{-1}$ duality transform, but we stress that it concerns only the high temperature phase. In the $\beta > 1$ phase, $f(\beta) = -2$ as a manifestation of freezing.  Freezing can be seen as a consequence of duality, as the latter implies $\beta = 1$ is a stationary point of $f$: $\partial_\beta f\vert_{\beta = 1} = 0$; then, by convexity and monotonicity, $f(\beta > 1)$ can only be a constant. FDC promotes this idea further by boldly suggesting that any duality invariant thermodynamic observable must freeze beyond the self-dual point $\beta=1$. It has been successfully applied to calculate free energy distribution \cite{fyodorov2009statistical}, moments of minimum position distribution \cite{fyodorov2010freezing,fyodorov2015moments}, and more involved quantities \cite{cao16maxmin}.  \\

For the sake of concreteness, let us first restrict our discussion to the distribution of the free energy. It has been observed that for all known exactly solved Gaussian logREMs the continuum partition function \eqref{eq:contZ} enjoys the following duality invariance
\footnote{See also works \cite{ostrovsky2016barnes,Ostrovsky2016A,ostrovsky2016riemann} where such property was called 'involution invariance' and shown to be a generic feature of moments of Barnes $\beta-$ distributions.}:
\begin{equation} \overline{Z_\beta^{-t/\beta}} \Gamma(1 + t / \beta)  = A\left(\beta + \beta^{-1}, t \right) \exp(t r_1(\beta)) \,, \label{eq:duality} \end{equation}
where $A(Q,t)$ is a model-dependent function of two arguments, while the factor $\Gamma(1 + t / \beta)$ is model \textit{independent}, and $r_1$ is a function of $\beta$ only; in fact, in all known cases, we have $r_1 = \beta^{-1} \ln\Gamma(1 - \beta^2) + c$ where $c$ is a numerical constant. For example, for the circular model \eqref{eq:dyson}, we have $A = \Gamma(1 + t \beta) \Gamma(1 + t/\beta)$ and $c = \beta \ln (2\pi)$ (clearly, $A$ is invariant by $\beta \to 1/\beta$, hence via power series expansion is only a function of $\beta + 1/\beta$ and $t$.

To discuss the rôle of the equation \eqref{eq:duality} in freezing, we reconsider \eqref{eq:Etf} (with the continuation \eqref{eq:comb_cont} for $C_{n,m}$) in the $\beta > 1$ phase, but regarding $m$ as a free parameter (\textit{i.e.}, forgetting for the moment \eqref{eq:moptimum}):
\begin{equation}
\overline{\exp(t\mathcal{F})}
=  e^{-t\ln M (b + b^{-1}) + t\widetilde{S_{[\text{UV}]}}} \,\, \overline{Z_{b}^{-\frac{t}{b}}}\,\, \frac{\Gamma\left(1 + \frac{t}{b}\right)}{\Gamma(1 + t/\beta)}
=  e^{-t\ln M (b + b^{-1})+(r_1 + \widetilde{S_{[\text{UV}]}})t} \,\, \frac{A\left(b + b^{-1}, t\right)}{\Gamma(1 + t / \beta)}\,,\quad b = m \beta \,. \label{eq:Fduality}
\end{equation}
where $\widetilde{S_{[\text{UV}]}}= S_{[\text{UV}]} + b^{-1} \ln \Gamma(1+m)$.
In the last equality we have used \eqref{eq:duality} with $\beta \to b$. It follows that, when $\beta > 1$, the solution $m = \beta^{-1} \Leftrightarrow b = 1$ of \eqref{eq:moptimum} is also a stationary point for the extensive (proportional to $\ln M$) part of the first moment, as well as for all higher cumulants of $\mathcal{F}$:
\begin{align}
  &\left[\partial_m  \lim \frac{\overline{\mathcal{F}}}{\ln M} \right]_{m = 1/\beta} = 0 \,,\,\, \left[\partial_m \overline{\mathcal{F}^k}^{(c)}\right]_{m = 1/\beta} = 0\, , \, \quad  k = 2, 3, \dots \,,\quad \beta > 1 \,. \label{eq:optimummoment}
\end{align}
The equation \eqref{eq:optimummoment} shows that the duality invariance property \eqref{eq:duality} ensures that the choice $m = \beta^{-1}$ of \eqref{eq:moptimum} satisfies an infinite series of selection criteria
\footnote{Note that for $\beta < 1$, $m = 1 / \beta > 1$ would not be a valid solution (it is known for general replica trick that $m \in (0,1]$ is the correct optimization range, see Appendix \ref{sec:frsb}).}. The first of them concerns the optimization of the extensive free energy and is familiar in the disordered system theories, while the others extend the RSB optimization principle to fluctuations.
As a consequence, the renormalised temperature $b = m \beta = 1$ freezes in the $\beta > 1$ phase, and so does the dual invariant quantity $A(b + b^{-1}, t)$ in \eqref{eq:Fduality}: this is an instance of FDC.

In hindsight, from a 1RSB viewpoint, FDC holds here because the quantity $$\overline{\exp(t(\mathcal{F}- r_1 -\widetilde{S_{[\text{UV}]}}))} = e^{-t\ln M (b + b^{-1})} A(b + b^{-1},t)$$ is both duality invariant in the $\beta < 1$, replica symmetric phase (where $b = \beta$), and has its temperature effectively renormalized as $\beta \rightarrow b = m \beta$ in the $\beta > 1$, RSB phase, with the parameter $m$ to be optimized. Duality invariance ensures that $m = 1 / \beta \Leftrightarrow b = 1$  is optimal in the thermodynamic limit $M\to \infty$, and thus insures freezing of the $b$-dependent, duality invariant quantities. Clearly the latter mechanism is not restricted to the free energy but applies in general. Therefore, we have shown that FDC would follow from a ``1RSB-duality scenario'', {stating that observables $O(\beta)$ will possess two properties simultaneously}: \texttt{(i)} when calculated using 1RSB with group size $m$, $O(\beta) = O'(b = \beta m)$ depends only on the renormalised temperature; \texttt{(ii)} in the $\beta < 1$ replica symmetric phase, $O(\beta) = O'(\beta)$ is duality invariant.	

At the moment we are unable to verify either FDC or 1RSB-duality scenario from the first principles. Nevertheless, in all known applications of FDC, the result can be recovered from a 1RSB calculation, and the above 1RSB-duality scenario holds. Indeed, the above discussion covers all (integrable) cases where the replicated partition function $\overline{Z^n}$ can be calculated explicitly, and can be extended to min-max correlation as was shown in \cite{cao16maxmin}. As to the other cases,  such as the distribution of
position of the global minimum \cite{fyodorov2010freezing,fyodorov2015moments} or Edwards-Anderson order parameter \cite{cao15gff}, one can argue they can either be reduced to the replicated partition function case (we shall see this for the minimum position in Sect. \ref{sec:minposition}), or to the variation thereof with respect to some parameter (the Edwards-Anderson case).
\\[.05cm]

The above analysis also reveals the limits of when to expect the freezing phenomena. We see that such phenomena are {\it in fine} reducible to the $\overline{Z^n}$ case, for which the UV contribution is well confined to a $O(1)$ first moment shift (eq. \eqref{eq:Fduality}). This will no longer be the case for the observables related to higher minima, to which we turn our attention now.

\subsection{Second minimum by 1RSB}\label{sec:derivation}
From this section on we consider the higher order statistics proper, starting with the 1RSB derivation of the main result for the second minimum \eqref{eq:main}. The first step is similar to that of the global minimum value case, eq. \eqref{eq:Ggen}: we start with writing $H_\beta$ \eqref{eq:defH} as the generating function (inverse Fourier transform) of replica averages:
\begin{equation}
  H_\beta(y_0, y) = -\sum_{n=0}^\infty \frac{(-e^{\beta y})^n}{n!} \sum_j  \left. \overline{\mathcal{W}^n(j, \Delta)}\right\vert^{y - y_0}_{\Delta = +\infty}  \,,\quad  \mathcal{W}(j, \Delta) \defeq \, e^{-\beta (V_j + \Delta)} + \sum_{k\neq j} e^{-\beta V_k}\,,\, f(x)\vert_{x=a}^b \defeq f(b) - f(a) \,. \label{eq:HGgen}
\end{equation}
Observe that $\mathcal{W}(j, \Delta)$ is the partition function of the potential obtained from $V_{j}$ (the log-REM) by shifting a  \textit{single} value $V_{j} \rightarrow V_{j} + \Delta.$ We shall call $(j,\Delta)$ a \textit{marker}, of which $j$ is the position and $\Delta$ the shift-value. Therefore $\mathcal{W}$ is a special case of \eqref{eq:Zshifted} in section \eqref{sec:1rsb}, so as argued there, for any $j$ fixed we can employ the 1RSB Ansatz in the replica sums {over} $j_1, \dots, j_n$. The latter form groups of $m$ replicas, each group occupying a block. Now \eqref{eq:HGgen} contains one more sum over the marker position $j$.  Yet, because of the presence of the difference $[\dots]\vert^{y - y_0}_{\Delta = +\infty}$, some $j_a$ must be equal to $j$ to give a non-vanishing contribution. This implies that one (and only one) group out of $n / m$ will be at the block containing $j$, and the sum over $j$ reduces to choosing one group amongst the $n/m$ and summing over the UV positions of the block that the chosen group occupies.  By permutation symmetry between the groups, we can assume that the marker is near the group $g=1$ and multiply the answer by the combinatorial factor $\frac{n}{m} C_{n,m}$.
As observed above, $\mathcal{W}(j,\Delta), \mathcal{W}(j, \infty)$ and $\mathcal{Z}$ differ only by one Boltzmann factor, which affects only the intra-group interaction of the marked ($g=1$) group. After collecting and factoring out all similar terms the (difference of) modified intra-group interaction isolated in  $D(\xi,\Delta)$ takes the form given below:
\begin{align}
 \sum_j  \left. \overline{\mathcal{W}^n(j, \Delta)}\right\vert_{\Delta = +\infty}^{y - y_0} \stackrel{M\rightarrow\infty}= &
C_{n,m} \frac{n}{m} \left(\frac{M}{N}\right)^{\frac{n}{m}}\int_{X} \prod_{g = 1}^{n/m} \left[ \abs{\dif^d \xi_g}  \exp(-m \beta c(\xi_g) - (1 + m^2 \beta^2)l(\xi_g)) \right] \prod_{g < g'} \exp(\beta^2 m^2 C(\xi_g, \xi_{g'})) \nonumber \\
\times & \left[\prod_{g=2}^{n/m} E(\xi_g)\right]  \left. D(\xi_{g=1},\Delta)\right\vert_{\Delta = +\infty}^{y- y_0}  \label{eq:Wn} \,, \\
D(\xi,\Delta) &= \sum_{i=1}^N \sum_{i_1, \dots, i_m = 1}^N
\prod_{l,l' = 1}^m \exp(\beta^2 \overline{V_{j(\xi, i_l)} V_{j(\xi, i_{l'})}}^c / 2) \prod_{i: i_{l} = i}
\exp(-\beta \Delta) \nonumber \\
 = &  \sum_{i=1}^N \overline{w_\beta^m}(i, \Delta)  \exp(\beta^2 m^2 (2\ln M - C_N) / 2) \label{eq:Dxi} \,.
\end{align}
 where $w_j(\beta, \Delta)$ was defined in eq. \eqref{eq:smallw}. In \eqref{eq:Wn}, we recall that $E(\xi)$, defined in \eqref{eq:Exi}, is the intra-group interaction of a group at block $\xi$ without marker. We now compare \eqref{eq:Wn} to \eqref{eq:Zn}.
 The two expressions share the same continuum integral part and most of combinatorial factors,
 though \eqref{eq:Wn} acquires an extra factor $n/m$. As for the intra-group interactions,  \eqref{eq:Zn} has a product of $n/m$ factors $E(\xi_g)$'s, whereas in \eqref{eq:Wn} one replaces the factor $E(\xi_{g=1})$ with $D(\xi_{g=1})$. In summary, we have
\begin{align}
\sum_j \left. \overline{\mathcal{W}^n(j, \Delta)}\right\vert_{\Delta = +\infty}^{y- y_0}
\nonumber =  \frac{n}{m} \overline{\mathcal{Z}^n} \left. \frac{ D(\xi,\Delta) }{ E(\xi) } \right\vert_{\Delta = +\infty}^{y - y_0} \,.
\label{eq:WnZn1}
\end{align}
Now we apply \eqref{eq:laplaceder} to express the above in terms of the generating functions \eqref{eq:Ggen} and \eqref{eq:HGgen}:
\begin{equation}
H_\beta(y_0, y) = \frac{-1}{\min(1,\beta)} \partial_y G_\beta(y) \left. \frac{ D(\xi,\Delta) }{ E(\xi) } \right\vert_{\Delta = +\infty}^{y - y_0} \,. \label{eq:main1}
\end{equation}
To bring \eqref{eq:main1} to the form \eqref{eq:main} we need to write down the last ratio factor in \eqref{eq:main1} explicitly. Combining  \eqref{eq:Dxi} and \eqref{eq:Exi}, we have:
\begin{align}
\left. \frac{ D(\xi,\Delta) }{ E(\xi) } \right\vert_{\Delta = +\infty}^{y - y_0} = \frac{ \left. \sum_{i=1}^N \overline{ w_\beta^m}(i, \Delta) \right\vert_{\Delta =+ \infty}^{y - y_0}}{\overline{z_\beta^m}} = K_\beta(y - y_0) \,, \label{eq:KisDoverE}
\end{align}
by recalling the definition of $K_\beta$ in \eqref{eq:Koffd}. This finishes the derivation of \eqref{eq:main}.


\subsubsection{Derivation of \eqref{eq:Kbzero}} \label{sec:kbzeroder}
We now derive the relation \eqref{eq:Kbzero} used in our numerical study. For this we set $y_0 = y$ in \eqref{eq:main}, and then integrate over $y$:
\begin{equation}
\int_{\R} \dif y H_\beta(y,y)  = \int_\R \dif y  \left[ - \partial_y G_\beta(y) \right] \times K_\beta(0) / \min(1,\beta) = K_\beta(0) / \min(1,\beta) \,, \label{eq:integratemain}
\end{equation}
since $G_\beta(-\infty) = 1$ and $G_\beta(+\infty) = 0$ by \eqref{eq:defG}. On the other hand, by \eqref{eq:HGgen}, we have
\begin{align}
\int_{\R} \dif y H_\beta(y,y) &= \sum_j \int_{\R} \dif y  \left[ \overline{\exp(-e^{\beta y} \mathcal{W}(j,+\infty))
- \exp(-e^{\beta y} \mathcal{W}(j,0))} \right] \nonumber \\
& = \frac{1}{\beta} \sum_j (\overline{\ln \mathcal{W}(j,0)} - \overline{\ln \mathcal{W}(j,+\infty)}) \,. \label{eq:integrateH}
\end{align}
Equating the RHS of \eqref{eq:integratemain} and \eqref{eq:integrateH} we obtain \eqref{eq:Kbzero}, since $\mathcal{W}(j,0) = \mathcal{Z}$ and $ \mathcal{W}(j,+\infty) = \mathcal{Z}_{\setminus j}$, due to \eqref{eq:HGgen}.

\subsubsection{Remote second minimum}\label{sec:remote2ndmin}
In Sect. \ref{sec:num}, we observed that the gap is exponentially distributed when measured between the global minimum $V_{\min}$  and
the remote second minimum $V_{\min,1}^\text{far}$ defined in \eqref{eq:fargap} (rather than the true second minimum $V_{\min,1}$).  For this we note that the joint distribution of $V_{\min}$ and $V_{\min,1}^\text{far}$ can be accessed by the following modified version of \eqref{eq:defH} with $y>y_0$ (see also \eqref{eq:Hinfty}):
\begin{equation}
H_{\beta}^{\text{far}}(y_0,y) = \sum_{j} \overline{(1 - \theta_{\beta}(V_{j} - y_0))\prod_{0 < \abs{i-j} \leq N} \theta_{\beta}(V_{i} - y_0) \prod_{\abs{i-j} > N} \theta_{\beta}(V_{i} - y)} \stackrel{\beta\rightarrow\infty}\longrightarrow \overline{\theta(y_0 - V_{\min})\,\theta(V_{\min,1}^{\text{far}} - y)}\,. \label{eq:Hfar}
\end{equation}
To treat this, we will follow the derivation just presented for $H_\beta$, and indicate only the necessary modifications. First, \eqref{eq:Hfar} generates the following replicated partition functions
\begin{align}
&H_{\beta}^{\text{far}}(y_0,y) = -\sum_{j} \sum_{n\geq 0} \frac{(-e^{\beta y})^n}{n!} \left.\overline{\mathcal{W}^n_{\text{far}}(j,\delta, y- y_0)}\right\vert_{\delta=+\infty}^{y-y_0} \,,\, \\
&\mathcal{W_{\text{far}}}(j,\delta,\Delta) =  e^{-\beta(V_j + \delta)} + \sum_{0 < \abs{k - j } \leq N} e^{- \beta (V_k + \Delta)} +  \sum_{\abs{k - j} > N} e^{- \beta V_k} \,. \label{eq:Wnfardef}
\end{align}
Compared to \eqref{eq:HGgen}, the potential values $V_k$ for $0 < \abs{k - j } \leq N$ are always shifted by $\Delta = y - y_0$, in addition to $V_j$ which is still shifted by $\delta = y - y_0$ and $\delta = +\infty$ (respectively in the two terms). Since $1 \ll N \ll M$ in eq. \eqref{eq:fargap}, when implementing 1RSB procedure, we can arrange the blocks in a way such that all the shifted sites form exactly one block, of which the site $j = j(\xi, i=0)$ summed over in \eqref{eq:Hfar} is at the centre. Note that this makes the division of the system into boxes dependent  on the marker position, while the division is fixed once and for all in the previous derivations. Working through the arguments leading to \eqref{eq:WnZn}, \textit{mutatis mutandis} \footnote{One encounters here the following technical subtlety: here, one sums over the marker's position $j = 1, \dots, M$, and constrains one replica group to be in the block centred at $j$ by the subtraction argument analogous to \eqref{sec:derivation}; each of the other replicas groups chooses one block from $N / M$. Hence, the $(M/N)^{n/m}$ factor in \eqref{eq:Wn} and \eqref{eq:Zn} will be replaced by $M \times (M/N)^{n/m - 1}$ here, giving rise to the factor $N$ in \eqref{eq:Dfarxidef} which replaces the sum $\sum_{i=1}^N$ in \eqref{eq:Dxi}. Now, the approach here can be used in the second minimum calculation above, and by doing that, we would obtain a different expression of \eqref{eq:Dxi}, obtained by replacing $\sum_{i=1}^N \dots(i) \to N \dots(i=0)$. However, we argue that such a difference becomes irrelevant in the $N \rightarrow \infty$ limit, in which the local logREM is statistically translation invariant, see \eqref{eq:localrem} and \eqref{eq:intra}.}, one arrives to a version of that equation, in which the intra-group energy of the marker's block (see \eqref{eq:Dxi}) is replaced by
\begin{equation}
D_{\text{far}}(\xi, \delta, \Delta) = N \overline{\left(e^{-\beta(u_0 + \delta)} + \sum_{i\neq 0} e^{-\beta (u_0 + \Delta)}  \right)^m}  \exp(\beta^2 m^2 (2\ln M - C_N) / 2) \,. \label{eq:Dfarxidef}
\end{equation}
In terms of the generating functions \eqref{eq:main1} takes the following form:
\begin{align}
&H_{\beta}^{\text{far}}(y_0,y) = \frac{-1}{\min(1, \beta)} \partial_y G_{\beta}(y) K^{\text{far}}_\beta(y - y_0) \,,\quad
K^{\text{far}}_\beta(\Delta) = \left. \frac{ D_{\text{far}}(\xi, \delta, \Delta) }{ E(\xi) } \right\vert_{\delta = +\infty}^{\Delta} \,.  \label{eq:mainfar}
\end{align}
 The remaining job is to calculate the last ratio. For this, observe from \eqref{eq:Dfarxidef} that \texttt{(i)}$D_{\text{far}}(\xi, \delta, \Delta)/D_{\text{far}}(\xi, \Delta, \Delta)$ is a function of $\delta - \Delta$; \texttt{(ii)} when $\delta = \Delta$, all potential values in the block are shifted by $\Delta$, so $D_{\text{far}}(\xi, \Delta, \Delta) / E(\xi) = N e^{-m\beta \Delta}$.  Combining the two steps, we have
 \begin{equation}
\frac{ D_{\text{far}}(\xi, \delta ,\Delta) }{E(\xi)}
= \frac{D_{\text{far}}(\xi, \Delta, \Delta)}{E(\xi)}\frac{ D_{\text{far}}(\xi, \delta ,\Delta) }{D_{\text{far}}(\xi, \Delta ,\Delta)} = e^{-m\beta \Delta} N F(\delta - \Delta) \,,
 \end{equation}
 where $F$ is some unknown function. Plugging this into \eqref{eq:mainfar}, we obtain $K^{\text{far}}_\beta(\Delta) = e^{-m\beta \Delta} N (F(0) - F(+\infty)) \,.$ Applying the analogue of \eqref{eq:gapstat2} gives $\overline{\delta(g_{1}^{\text{far}} - \Delta)} = \partial_\Delta^2 K^{\text{far}}_\infty(\Delta) =  \exp(-\Delta) N (F(0) - F(+\infty))$. Normalisation of the probability fixes the constant $N(F(0) - F(+\infty)) = 1$, giving
 \begin{equation}
 K^{\text{far}}_\beta(\Delta) = e^{-m\beta \Delta} \,. \label{eq:Kfar}
 \end{equation}
where  the parameter $m$ as a function of the temperature is defined in the equation \eqref{eq:moptimum}.
In the zero-temperature limit, this recovers the simple exponential gap distribution \eqref{eq:exponential} for the gap with a remote second minimum. As one could have expected, for such a situation the UV correction factor becomes UV-data independent for any temperature. {From that angle such quantity reflects only uncorrelated REM-like features of the model, rather than the full logREM structure. Note that  \eqref{eq:Kfar} provides an example of freezing that seems not to be accompanied by any obvious duality invariance,  the feature shared by the generating function (\ref{eq:defG}) of the uncorrelated REM \cite{fyodorov08rem}.}

\section{Generalization to higher orders}\label{sec:fullminima}
 To generalize the calculations of the previous section to higher order statistics, we shall consider the following observable, which generalizes to the case of  $k \geq 1$ markers the function $H_\beta(y_0, y)$ defined for $k = 1$ in \eqref{eq:defH}:
 \begin{align}
 H_\beta\left((j_0, y_0), \dots, (j_{k-1}, y_{k-1}),y \right) &= \overline{\prod_{s=0}^{k-1} (1 -  \theta_\beta(V_{j_s} - y_s)) \prod_{i \neq j_s}  \theta_\beta(V_{j_s} - y)} \,, \label{eq:Hgeneral}\\
 & \stackrel{\beta\rightarrow\infty}\longrightarrow
 \mathbb{P}(V_{j_0} < y_0,\dots,V_{j_{k-1}} < y_{k-1}, \text{ all other } V_{j} > y) \,.
 \end{align}
 Here and below, we assume $y_0 < y_1 < \dots < y_{k-1} < y$. Taking derivatives with respect to $y_1, \dots, y_{k-1}$ and suppressing the argument for compactness, we obtain:
 \begin{equation}
 \partial_{y_0,\dots,y_{k-1}} H_{\infty} =
 \overline{\delta(V_{j_0} - y_0) \dots \delta(V_{j_{k-1}} - y_{k-1}) \, \theta(V_{\min, k} - y)}
 \end{equation}
 where $V_{\min, j}$ stands for the value of the $(j+1)$-th ordered minimum, with $V_{\min,0} < V_{\min,1} < V_{\min,2} < \dots$ (as for Gaussian sequences the probability of two minima values to coincide is strictly zero); in particular, with such notations the global minimum is $V_{\min} = V_{\min,0}$. In the above equation, the position of $(k+1)$-th minimum is not fixed, while that of the first, second, \dots, and $k$-th minima are fixed at $j_0, \dots, j_{k-1}$ (with the order determined by $y_0, \dots, y_{k-1}$). For the moment let us perform the summation over position indices and concentrate on the minima values (see Sect. \ref{sec:minposition} for discussion on minima positions):
 \begin{align}
  \sum\limits^{*}_{j_0, \dots, j_{k-1}} \partial_{y_0,\dots,y_{k-1}}  H_{\infty} =
 \overline{\delta(V_{\min} - y_0) \dots \delta(V_{\min,k-1} - y_{k-1}) \, \theta(V_{\min, k} - y)}
 \label{eq:pdfgeneral2}
 \end{align}
 where $\sum\limits^{*}_{j_0,\dots, j_{k-1}}$ sums over distinct positions $(j_0, \dots, j_{k-1})$.

 \subsection{1RSB calculation of \eqref{eq:Hgeneral}}\label{sec:derivationgeneral}
In this section we demonstrate how to evaluate \eqref{eq:Hgeneral} (with marker positions summed over) within 1RSB framework. Similarly to \eqref{eq:HGgen}, \eqref{eq:Hgeneral} generates partition sums with markers at positions $j_0, \dots, j_{k-1}$:
\begin{align}
&H_\beta\left((j_0, y_0), \dots, (j_{k-1}, y_{k-1}),y \right) = (-1)^k \sum_{n=0}^{\infty} \frac{(-e^{\beta y})^n}{n!} \left. \overline{\mathcal{W}^n((j_0,\Delta_0), \dots, (j_{k-1},\Delta_{k-1}))}\right\vert_{\Delta_s  = \infty}^{y - y_s} \,, \label{eq:HGen_gen}\\
& \mathcal{W}((j_0,\Delta_0), \dots, (j_{k-1},\Delta_{k-1})) = \sum_{s=0}^{k-1} \exp(-\beta (V_{j_s} + \Delta_s)) +
\sum_{j \notin \{j_s\}} \exp(-\beta V_j) \,,\, \label{eq:bigWdef} \\
& f(x_1, \dots, x_k)\vert_{x_s=a_s}^{b_s} = \sum_{\sigma \in \{0,1\}^{k}} (-1)^{\sum \sigma_s} f(b_s + (a_s - b_s)\sigma_s) \label{eq:NewtonLeib}
\end{align}
Note that $\mathcal{W}$ is identical to the partition function in \eqref{eq:Zshifted}. So, as we justified in Sect. \ref{sec:1rsb}, we can treat the sum over replicas using 1RSB, \textit{i.e.}, considering only the dominant configurations where the replicas form groups of size $m = \min(1, 1/\beta)$ (eq. \eqref{eq:moptimum}). We then follow exactly the recipe \eqref{eq:sumtoint} to replace the sum over replica positions by integrals over group macroscopic coordinates times sums over microscopic intra-group positions.

\begin{figure}
	\includegraphics[scale=.4]{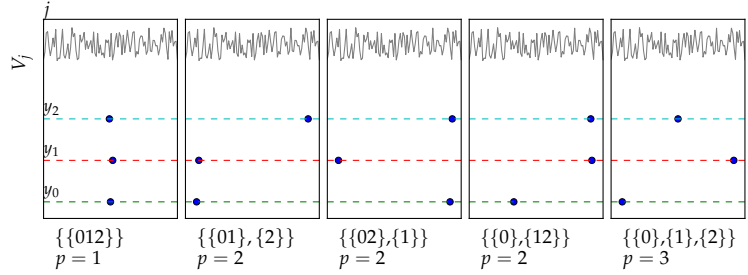}
	\caption{Illustration of the $5$ partitions of $\{y_0, y_1, y_2\}$, corresponding to $k = 3$. The figure of each partition illustrates the corresponding position-value configuration: two minimum values $y_s$ and $y_{s'}$ are in the same part if their position are close (technically, in the same block). Below each figure is written the corresponding partition (with shorthand notation $s \rightarrow y_s$). }
	\label{fig:part3}
\end{figure}

As before, each marker position must coincide with some replica positions to give non-vanishing contribution to $H_\beta$, due to the cancellations in \eqref{eq:NewtonLeib}.This imposes severe constraints on the sum over marker positions in \eqref{eq:HGen_gen}; in particular, at the macroscopic level it is reduced to summing over all possible ways of assigning each marker to a replica-group, which is described by a mapping $\{0, \dots, k-1\} \rightarrow \{1, \dots, n/m\}$. Since we will eventually continue $n$ and $m$ to non-integer values, we need to work out the combinatorics explicitly. For this, we recall that any above mapping can be constructed by partitioning $\{y_0, \dots, y_{k-1}\}$ into $p$ parts followed by the assignment of the $p$ parts to \textit{distinct} groups; the latter results in a combinatorial factor $(n/m)(n/m-1)\dots (n/m - p +1) = (n/m)_p$ in terms of the Pochhammer's symbol
\footnote{This is combinatorics of $ (n/m)^k = \sum_p S_{k,p} (n/m)(n/m-1)\dots (n/m - p +1)$, where $S_{k,p}$ is the number of partitions of the set $\{y_0, \dots, y_{k-1}\}$ into $p$ nonempty parts (without order), known as the Stirling numbers of the second kind. The total number of partitions of $k$ elements is known as the Bell number.}. So the replica average in the RHS of \eqref{eq:HGen_gen} takes the form of a sum $\sum_{\mathbb{P}(k)} (n/m)_p [\dots]$, where $\sum_{\mathbb{P}(k)}$ denotes the sum over all partitions of $\{y_0, \dots, y_{k-1}\}$, and $[\dots]$ denotes the contribution of each partition. The notion of a partition we use here should not be confused with the standard partitions of an integer: in the latter, the \textit{elements} as well as the parts are indistinguishable objects, while in our case, only the parts are indistinguishable, but the elements are distinguishable. For example, when $k = 3, p = 2$  (see Fig. \ref{fig:part3} for illustration), $\{\{1, 2\},\{3\}\}$ and $\{\{1, 3\},\{2\}\}$ are \textit{distinct} partitions (although they give the same \textit{integer} partition $3 = 2 + 1$). In general, we describe a partition as follows: $p$ denotes the number of parts, $(k_1, \dots, k_p) $ denote the numbers of elements in each part, so that $k_1 + \dots + k_p = k$; for each part, let $\{y_{q,0}< \dots <y_{q, k_q-1}\}$ be the list in ascending order of its elements; since the parts are indistinguishable, we also order them by their first element: $y_{1,0} < y_{2,0} \dots <y_{p,0}$. In the following, the set of all partitions of $k$ elements will be denoted $\mathbb{P}(k)$. Clearly, knowing a partition of $\{y_0,  \dots, y_{k-1}\}$ is equivalent to that of $\{0, \dots, k-1\} = \{j_{q,s}: q = 1, \dots, p, s = 0, \dots, k_q -1 \}$, and for any set of variables $\{x_0, \dots, x_{k-1}\}$ indexed similarly, we will adopt the short-hand $x_{q,s} = x_{j_{q,s}}$.

Let us now consider the contribution of a given partition. Similarly to the second minimum case, the presence of markers does not change the Coulomb gas integral of the IR sector, but does modify the intra-group interactions. Moreover, since more than one marker can be at one block we need to introduce the higher order generalizations of \eqref{eq:smallw} to express the ratios of intra-group energies:
\begin{equation}
 w_\beta((i_0,\Delta_0), \dots, (i_{k-1}, \Delta_{k-1})) \defeq \, \sum_{s=0}^{k-1} e^{-\beta(u_{i_s} + \Delta_s)}
+ \sum_{i\notin \{i_s\}} e^{-\beta u_i}  \,. \label{eq:smallw_general}
\end{equation}
With the above ingredients at our disposal, it is not hard to generalize the calculation of the second minimum and obtain the following:
\begin{align}
&\sum\limits^{*}_{j_0, \dots, j_{k-1}} \left. \overline{\mathcal{W}^n((j_0,\Delta_1), \dots, (j_{k-1},\Delta_k))}\right\vert_{\Delta_s  = \infty}^{y - y_s}
 \stackrel{M\rightarrow\infty}= \overline{\mathcal{Z}^n} \sum_{\mathbb{P}(k)} (n/m)_{p}
\prod_{q=1}^p K_\beta(y-y_{q,0}, \dots, y-y_{q,k_q-1}) \,,\,  \label{eq:WnZn}\\
& K_\beta(\Delta_0, \dots, \Delta_{\ell -1})  \defeq  \lim_{N\rightarrow\infty} \left. \sum\limits^{*}_{i_0, \dots, i_{\ell-1} \leq N} \frac{\overline{w_\beta^{m}((i_0,\delta_0), \dots, (i_{\ell -1}, \delta_{\ell - 1}))}}{\overline{z^m_\beta}} \right\vert_{\delta_s  = \infty}^{\Delta_s}  \,,\quad \Delta_0 > \dots > \Delta_{\ell-1} \geq 0 \,,\, \ell = 1, 2, \dots \, \label{eq:defKgeneral} \\
& (x)_p = x(x-1)\dots (x-p+1) \,.
\end{align}
Note that \eqref{eq:defKgeneral} defines an infinite series of UV correction factors depending on $\ell = 1, 2, \dots$ arguments and  generalizing \eqref{eq:Koffd}, to which  \eqref{eq:defKgeneral} reduces for $\ell = 1$. The RHS of \eqref{eq:defKgeneral} is symmetric in its arguments $\Delta_0, \dots, \Delta_{\ell-1}$; it will prove helpful to assume the ordering $\Delta_0 > \dots > \Delta_{\ell-1}$, in agreement with the ordering of the arguments appearing in \eqref{eq:WnZn}. In terms of the generating functions \eqref{eq:Ggen} and \eqref{eq:HGen_gen} the equation \eqref{eq:WnZn} is equivalent to the following relation
\begin{align}
&\sum\limits^{*}_{j_0\dots j_{k-1}} H_{\beta}((j_0, y_0), \dots, (j_{k-1}, y_{k-1}), y) =  \sum_{\mathbb{P}(k)} (-1)^{p+k} \mathcal{D}_{\beta,p} G_{\beta}(y) \prod_{q=1}^p K_{\beta}(y-y_{q,0}, \dots, y-y_{q,k_q-1}) \,,\, \label{eq:maingeneral}\\
&\mathcal{D}_{\beta,p} = -b^{-1}\partial_y(1- b^{-1}\partial_y)\dots (p-1-b^{-1}\partial_y) \,,\quad b = \min(1, \beta) \,. \label{eq:Dpb}
\end{align}
Here we used the Laplace transform \eqref{eq:laplaceder} that helps to transform $(n/m)_p = (n \beta / b)_p$ in \eqref{eq:WnZn} into $(b^{-1}\partial_y)_p = (-1)^{p} \mathcal{D}_{\beta,p}$ (the signs $(-1)^p$ are inserted for convenience later).
This generalizes \eqref{eq:main} (recovered for $k=1$, $p=1$) and is the main result of this work in full generality.

We first consider implications of \eqref{eq:maingeneral} in the $\beta<1$ phase with the parameter $m = 1$. Plugging \eqref{eq:smallw_general} into \eqref{eq:defKgeneral}, we have
\begin{align}
K_{\beta < 1} (\Delta_0, \dots, \Delta_{\ell-1})& = \lim_{N\rightarrow\infty} \sum\limits^{*}_{i_0, \dots, i_{\ell-1}} \left.\frac{ \sum_{s=0}^{\ell-1} \overline{ \exp(-\beta(u_{i_s} + \delta_s))} + \sum_{i\notin \{i_s\}} \overline{\exp(-\beta u_{i_s})} }{\sum_{i=1}^N \overline{\exp({-\beta u_i})}}  \right\vert_{\delta_s = + \infty}^{\Delta_s}
\end{align}
Note that the formula \eqref{eq:NewtonLeib} has a nice feature, namely that $f(x_0, \dots, x_{\ell-1})\vert_{x_s=a_s}^{b_s} = 0$ whenever $f$ does not depend on \textit{all} of its arguments. So the term $\sum_{i\notin \{i_s\}}$ in the above equation always vanishes since it does not depend on any $\delta_s$. Each of the other terms depends on one (and only one) of the variables $\delta_s$, and thus vanishes when $\ell > 1$. So we have
\begin{equation}
\ell > 1\Rightarrow K_{\beta < 1} (\Delta_0, \dots, \Delta_{\ell-1}) = 0 \,.
\end{equation}
This implies that only the partition with $p = k$ and $\forall k_q = 1$  survives in \eqref{eq:maingeneral} (e.g. for $k=3$ it is the rightmost one in Fig. \ref{fig:part3}) which makes the latter to simplify to
\begin{align}
\sum\limits^{*}_{j_0\dots j_{k-1}} H_{\beta < 1}((j_0, y_0), \dots, (j_{k-1}, y_{k-1}), y) =
\mathcal{D}_{\beta,k} G_{\beta}(y) \prod_{s=0}^{k-1} e^{\beta(y_s - y)} \,.
\end{align}
This relation generalises \eqref{eq:H1highT}, and is also ({super}) universal, \textit{i.e.} shows neither IR nor UV data dependence.

\subsection{Extreme values process}\label{sec:lowTgen}
 Let us now discuss the implications of \eqref{eq:maingeneral} at zero temperature. In all the $\beta > 1$ phase the differential operator $ \mathcal{D}_{\beta > 1, p}$ \eqref{eq:Dpb} becomes
\begin{equation} \mathcal{D}_p \defeq \mathcal{D}_{\beta > 1, p} = -\partial_y(1 -\partial_y ) \dots (p-1-\partial_y) \,. \label{eq:Dp} \end{equation}

To obtain minima value distribution we should combine \eqref{eq:pdfgeneral2} to \eqref{eq:maingeneral}, hence the need to calculate the derivatives $\partial_{y_0\dots y_{k-1}} \dots $. Noting that the product $\prod_{q=1}^p K_{\beta}(y-y_{q,0}, \dots, y-y_{q,k_q-1})$ contains every $y_{k-1}$ and only once for each, we obtain
\begin{align}
&\overline{\theta(V_{\min,k} - y) \prod_{s=0}^{k-1} \delta(V_{\min,s} - y_s)} =  \sum_{\mathbb{P}(k)}(-1)^p \mathcal{D}_p G_{\infty}(y) \prod_{q=1}^p \partial_{0\dots k_q-1} K_\infty(y-y_{q,0}, \dots, y -y_{q,k_q-1}) \,,\,  \label{eq:difmaingeneral}
\end{align}
where $\partial_{0\dots \ell-1} = \partial_{\Delta_0}\dots  \partial_{\Delta_{\ell-1}}$. The last mixed partial derivative can be evaluated more explicitly. For this we first calculate $\beta \rightarrow \infty$ limit of $K_\beta$ from \eqref{eq:defKgeneral} and \eqref{eq:smallw_general}:
\begin{align}
&K_\infty(\Delta_0, \dots, \Delta_{\ell-1}) =  \frac{\overline{\exp(-\min(u_{\min} + \Delta_0, \dots, u_{\min,\ell-1} + \Delta_{\ell-1}, u_{\min,\ell}))} + R}{\overline{\exp(-u_{\min})}}  \nonumber \\
=& \frac{\overline{\exp(-u_{\min})\exp(-\min(\Delta_0, u_{\min,1}-u_{\min} + \Delta_{1},\dots, u_{\min,\ell-1} -u_{\min}+ \Delta_{\ell-1}, u_{\min,\ell}-u_{\min}))} + R }{\overline{\exp(-u_{\min})}} \label{eq:Kinftyumin}
\end{align}
where $u_{\min} = u_{\min,0}, u_{\min,1}, \dots, u_{\min,\ell} $ are the first $(\ell+1)$ minima of the UV block and $R$ denotes the $2^{\ell}-1$ other terms generated by \eqref{eq:NewtonLeib},  which will be eventually annihilated by the mixed partial derivative $\partial_{0\dots \ell-1}$ (nonetheless, it will be useful to write them out explicitly for other purposes, see Appendix \ref{sec:gapstatistics}, eq. \eqref{eq:Kj}).
The situation is similar to \eqref{eq:dKinfty}: although one cannot cancel out the denominator, one can define a \textit{biased minima process} $(v_{\min,s})_{s=0}^{\infty}, v_{\min} = v_{\min,0} < v_{\min,1} < \dots$, uniquely up to a translation, by the following property
\begin{equation}
\overline{f(v_{\min,1}-v_{\min,0}, v_{\min,2}-v_{\min,0}, \dots,)} \defeq  \frac{\overline{e^{-u_{\min}}f(u_{\min,1}-u_{\min,0}, u_{\min,2}-u_{\min,0}, \dots,)}}{e^{-u_{\min}}} \,,\,  \label{eq:defbiased}
\end{equation}
for any reasonable function of $f$ of arbitrary number of variables. Note that \eqref{eq:defbiased} is generalizing \eqref{eq:gapbiased} to the multi-variable case; we will discuss it in more detail in \ref{sec:biased}. In terms of $(v_{\min,s})$ \eqref{eq:Kinftyumin} is expressed as
\begin{equation}
K_\infty(\Delta_0, \dots, \Delta_{\ell-1}) =
\overline{\exp(-\min(\Delta_0, v_{\min,1}-v_{\min} + \Delta_{1},\dots, v_{\min,\ell-1} - v_{\min}+ \Delta_{\ell-1}, v_{\min,\ell}-v_{\min}))} + R \label{eq:Kinftyvmin} \,.
\end{equation}
 Note that \eqref{eq:Kinftyvmin} is only valid if $\Delta_0 > \Delta_1 > \dots > \Delta_{\ell-1}$ are ordered, and its RHS is \textit{not} symmetric in $\Delta_0, \dots, \Delta_{\ell-1}.$ Nevertheless, it \textit{is} a symmetric function of variables
\begin{equation} x_s \defeq v_{\min,s}-v_{\min} + \Delta_s \,,\quad  s = 0,1,\dots,\ell-1\,. \label{eq:defshifts}\end{equation}
In particular, $x_0 = \Delta_0$. This ensures that $\partial_{0,\dots,\ell-1} = \partial_{x_0,\dots,x_{\ell-1}}$ is well-defined independently of the differentiation order, so we can select any suitable order:
\begin{align}
&\partial_{0 \dots {\ell-1}} \exp(-\min(x_0, \dots, x_{\ell - 1}, v_{\min,\ell} - v_{\min}))
\stackrel{\partial_{\ell-1}}= \partial_{0\dots \ell-2} \left[- e^{-x_{\ell-1}} \theta(v_{\min,\ell}-v_{\min} - x_{\ell-1}) \prod_{s=0}^{\ell-2}\theta(x_s  - x_{\ell-1}) \right] \nonumber\\
=& - e^{-x_{\ell-1}} \theta(v_{\min,\ell}-v_{\min} - x_{\ell-1}) \partial_{0\dots \ell-2}  \prod_{s=0}^{\ell-2}\theta(x_s  - x_{\ell-1})
=- e^{-x_0} \theta(v_{\min,\ell}-v_{\min} - x_{\ell-1})\prod_{s=0}^{\ell-2}\delta(x_s  - x_{\ell-1})  \,. \label{eq:partialKofx}
\end{align}
Applying the above identity with \eqref{eq:defshifts} to \eqref{eq:Kinftyvmin} we obtain
\begin{align}
&\partial_{0\dots \ell-1} K_\infty(\Delta_0, \dots, \Delta_{\ell-1}) = - e^{-\Delta_0} D(\Delta_{0},\dots, \Delta_{\ell-1}) \label{eq:partialKisD} \,,\, \\
& D(\Delta_{0},\dots, \Delta_{\ell-1}) \defeq \overline{\theta(v_{\min,\ell}-(v_{\min,\ell-1}+\Delta_{\ell-1})) \prod_{s=0}^{\ell-2} \delta((v_{\min,s} + \Delta_s) - (v_{\min,\ell-1} + \Delta_{\ell-1}))}  \label{eq:decoration}\\
= & \mathrm{Prob.}(v_{\min,\ell} - v_{\min,0} - \Delta_0 = v_{\min,\ell} - v_{\min,1} - \Delta_1 = \dots  = v_{\min,\ell} - v_{\min,\ell-1} - \Delta_{\ell-1} > 0) \,. \label{eq:Dprobability}
\end{align}
Here $D(\Delta_{0},\dots, \Delta_{\ell-1})$, $ \Delta_0 > \dots >\Delta_{\ell-1}$ is an observable of the process $(v_{\min,s})$ and is the probability (density) of the event described by the last line. In particular,
\begin{equation}
D(\Delta_0, \dots, \Delta_{\ell-2},0) = \overline{\prod_{s=0}^{\ell-2} \delta(v_{\min,\ell-1} - v_{\min,s} - \Delta_{s})}
= \overline{\prod_{s=0}^{\ell-2} \delta((v_{\min,s+1} - v_{\min,s}) - (\Delta_{s} - \Delta_{s+1}))} \label{eq:Dlastzero}
\end{equation}
gives the pdf of the first $\ell - 1$ gaps of $(v_{\min},s)$.
 Thus the knowledge of $D(\Delta_0, \dots, \Delta_{\ell-1})$ for all $\ell = 1, 2, \dots$ determines the process $(v_{\min,s})$ completely. In fact the information in the set of $D(\Delta_0, \dots, \Delta_{\ell-1})$ is excessive, since \textit{e.g.}, by applying $\sum_{s=0}^{\ell-1}\partial_{\Delta_s}$ to \eqref{eq:decoration} and comparing with \eqref{eq:Dlastzero} (with $\ell \rightsquigarrow \ell + 1$), we obtain a series of functional relations between $D(\dots)$'s
\begin{equation}
\sum_{s=0}^{\ell-1} \partial_{\Delta_{s}} D(\Delta_0, \dots, \Delta_{\ell-1})  = -D(\Delta_0, \dots, \Delta_{\ell-1},0) \,,\quad \ell = 1, 2, \dots\,.\label{eq:Drelation}
\end{equation}

Now plugging \eqref{eq:partialKisD} back into \eqref{eq:difmaingeneral}, we obtain the final expression for the joint distributions of minimum values, quoted in \eqref{eq:pdfvalue0}
\begin{align}
&\overline{\theta(V_{\min,k} - y)\prod_{s=0}^{k-1}\delta(V_{\min,s} - y_s) } = \sum_{\mathbb{P}(k)} \mathcal{D}_p G_{\infty}(y) \prod_{q=1}^p \left[e^{y_{q,0}-y} D( y - y_{q,0},\dots,  y - y_{q,k_{q}-1})\right] \,. \label{eq:pdfvalue}
\end{align}
As a check, let us retrieve the special case \eqref{eq:pdfmin1} from \eqref{eq:pdfvalue}. A possible way to proceed starts with setting $k = 1$ so \eqref{eq:pdfvalue} becomes $\overline{\theta(V_{\min,k} - y) \delta(V_{\min} - y_0)} = -G'_{\infty}(y) e^{y_0 - y} D(y - y_0) =  G'_{\infty}(y) K_\infty'(y-y_0)$ by \eqref{eq:partialKisD}; applying then $-\partial_y$ we get directly \eqref{eq:pdfmin1}.
The second, perhaps more instructive, way is to set $k = 2$ and $y = y_{k-1}$. Then there are two partitions of $\{y_0, y_1\}$: $\{\{y_0, y_1\}\}$ and $\{\{y_0\}, \{y_1\}\}$, so \eqref{eq:pdfvalue} can be calculated explicitly (we denote $\Delta = y_1 - y_0$, and use repeatedly \eqref{eq:decoration}):
\begin{align}
&\overline{\delta(V_{\min} - y_0) \delta(V_{\min,1} - y_{1})} = \mathcal{D}_1 G_{\infty}(y_1) e^{-\Delta} D(\Delta, 0) +
\mathcal{D}_2 G_{\infty}(y_1) e^{-\Delta} D(\Delta) D(0) \nonumber\\
= & -G'_{\infty}(y_1) e^{y_0 - y_1} \,\,\overline{\delta(v_{\min,1}-v_{\min} - \Delta)} + (G''_\infty(y_1) - G'_\infty(y_1))
e^{y_0 - y_1} \, \overline{\theta(v_{\min,1}-v_{\min} - \Delta)}  \nonumber\\
= & -G'_\infty(y_1) e^{-\Delta}\,\,\overline{(\delta(v_{\min,1}-v_{\min} - \Delta) + \theta(v_{\min,1}-v_{\min} - \Delta))} + G''_\infty(y_1) e^{-\Delta} \, \overline{\theta(v_{\min,1}-v_{\min} - \Delta)} \,,
\end{align}
which is equivalent to \eqref{eq:pdfmin1} combined with \eqref{eq:gapstat1} and \eqref{eq:gapstat2}.

At this point an initiated reader may have recognized that \eqref{eq:pdfvalue} is precisely the order statistics of a randomly shifted decorated Gumbel Poisson point process (SDPPP), with the decoration process given by \eqref{eq:defbiased} (we recall the definition of SDPPP in Appendix \ref{sec:dppp} and why its extreme value statistics coincides with \eqref{eq:pdfvalue}). Thus, by arriving at \eqref{eq:pdfvalue} we have achieved one of the main conceptual goals of this work. Namely, we demonstrated that the
 1-step replica symmetry breaking mechanism, properly understood and generalized, implies that the sequence of ordered minima of a general Euclidean-space logREM is described by SDPPP. To our knowledge, the approach is new and is a first clear demonstration of the ``1RSB $\Rightarrow$ SDPPP'' relation. Our result, although not unexpected, does not seem to be addressed from such a viewpoint and in such generality in the literature; the characterisation of the decoration process \eqref{eq:defbiased} seems, to the best of our knowledge, new.

 It is apparent from \eqref{eq:pdfvalue} that the order and gap statistics are primary objects of interest in our approach, and
 we feel they were not paid due attention in the existing SDPPP literature. To fill in this gap we shall study some implications  of \eqref{eq:pdfvalue} in the Appendix \ref{sec:sdppp}, which contains the derivation of \eqref{eq:gapgenerating0} and \eqref{eq:marginalgenerating}.

\section{Discussions on the UV sector}\label{sec:uvtechnical}
\subsection{A toy model}\label{sec:gREM}
One of the most essential points we hope to have clarified in our article is the way to take UV limit data into account when performing replica calculations. Although the contribution of UV sector is captured by  \eqref{eq:defKgeneral} in a somewhat sophisticated form, the predictions are amenable to direct numerical verifications, see Sect. \ref{sec:num}.  Therefore, it is instructive too see how eq. \eqref{eq:defKgeneral} can arise in an \textit{ab initio} calculation without any use of the replica-trick. For this purpose we consider below the following toy model. It is defined by $M$ values of the random potential that form blocks of a fixed size $N$, $V_j = V_{(g,i)},\,\, g = 1, \dots, M / N,\, i = 1, \dots, N$, such that different blocks are statistically independent and the correlations inside a block  mimic those of the logREM:
\begin{equation}
\overline{V_{(g,i)}} \defeq 0 \,,\quad \overline{V_{(g,i)}V_{(h,i')}} \defeq \delta_{g,h} (f_{ii'} + 2 \ln M) \,,
\end{equation}
where $f_{ii'}$ is a symmetric positive definite $N \times N$ matrix, with $N$ being fixed for simplicity.

This model retains the UV-limit correlations of general logREMs, but has trivial ``IR'' correlations. So the notion of the local logREM $(u_i)_{i=1}^N$ and its partition functions, defined in \eqref{eq:localrem} and \eqref{eq:smallw_general}, still makes sense here:
\begin{equation}
\overline{u_i} \defeq 0 \,,\quad \overline{u_i u_{i'}} \defeq f_{ii'} \,,\quad z \defeq \sum_{i=1}^N e^{-\beta u_i} \,,\quad w((\Delta_0, i_0), \dots, (\Delta_{k-1},i_{k-1})) \defeq
\sum_{l=0}^{k-1} e^{-\beta (u_{i_l} + \Delta_l)}  + \sum_{i\neq i_l} e^{-\beta u_i} \,. \label{eq:smallremtoy}
\end{equation}
The toy model is exactly $M/N$ independent copies of local logREMs, to each of which we added a constant centered Gaussian  variable shift $C_g$ of variance $2 \ln M$; shifts of different copies are independent. The independence among different blocks makes it possible to perform calculations without replicas.

For this we need some technical result standard in the study of uncorrelated REM. Let $C$ be a centred Gaussian of variance $2 \ln M$, and consider the following average over $C$:
\begin{equation} \gamma(x) \defeq \overline{\exp(-e^{\beta(x-C)})}^C =  \int_\R \frac{\dif c}{\sqrt{4\pi \ln M}} \exp\left(-\frac{c^2}{4\ln M} \right) \exp\left(-e^{\beta(x-c)}\right) \,. \label{eq:defREMgamma} \end{equation}
For convenience of the readers the asymptotic behaviour of \eqref{eq:defREMgamma} is analysed in Appendix \ref{sec:REMFyo}. The results can be summarized as follows:
\begin{align}
\gamma(x) \simeq 1 - \frac{e^{m \beta(x - f)}}{M} \,,\, f = \begin{dcases}
 -(\beta^{-1} + \beta) \ln M + O(1) \,, & \beta < 1 \,,\\
 -2\ln M + \frac{1}{2} \ln \ln M + O(1) \,, & \beta > 1 \,.
\end{dcases} \label{eq:gREM}
\end{align} 
The above equations hold asymptotically as $M\rightarrow+\infty$ with $x - f \sim O(1)$ fixed, and $ m  = \min (1, 1 / \beta)$ is the same as defined in \eqref{eq:moptimum}.

Coming back to our toy model, we start with the observable for the minimum value\eqref{eq:defG}. By independence, it is given by $M/N$-th power of the contribution of an individual block (recall $\theta_\beta(x) = \exp({-e^{-\beta x}})$):
\begin{equation}
G_\beta(y) = g_\beta(y)^{\frac{M}{N}} \,,\quad g_\beta(y) \defeq \overline{\prod_{i=1}^N {\theta_\beta(V_{g,i} - y)}}  = \overline{\prod_{i=1}^N {\theta_\beta(u_{i} + C_g - y)}} = \overline{\exp(-e^{\beta (y-C_g)} z)} \,. \label{eq:Gtoy}
\end{equation}
Note that $g_\beta(y)$ is independent of the block index $g$, since all blocks are identical. The average $\overline{[\dots]}$ here goes over the local logREM and over the local $C_g$. Clearly, by \eqref{eq:defREMgamma}, $g_\beta(y) = \overline{\gamma(y + \beta^{-1} \ln z)}$ (where the over-line in the RHS refers to the random variable $z$ alone). Applying \eqref{eq:gREM}, we reduce \eqref{eq:Gtoy} in the limit $M \rightarrow \infty$ to
\begin{equation}
G_\beta(y) = \exp\left(- e^{\beta m(y - s)} \overline{z^{m}} N^{-1}\right) \,.  \label{eq:GbetaREM}
\end{equation}
In the $\beta\rightarrow\infty$ limit this implies that the the minimum value for the toy model is distributed, as expected, according to a shifted Gumbel law, with the shift depending on the minimum value of the local REM potential $\overline{z^{1/\beta}} \rightarrow_{\beta \to +\infty} \overline{\exp(-\min_i u_{i})}$.
Note that in comparison with logREMs \eqref{eq:FM} the shift for the toy model displays a different coefficient
 in front of $\ln\ln M$ term, consistent with the fact that our toy model is \textit{not} a logREM, according to the definition \eqref{eq:logremdef}.

Now to probe the higher order statistics, we calculate the observable \eqref{eq:Hgeneral}, summed over marker positions. Observe first that the sum over markers positions in the LHS is factored into a sum over partitions indicating the occupation of the blocks, and the combinatorial factor $\left(\frac{M}{N}\right)_p$ that chooses $p$ blocks for the parts. The blocks without markers each yield the same contribution as in the RHS of \eqref{eq:Gtoy}, while the blocks with markers contribute a similar factor, with $z$
replaced with the local partition function with a few value shifts in the arguments, \textit{i.e.}, $w((\Delta_0, i_0), \dots,(\Delta_{k-1},i_{k-1}))$. Finally, we get
\begin{align}
\sum_{j_0, \dots, j_{k-1}} H_\beta(y_0,j_0, \dots, y_{k-1}, j_{k-1}, y) =& (-1)^k \sum_{\mathbb{P}(k)} \left(\frac{M}{N}\right)_p
g_\beta(y)^{\frac{M}{N}-p} \prod_{q=1}^{p} k(y-y_{q,0}, \dots, y-y_{q,k_q - 1}) \label{eq:Htoy} \\
k(\Delta_0, \dots, \Delta_{k-1}) =& \left. \sum_{i_0, \dots, i_{k-1}} \overline{\exp(-e^{\beta (y-C)}  w((i_0,\delta_0), \dots, (i_{k-1},\delta_{k-1})))}\right\vert_{\delta_j=+\infty}^{\Delta_j} \,. \label{eq:Ktoy}
\end{align}
Note that the above equations are exact for finite $M$. In the $M\rightarrow \infty$ limit,  by applying \eqref{eq:defREMgamma} and \eqref{eq:gREM} with $x = y + \beta^{-1}\ln w((i_0,\delta_0), \dots, (i_{k-1},\delta_{k-1}))$ to \eqref{eq:Ktoy} and using \eqref{eq:NewtonLeib}, we can simplify them to the following form:
\begin{align}
\sum_{j_0, \dots, j_{k-1}} H_\beta(y_0,j_0, \dots, y_{k-1}, j_{k-1}, y) &= (-1)^k
\sum_{\mathbb{P}(k)} G_\beta(y) \prod_{q=1}^{p} \tilde k(y-y_{q,0}, \dots, y-y_{q,k_q - 1})  \,, \label{eq:Htoy1} \\
\tilde k(\Delta_0, \dots, \Delta_{k-1}) =& - \left. \sum_{i_0, \dots, i_{k-1}}  e^{\beta m(y - s)} N^{-1}  \overline{w((i_0,\delta_0), \dots, (i_{k-1},\delta_{k-1}))^{m}} \right\vert_{\delta_j=+\infty}^{\Delta_j} \,. \label{eq:Ktoy1}
\end{align}
To compare with \eqref{eq:maingeneral}, we shall apply the differential operator $\mathcal{D}_p = -\partial_y(1 -\partial_y)\dots (p-1-\partial_y)$ (eq. \eqref{eq:Dpb}) to $G_\beta(y)$ (eq. \eqref{eq:GbetaREM}). One can check by recursion that
\begin{equation}
\mathcal{D}_{\beta,p} G_\beta(y) =  \left( \overline{z^{m}} N^{-1}\right)^p \exp\left(- e^{\beta m(y - s)} \overline{z^{m}} N^{-1} + p \beta m (y-s) \right) = G_\beta(y)
\left( \overline{z^{m}} N^{-1} e^{\beta m(y-s)} \right)^p \,.
\end{equation}
Combining this with \eqref{eq:Htoy1} and \eqref{eq:Ktoy1},  one shows with some rearrangement that
\begin{align}
\sum_{j_0, \dots, j_{k-1}} H_\beta(y_0,j_0, \dots, y_{k-1}, j_{k-1}, y) &= (-1)^{p+k}
\sum_{\mathbb{P}(k)} \mathcal{D}_{\beta,p} G_\beta(y) \prod_{q=1}^{p} K(y-y_{q,0}, \dots, y-y_{q,k_q - 1}) \,, \\
K(\Delta_0, \dots, \Delta_{k-1}) =& \left. \sum_{i_0, \dots, i_{k-1}} \frac{\overline{w((i_0,\delta_0), \dots, (i_{k-1},\delta_{k-1}))^{m}}}{\overline{z^m}} \right\vert_{\delta_j=+\infty}^{\Delta_j} \,.
\end{align}
These equations are identical to \eqref{eq:maingeneral}, with the same UV sector contribution \eqref{eq:defKgeneral}, except that the UV block size $N$ is fixed here. This result is obtained without any use of the replica trick, by a well-controlled calculation.on a variant of uncorrelated REM that mimics the correlations of logREMs at the UV scale, so the coincidence of the results confirms the validity of the replica approach to UV sector.

\subsection{Biased minima: local-global equivalence}\label{sec:biased}
In a log-correlated model the scale separation between UV and IR is only emergent in the thermodynamic limit. Note that, in contrast to the toy model considered above, for log-correlated models the size of UV block $N$ is not fixed and one should consider it infinite in the thermodynamic limit as well. As we have however seen in Sect. \ref{sec:modeldef}, in general, the covariance matrix of the local logREM has a logarithmic behavior at large distance.

Therefore, when $N \rightarrow \infty$ the local minimum $v_{\min}$ will have the same log-REM signatures as $V_{\min}$. Moreover, the minima in the local REM can be separated by distances diverging in the thermodynamic limit, so the gap process of local minima  $(u_{\min,k} - u_{\min})_{k\geq1}$ can be in practice hard to distinguished from that of the global minima $(V_{\min})_{k\geq1}$ of some logREM. The worst scenario seems to arise for the case of the interval model (with $c_{1,2} =  d_{1,2} = 0$, see Sect. \ref{sec:interval}), where the local logREM looks identical to the global one with an appropriate choice of $C_N$. These facts may cast legitimate doubts on the consistency of our strategy of separating UV contributions.

To this end, we are able to demonstrate that statistical properties of the biased minima are unchanged by a local-global replacement $(u_{\min,k}) \leadsto (V_{\min,k})$. To show this, it suffices to calculate
\begin{equation}
\widehat{D}(\Delta_0, \dots, \Delta_{k-2}, \Delta_{k-1} = 0) = \frac{\overline{\prod_{s=0}^{k-2} \delta(V_{\min,s} + \Delta_s - V_{\min,k-1} )\exp(-V_{\min})}}{\overline{\exp(-V_{\min})}} \,, \label{eq:hatD}
\end{equation}
which differs from \eqref{eq:decoration} by the local-global replacement (and by setting $\Delta_{k-1} = 0$, which causes no loss of generality since we cover all $k$'s). The statistics of $(V_{\min,k})$ is given by \eqref{eq:pdfvalue}
(where we take now $y=y_{k-1}$) which implies
\begin{align}
 &\widehat{D}(\Delta_0, \dots, \Delta_{k-2},\Delta_{k-1} = 0) = \left(\overline{\exp(-V_{\min})} \right)^{-1} \int_{\R}\dif y e^{-y+\Delta_0} \sum_{\mathbb{P}(k)} \mathcal{D}_p G_{\infty}(y) \prod_{q=1}^p e^{-\Delta_{q,0}} D(\Delta_{q,0},\dots, \Delta_{q,k_q - 1}) \label{eq:numeratorhatD}
\end{align}
Here, $y$ stands for $V_{\min,k-1}$ and is integrated over. To proceed further, recall by \eqref{eq:Dp} that $\mathcal{D}_p = -\partial_y (1-\partial_{y}) \dots (p-1-\partial_y)$. Note that for any function $f(y)$ satisfying  $f(y) e^{-y} \rightarrow 0$ when  $y \rightarrow \pm\infty$, we have $\int_{\R} e^{-y} (1 - \partial_y) f(y) = - \int_{\R} \partial_y (e^{-y} f(y)) = 0$. Applying this to $f = - \partial_y (2 - \partial_y)\dots (p-1-\partial_y) G_{\infty}$, and assuming fast enough decay of $G_{\infty}(y)$ with $y\to \pm \infty$ (we will come back to this crucial point below), we obtain
\begin{equation} p > 1 \Rightarrow \int_{\R} e^{-y} \mathcal{D}_p G_{\infty}(y) = 0  \,. \label{eq:expDpG}  \end{equation}
So the sum \eqref{eq:numeratorhatD} has in fact only one surviving term given by the trivial partition, for which we have $p = 1$ and $\Delta_{1,s} = \Delta_s$, $s = 0, \dots, k-1$, with $\Delta_{k-1} = 0 $. Now by \eqref{eq:Ginfty} we have $ \int_{\R} e^{-y} \mathcal{D}_1 G_{\infty}(y)  = \overline{\exp(-V_{\min})}$. Incorporating these into the RHS of \eqref{eq:numeratorhatD} simplifies it to
\begin{equation}
\widehat{D}(\Delta_0, \dots, \Delta_{k-2},0) = D(\Delta_0, \dots, \Delta_{k-2}, 0) \,. \label{eq:DequaltoD}
\end{equation}
This means precisely that statistics of the biased minima can be obtained from either local logREM minima following \eqref{eq:decoration}, or from the original logREM minima following \eqref{eq:hatD}, with the results being identical. Moreover, the calculation tells us that the biased minima process contains solely UV information: indeed, the only contributing partition corresponds to the configurations in which all the minima in consideration are in one UV block. Therefore, even if one supposes that further scale separation emerged inside the blocks themselves, so that the local minima $(u_{\min,k})$ acquire an addition partitioning structure of the global minima, only the trivial partition would contribute to the biased minima process $(v_{\min,k})$. This fact makes it clear that the assumption of an additional scale separation inside the UV blocks is {\it de facto} immaterial.

As promised, we will now justify the assumption of a fast decay of  $G_{\infty}(y\rightarrow \pm \infty)$. It is clear that the sufficient and necessary condition of \eqref{eq:expDpG} is $e^{-y} \partial_y G_{\infty}(y) \rightarrow 0$ as $y \rightarrow \pm \infty$. The right tail $+\infty$ poses no problem, but the left tail does: it is well known that the asymptotic Carpentier-Le Doussal (CLD) tail $\partial_y G_{\infty}(y) \sim A y e^y$ holds for generic logREM \cite{carpentier2001glass}, so $e^{-y} \partial_y G_{\infty}(y) \propto y$ does not vanish. However, the CLD tail holds down to $y = -\infty$ only in the thermodynamic $M\rightarrow\infty$ limit; for any system of finite size $M$, when $y / (2 \ln M) \gg 1$, the tail crosses over to a Gaussian tail $\sim B \exp(-C y^2)$\footnote{To understand this, it is instructive to have a brief look at a similar tail
for a uncorrelated REM, which has the shape $e^{-y}$. To this end we calculate $\overline{\theta(V_{\min} - y)}$ for a uncorrelated REM. Letting $y = -2 \ln M + y_0$, we have $\overline{\theta(V_{\min} - y)} = \exp(- \exp(y_0 - y_0^2 / 4 \ln M ) ) \sim 1 - \exp(y_0 - y_0^2 / 4\ln M)$ as $y_0 \rightarrow  -\infty$. If $y_0 / \ln M$ does not vanish  in the $M\rightarrow\infty$ limit, the quadratic term cannot be neglected.}, which still decays to $0$ after multiplying by $e^{-y}$. This argument justifies eq. \eqref{eq:expDpG}, and provides a further insight: the exponential weight $\exp(-u_{\min})$ of the biased minima process probes configurations with \textit{atypically} negative values beyond the logREM scaling regime $y = - 2 \ln M + o(\ln M)$. Given the correlation structure of logREM, multiple occurrence of values much more negative than the typical scale $- 2 \ln M$ is much more likely to happen at nearby points compared to that of $ \sim  - 2 \ln M$ values. This reinforces the claim that the biased minima process concerns exclusively the UV data, and explains the numerical observations made in Fig. \ref{fig:gapstats} (Left):  its numerical measure has faster convergence with respect to $N$, but the relevance of atypical events means that higher statistics are necessary.

\section{Conclusion and outlook}
In this paper we systematically applied the replica symmetry breaking (RSB) formalism to the extreme order statistics of a generic (see Sect. \ref{sec:modeldef} for definition of scope) logarithmic random energy models defined on Euclidean spaces. As is well-known, for
all such models the one-step replica-symmetry broken solution (1RSB) characterized by the familiar overlap distribution $P(q) = m \delta(q) +(1 - m)\delta(q -1)$,
$m = min(\beta, 1)$ (see Sect. \ref{sec:1rsb} and Appendix \ref{sec:frsb}, in particular \eqref{eq:Pq} and \eqref{eq:moptimum}) captures correctly the thermodynamics of the model. Our main finding is that
after appropriate ramifications the same 1RSB mechanism allows one to efficiently analyse fluctuations in the
ordered sequence of the deepest minima in the model. These fluctuations turn out to be crucially dependent on the IR and UV scaling limits (Sect. \ref{sec:modeldef}) of particular models, and in a peculiar way (see Sect. \ref{sec:fullminima}).
Namely, the process of minimal values form a randomly shifted and decorated Gumbel Poisson point process (see Appendix \ref{sec:dppp}). The random shift depends only on the IR data (up to a translation, it has the same law as the critical temperature free energy).  The decoration process depends only on the UV data, and is identified with the biased minimum process described in Section \ref{sec:biased}. This picture is in full agreement with that obtained in \cite{brunet2011branching} and rigorously in \cite{aidekon2013branching,arguin2013extremal} for hierarchical logREMs. Motivated by recent work of  \cite{biskup2016extreme,biskup2016full}, we also extend the results to include minimum positions in Appendix \ref{sec:minposition}.

We worked out in a considerable detail the implications of this picture for the form of the joint distribution of the first and second minima in Section \ref{sec:mainresult}, and provided two numerical tests for the circular model in Section \ref{sec:num}. For the first gap, see Fig. \ref{fig:gapstats}, we tested direct numerical simulations of the original model \textit{vs.} our analytical prediction (combined with numerically evaluated UV contributions), and {also analyzed a situation} where the decoration process is trivial and the first gap is exponentially distributed. For the fluctuation of the second minima, see Fig. \ref{fig:gapstats}, our prediction \eqref{eq:prediction1} (in general \eqref{eq:pdf2ndmin}) constrains its distribution to a one-parameter family, where the parameter is given by the mean value of the first gap. We also derive a generalization of \eqref{eq:pdf2ndmin} to arbitrary order statistics: the distribution of the $k$-th minimum is determined by that of the first minimum and the mean value of $k-1$ first successive gaps, in a universal way encoded in the generating function \eqref{eq:marginalgenerating}.  

We also clarified the relation of 1RSB approach to that based on freezing, in particular, the freezing-duality conjecture, see Section \ref{sec:FDC}. The main message is that FDC holds true for observables probing exclusively IR data. When the UV data is relevant, non-trivial temperature dependence takes place in the $\beta > 1$ phase.

What is left for future work is the remaining major challenge of performing analytical calculations of the decoration/biased minimum process. Although this process is fully characterized for BBM in \cite{aidekon2013branching,arguin2013extremal} and for discrete GFF in \cite{biskup2016full}, we are not aware of any quantitative prediction of the gap distributions, besides the trivial decoration process $v_{\min} = 0, v_{\min,1} = v_{\min,2} = \dots = + \infty$, which is obtained when the minima are constrained to be far away from each other (see Sect. \ref{sec:num} for precise definition). At least two approaches may be anticipated towards that goal: performing perturbation expansion around the trivial case to obtain approximate solutions, or relating the decoration process to some discrete integrable model (recall that explicit predictions of the minimum value distribution have so far relied on exact evaluation and subsequent analytical continuation of the Coulomb gas integrals \eqref{eq:contZ}). Implementing either of the approaches would yield new and interesting results.


\textit{Acknowledgements.} We thank P.-L. Arguin for helpful explanations on the decorated Poisson point process. XC is grateful to S. Franz and A. Rosso for stimulating discussions at early stages of this work. XC and PLD thank the hospitality of KITP (under Grant No. NSF PHY11-25915), where part of this work was done. The research at King's College London was financially supported by
EPSRC grant EP/N009436/1 ``The many faces of random characteristic polynomials''.

\appendix
\section{Full RSB of logREM: 1 step and beyond}\label{sec:frsb}
In this Appendix we give another justification of the 1RSB Ansatz used in the main text by performing a general full RSB analysis of the logREMs. We shall show that, assuming that full RSB Ansatz applies, the general definition \eqref{eq:logremdef} of logREM suffices to show that the optimal solution is confined to the 1RSB sector described in the main text. The formalism here generalizes naturally to the multi-step variants of logREMs \cite{fyodorov2007explicit,fyodorov2008multiscale}, which we shall also discuss. Note that for the special case of log-correlated models in $d\gg 1$-dimensional space such statement was verified in \cite{fyosom2007}. Here we will follow very different line of reasoning.  Comparing and combining the two is an interesting task that we leave for future work.

RSB is best described in terms of \textit{overlaps} between positions. In logREMs, we will define them via
\begin{align} q(j,k) \defeq \overline{V_{j,M} V_{k,M}}^{c} / (2\ln M)  \,. \label{eq:overlapdis} \end{align}
Note in particular that \eqref{eq:variance} implies $q(j,j) = 1$ as $M\rightarrow \infty$. In general $0 \leq q \leq 1$, and the two limiting values correspond to IR and UV limits (see Sect. \ref{sec:synopsis}), respectively. Distinct macroscopic positions have zero overlap: indeed,  when $\xi_{j,M} \rightarrow \xi $ and $ \xi_{k,M} \rightarrow \eta$ with $\xi \neq \eta$ we have  $q(j,k) \rightarrow  C(\xi, \eta) / (2 \ln M) \rightarrow 0$ (see \eqref{eq:irdata}). On the other hand, points separated by a distance of finite number lattice spacings have overlap $1$, see \eqref{eq:intra}. In hierarchical logREMs the overlap coincides with its literal interpretation: the length of common part between two paths/walkers. To put the analysis here in a more general context we shall perform it for a larger class of models containing logREMs, defined by generalizing \eqref{eq:logremdef} in terms of the overlap to the following form:
\begin{equation}
\forall k, \abs{\{ \, j \,: q(j,k) > q  \}} \sim M^{1 - f(q)},\,\,  q \in [0,1]\,;\quad f(0) = 0,\, f(1) = 1, \quad f'(q) ,\, f''(q)\geq0 \,,\, \label{eq:logremdefrsb}
\end{equation}
where we recall that $\abs{\{[\dots]\}}$ stands for the number of elements. The logREM class is retrieved with $f(q) = q$.  Among the general assumptions on $f(q)$, the monotonicity $f'(q) \geq 0$ is by definition. $f(0)=0$ ensures that $\overline{V_j V_k}^c / \ln M$ stays non-negative as $M\rightarrow\infty$ \footnote{In this work we are taking the UV-divergence point of view, in which in the $M\rightarrow +\infty$ limit, the total system size and the covariance at system-scale remain finite, but the lattice cut-off goes to $0$, and the short-distance covariance diverges to $+\infty$. This should be noted when comparing our results with works that take the IR-divergence (infinite system size) point of view. \label{foot9}}. Finally, $f(1) = 1$ preserves $\overline{V_j^2} = 2 \ln M$ \eqref{eq:variance}; finally the reason of the convexity assumption will be clear below in \eqref{eq:moptimum1gen}.

\subsection{Optimization of the RSB order parameter}
Now, we can rewrite the replica sum $\overline{\mathcal{Z}^n}$ \eqref{eq:laplace} as an integral over overlaps between pairs of replicas, which form the overlap matrix $Q_{ab}$:
\begin{align}
\overline{\mathcal{Z}^n} &=  \sum_{(j_a)} \overline{\exp\left( - \sum_{a} \beta V_{j_a} \right)} = \int \mathcal{D} Q  \exp(-\ln M \, H[Q]) \label{eq:ZnHQ} \\  H[Q] &=  -E[Q] - S[Q]+ o(\ln M) \,,\quad E[Q] = \sum_{a,b} \beta^2 Q_{ab} \,,\quad S[Q] = \frac{1}{\ln M} \ln \sum_{(j_a)} \prod_{a < b} \delta(q(j_a, j_b) - Q_{ab})  \label{eq:SQ}
\end{align}
Here $\mathcal{D} Q = \prod_{a < b} \dif Q_{ab}$ and the integral is taken over all matrices satisfying $Q_{aa} = 1$ and $Q_{ab} = Q_{ba} \in [0,1]$. Comparing to the treatment in the main text all the UV and IR data give $o(\ln M)$ corrections which are not explicitly written here. This is done in order to focus on the extensive quantities ({note, however, that those corrections are the most essential ingredient when one studies fluctuations}). In the ``action'' $H[Q]$, the term $E[Q]$ comes from $\overline{\exp(\sum -\beta V_{j_a})}$ (by Wick theorem) and $S[Q]$ is the entropy term which counts the number of replica configurations with the given overlap. To calculate the action $H[Q]$ we will resort to the assumptions of the RSB Ansatz. The first assumption is that $Q_{ab}$ has the Parisi's RSB form, determined by the overlap distribution function $P_>(q)$:  
\begin{align}
\abs{\{ b \neq a: Q_{ab} > q\}} = (n - 1) P_>(q) \,. \label{eq:Pq}
\end{align}
This is the (continuous) full-RSB Ansatz. Note that $P_>(q)$ depends on $n$ as well as $q$. It is known (and can be shown by the standard replica trick) that $P_>(q)\vert_{n\rightarrow0}$ gives the probability that two particles thermalized according to the Boltzmann-Gibbs measure generated by the same random potential occupy positions with overlap $> q$. So $P'_>(q)\leq 0$ for $n \rightarrow 0$.

In terms of the overlap distribution $P_>$, the energy part of the action is
\begin{equation} E[Q] = \beta^2 n \left( 1  + (n-1) \int_0^1 \dif q  (-\partial_q P_>(q)) q \right) =
\beta^2 n \left( 1  + (n-1) \int_0^1 \dif q P_>(q) \right) \,. \label{eq:EQ}\end{equation}
Note that the energy part depends on the definition of the overlap {in terms of the covariance} \eqref{eq:overlapdis}, but not directly on $f(q)$.

Now we consider the entropy part $S[Q]$. To simplify, let us first restrict our attention to the discrete RSB (and then take the continuous limit), where $P_> (q)$ is a staircase-like function:
\begin{equation}
P_>(q) = p_i \,, \, q \in (q_i, q_{i+1}) \,,\quad  0 \leq i \leq s \,,\quad  1 = p_0 > p_1 > \dots > p_s > p_{s+1} = 0 \,,\quad 0 = q_0 < q_1 < \dots < q_s \leq q_{s+1} = 1 \,, \label{eq:rsbdiscrete}
\end{equation}
$s$ being the number of \textit{steps}. The discrete RSB has an equivalent description in terms of hierarchy of replica groups. To spell it out, let us choose an arbitrary replica index $a$ as a reference. Let $i =0,\dots,s+1.$ Among the $n-1$ other replicas $(n-1)p_i$ of them have overlap $\geq q_i$ with $a$ (for $n$ large the inequality is saturated by \eqref{eq:rsbdiscrete}). Therefore, the \textit{$i$-group} formed by these elements and the added replica $a$ is of the size
 \begin{equation}
 m_i = m(p_i) \,,\quad  m(p) \defeq 1 + (n-1)p  \,.  \label{eq:mofP}
 \end{equation}
 For $i = s + 1$ the group consists of a single replica: $m_i=1$; while for $i = 0$ the group contains all the replicas. Now, by \eqref{eq:logremdefrsb}, the members of the $i$-th group occupy a volume of $M^{1-f(q_i)}$ (number of points $j$ such that $q(j,j_a)>q_i$). So seen as a whole, the group's position is determined only up to $M^{1-f(q_i)}$. On the other hand, for any $i > 0$ the level-$i$ group is contained in a level-$(i-1)$ group. The entropy contribution of a level-$i$ group is the log-volume  available for different positioning of the group as a whole (\textit{i.e.} the volume of the level-$(i - 1)$ group containing it), minus the log of its position resolution (\textit{i.e.} its own volume; what happens at smaller scales will be taken care of in contributions of smaller groups):
\begin{equation} S(\text{one level-}{i}-\text{group}) = \ln  M^{1-f(q_{i-1})} - \ln M^{1-f(q_{i})}= (f(q_i) - f(q_{i-1})) \ln M \,. \label{eq:Scontribution} \end{equation}
$a$ being arbitrary, what we said holds for any $i$-group; in particular, there are $n/m_i$ of them in total. So we can calculate $S[Q]$, given by the sum of \eqref{eq:Scontribution} over \textit{all groups}, as follows:
\begin{equation} S[Q] = \frac{n}{m_1} (f(q_1)-f(q_0)) + \frac{n}{m_2} (f(q_2) - f(q_1)) + \dots + \frac{n}{m_{s+1}} (f(q_{s+1})-f(q_s)) = n \int_0^1 \dif q f'(q) (1 + (n-1) P_> (q) )^{-1} \,. \label{eq:SQintegral}\end{equation}
Now we can take the $s\rightarrow\infty$ limit where $P_>(q)$ tends to any {non-increasing} function; in that case the RHS still makes sense. Combining with \eqref{eq:EQ}, we have
\begin{align}
H[Q] = - n \int_0^1 \dif q \left( \beta^2 m(P_> (q)) + f'(q) m(P_> (q))^{-1} \right)  \,, \label{eq:HQsimple}
\end{align}
where $m(\dots)$ is defined in \eqref{eq:mofP}. Considering $\ln M$ as a large parameter we can employ the saddle-point method to evaluate
the integral in \eqref{eq:ZnHQ}, and thus have to minimize $H[Q]$ with respect to $P_>(q)$. By \eqref{eq:mofP}, we can equivalently optimize with respect to $m = m(P_>(q))$ for $n$ integer, and then continue the answer to $n \rightarrow 0_+$. This operation involves two subtleties, both well known in general replica calculations \cite{mezard87beyond}. The first concerns the range of the group size $m$. Naïvely, $m \in [1, n]$ when $n$ is an integer; however, when $n$ is continued to $n \rightarrow 0_+$, we demand $m$ be still between $1$ and $n$, \textit{i.e.}, $m \in (0,1]$. The second subtlety is that after the continuation the minimization becomes maximization. To see this one may use  \eqref{eq:mofP} and integration by parts to rewrite eq. \eqref{eq:HQsimple} as
 \begin{align}
 H[Q] &= -n\beta^2 - n(n-1) \int_0^1 \dif q \beta^2 P_> (q) - n \int_0^1  (1 + (n-1)P_> (q))^{-1} \dif f(q) \nonumber \\
  & =  -n\beta^2 - n(n-1) \int_0^1 \dif q \beta^2 P_> (q) - n -  n(n-1)\int_0^1  (1 + (n-1)P_> (q))^{-2} f(q)\dif q \,,
 \end{align}
where we recalled $f(0) = 0, f(1)=1$ \eqref{eq:logremdefrsb} and $P_>(1) = 0$ from \eqref{eq:Pq}. Then we observe that the $P_>(q)$ dependent-part is proportional to $n(n-1)$, which changes sign when going from $n > 1$ to the range $0<n<1$.

With these technical points in mind, the actual optimization of \eqref{eq:HQsimple} is rather straightforward. In fact, the problem amounts to the maximization of $h[m] = -n (\beta^2 m + f'(q) m^{-1} )$ with respect to $m\in (0,1]$, for each $q \in [0,1]$. The solution is
\begin{equation}
m(P_>(q)) = \min\left(1, \beta^{-1} \sqrt{f'(q)}\right)  \,,\,
\label{eq:moptimum1gen}
\end{equation}
Note that the convexity assumption $f''(q) \geq 0$ ensures $\dif m / \dif q \geq 0$, so that $P'_>(q) \leq 0$ for $n \rightarrow 0$ (since $m(P) = 1 + (n-1)P$). Applying \eqref{eq:moptimum} to \eqref{eq:HQsimple} and \eqref{eq:ZnHQ}, and using the saddle point approximation, we have the extensive free energy
\begin{equation}
\lim_{M\rightarrow\infty}\mathcal{F}/\ln M = \beta(q_\beta-1) + \beta^{-1}(f(q_\beta) - 1) - 2 \int_{0}^{q_\beta}  \sqrt{f'(q)}  \dif q \,,\, f'(q_\beta) = \beta^2 \,. \label{eq:freeenergygen}
\end{equation}
We note that such  a formula for free energy appears also for the generalized Random Energy Model \cite{derrida1985generalization,derrida1986solution,bovier2004grem}.

\subsection{LogREM}
In this subsection we set $f(q) = q$ to specify the above results to logREM. By \eqref{eq:moptimum1gen}, $m(P_>(q)) = \min\left(1, \beta^{-1}\right)$ is a constant (with respect to $q$), which coincides with \eqref{eq:moptimum}. Eq. \eqref{eq:freeenergygen} recovers also the logREM free energy density \eqref{eq:FM}. Finally, applying \eqref{eq:mofP} with $n \rightarrow 0_+$, we obtain
\begin{equation} P_>(q \in (0,1))\vert_{n = 0} = 1 - \min(1, \beta^{-1}) \,. \end{equation}
 The above result is then the well-known zero-one law of overlap in logREM: the two particles have zero overlap with probability $\min(\beta^{-1}, 1)$, and have unity overlap with probability $\max(1 - \beta^{-1}, 0)$. It was first calculated on hierarchical logREMs \cite{derrida1988polymers}. Later it is argued to hold for Euclidean ones in \cite{carpentier2001glass}, Sect. III.E.2. More recently, it has been proved quite generally \cite{arguin2014poisson}. The RSB analysis here does not assume hierarchical structure of the logREM. Nevertheless, inherent to any RSB Ansatz is the assumption of  hierarchical organisation (overlap structure) of the replicas.

Let us identify the logREM result with the 1RSB Ansatz described in Sect. \ref{sec:1rsb}. When $\beta < 1$, $P_> (q)$ {is of the form} \eqref{eq:rsbdiscrete} with $s = 0$, $q_0 = 0$: this is the \textit{replica symmetric} solution such that distinct replicas have overlap $0$, \textit{i.e.}, their separations {are of the order} of IR scale. When $\beta > 1$, we have $s = 1$, $q_0 = 0, q_1 = 1$ and $p_0 = \frac{m-1}{n-1} = \frac{1/\beta - 1}{n-1}$ (note that $p_0 \in [0,1]$ only when $n\rightarrow 0$). This is the 1-step RSB solution: the replicas form groups of size $m$ given by \eqref{eq:moptimum}. Distinct groups have overlap $0$, \textit{i.e.}, are separated by system size, while replicas of the same group have their mutual overlap equal to unity, so their separation is of UV (lattice spacing) scale.

\subsection{Multi-scale generalization}\label{sec:multistep}
Going beyond logREM, let us discuss the physical significance of general $f(q)$, by relating to the works \cite{fyodorov2007explicit,fyodorov2008multiscale}, which considered \textit{multi-scale} generalisations of Euclidean logREMs. We recall that these models have the following covariance matrix (see eq. (13) of \cite{fyodorov2008multiscale}) :
\begin{equation}
\overline{V_{j,M}V_{k,M}}^c =  - 2 \ln M \, \Phi\left( \frac{\ln (\abs{i-j} + 1)}{\ln M}\right)\,,  \label{eq:covariancemulti}
\end{equation}
where $\Phi$ satisfies $\Phi'(y)\geq 0$, $\Phi''(y) \geq 0$. We demand moreover $\Phi(0)=-1$ and $\Phi(1)=0$ so that $\overline{V_{j,M}}^{c} = 2\ln M$ \eqref{eq:variance} still holds, and that points separated by the system size are almost non-correlated . One can check that for logREM, $\Phi(y) = y-1$. Using \eqref{eq:overlapdis}, it is direct to show that \eqref{eq:logremdefrsb} is satisfied with
\begin{equation}
-q = \Phi(1 - f(q)) \,, \label{eq:fqPhiy}
\end{equation}
\textit{i.e.}, they are essentially inverse of each other. Note that the properties of $\Phi(y)$ above are equivalent to those of $f(q)$. \eqref{eq:moptimum1gen} implies that every value of $\sqrt{f'(q)}$ is a critical inverse temperature ,\textit{i.e.}, a non-analyticity of $m$.

It is helpful to consider first the case of \textit{discrete $s$-step RSB}, which can be retrieved by a piece-wise linear function determined by its cusps:
\begin{equation}
f(q_i) = f_i \,,\,  0 = q_0 < q_1 < \dots < q_{s} = 1 \,,\,0 = f_0 < f_1 < \dots < f_{s} = 1 \,,\, \beta_i \defeq\sqrt{\frac{f_i - f_{i-1}}{q_i - q_{i-1}}} < \beta_{i+1} \,. \label{eq:fkstep}
\end{equation}
Note that $\sum_{i=1}^s \beta_i^2 (q_i - q_{i-1}) = \sum_{i=1}^s (f_i - f_{i-1}) = 1,$ generalizing \eqref{eq:betac} of the 1RSB-logREM. Using \eqref{eq:fqPhiy}, this gives a piece-wise linear $\Phi$ with singular points $\Phi(1 - f_i) = -q_i$. A concrete realisation is given by a $s$-scale logREM  (compare with eq. (15) of \cite{fyodorov2008multiscale})\footnote{{Note, that the paper \cite{fyodorov2008multiscale} used IR rather than UV point of view, see the previous footnote \ref{foot9}}}):
\begin{equation}
\overline{V_{j,M}V_{k,M}}^c = \sum_{i=1}^s 2g_i  \ln \frac{M^{f_i}}{1 + \abs{j-k} M^{f_i - 1}} \,,\,
g_i = \beta_i^{-2} - \beta_{i+1}^{-2} \,, \, i = 1, \dots, s\,. \label{eq:covariancekstep}
\end{equation}
where we define $\beta_{s+1}^{-2} = 0$ by convention. One may check that \eqref{eq:covariancekstep} and \eqref{eq:fqPhiy} imply \eqref{eq:fkstep}. Note that $\sum_{i=1}^s g_i = \beta_1^{-2}$. $V_{j,M}$ can be seen as the independent sum of $s$ ``logREM''s, of which the $i$-th is smooth below the scale $M^{1 - f_i}$ and log-correlated above. As the temperature $T = 1/\beta$ decreases to $< 1/\beta_1$, the particle starts to be confined in a block of length $M^{1- f_1}$, inside of which the first ``log-REM'' becomes smooth and no longer affects the thermodynamics. The same repeats at scale $M^{1-f_2}$ as one continues to cool down to $T = 1/\beta_2$, and so on. After the $s$-th (last) transition, the confinement scale becomes $M^{1-f_s} = 1$, \textit{i.e.}, the UV scale.

Now, the generic \textit{continuous RSB} can be seen as the limit in which $s\rightarrow \infty$ and the transition points $\beta_i$ and $\beta_{i+1}$ are separated by $O(1/s)$. One has a replica symmetry phase for $0 \leq \beta \leq  \beta_{\min} = \sqrt{f'(0)}$, beyond which we have full RSB phase\textit{s} with a continuum of phase transitions, until $\beta = \beta_{\max} = \sqrt{f'(1)}$. Finally, if $\beta_{\max} < +\infty$, for $\beta > \beta_{\max}$ one has a frozen phase: from \eqref{eq:freeenergygen} free energy density is
\begin{equation}
\mathcal{F}= -2 \ln M \int_0^1 \sqrt{f'(q)} \dif q + o(\ln M) \,.
\end{equation}
The free energy density becomes independent of $\beta$ and gives also the leading behaviour of the minimum. Leaving detailed analysis to future work, we expect the full minima process to be a randomly shifted decorated Ruelle cascade \cite{ruelle1987mathematical}, where Ruelle cascade is the analogue of Gumbel Poisson point process for full RSB, and the random shift and the decoration behave as in the 1-step case considered in this work. However, the frozen phase can disappear if $\beta_{\max} = +\infty$, which is the case for example when there is a power law $f(q) \sim 1 - (1 - q)^{\alpha},  q \rightarrow 1$ with $0 < \alpha < 1$. In this case the zero temperature becomes critical, and describing the minima process is an interesting question. In fact, even the absolute minimum is expected \cite{fyodorov2008multiscale} to have different universal behaviour from logREMs, \textit{e.g.}, the $\frac{3}{2} \ln \ln M$ correction is expected to be no longer true. {To this end one may mention also a mathematical work  \cite{Arguinoimet2015} investigating related questions, though presented from a somewhat different perspective.}

\section{Edwards-Anderson's order parameter by 1RSB}\label{sec:EA}
The Edwards-Anderson's (EA) order parameter for the circular model (see Sect. \ref{sec:circularmodel} for its basics) has been calculated in \cite{cao16maxmin} using the freezing-duality conjecture (FDC). Its numerical check made the most convincing case for FDC; yet, the way of applying FDC there was quite tricky. Here we reproduce the result using 1RSB.  In fact, we shall consider a more general class of models that we call \textit{spherical} logREMs. They are Euclidean logREMs whose geometric manifold $X$ (see Sect. \ref{sec:irdata}) is contained in the unit sphere of a Euclidean space
\begin{equation}\forall  \xi \in X \,,\, (\xi.\xi) = 1 \,, \end{equation}
where $(\xi.\xi')$ denotes the Euclidean inner product. This is a vast class of models, since there is no further restriction on IR data \footnote{Provided one stays away from other transitions, \textit{e.g.}, binding, that invalidates the 1RSB Ansatz here}.
For example, our analysis applies to the 2d GFF-logREM on the sphere, or variants of circular models related to Morris integral.

We consider the following replica sum
\begin{align}
\mathcal{X}_\beta(n)= \overline{\mathcal{Z}^n \abs{\langle \xi \rangle}^2} = \overline{\sum_{j_1, \dots, j_n} e^{-\sum_{a} \beta V_{j_a}} (\xi_{j_1}.\xi_{j_2})} \,. \label{eq:EAsum}
\end{align}
Here $\langle  \mathcal{O} \rangle = \sum_j \mathcal{O}_j e^{-V_j} / \mathcal{Z}$ is the thermal average. Recall that according to the 1RSB Ansatz, the sum \eqref{eq:EAsum} is dominated by configurations where the replicas form groups of size $m = \min(\beta^{-1}, 1)$: each group occupying a $O(1)$ block corresponding to a continuum position $\xi_g$ ($g = 1, \dots, m$) on the {spherical manifold $X$} , and the sum $\sum_{j_a}  \rightarrow \sum_{\text{comb.}} \int \dif\mu(\xi_1) \dots \dif \mu(\xi_{n/m}) $ is replaced by a sum over replica grouping configurations (involving
a combinatorial factor) and a continuum integral over group positions (times UV contributions). For the simple replica partition function $ \overline{\mathcal{Z}^n}$ \eqref{eq:Zn}, the former sum gives merely a factor $C_{n,m}$ \eqref{eq:Cnm}, the only novelty of $\mathcal{X}_\beta(n)$ with respect to $\overline{\mathcal{Z}^n}$ is that there are two kinds of grouping configurations: for $\frac{m-1}{n-1} C_{n,m} $ configurations, the replicas $1$ and $2$ can be in the same group, so $\xi_{j_1} = \xi_{j_2} \Rightarrow (\xi_{j_1}. \xi_{j_2}) = 1$ and the continuum integral is that of $\overline{\mathcal{Z}^n}$; for the remaining $\frac{n-m}{n-1} C_{n,m}$ configurations a new continuum integral defined in \eqref{eq:EAintegral} (with $n/m$ variables and with renormalized temperature $b=\beta m$) replaces that of $\overline{\mathcal{Z}^n}$. The combinatorial factors above can be understood by a ``probabilistic'' argument: if all $C_{n,m}$ groupings are equi-probable, the probability that replicas $1$ and $2$ be in the same group is $\frac{m-1}{n-1}$.  Note also that the inserted observable $(\xi_{j_1}.\xi_{j_2})$ does not probe any UV structure (\textit{i.e.}, the scalar product is unity if the two replica belong to the same group), so the UV contributions to $\mathcal{X}_\beta(n)$ is identical to that of $\overline{\mathcal{Z}^n}$. Summarizing, we have the following:
\begin{align}
\mathcal{X}_\beta(n) & =\overline{\mathcal{Z}^n} \left( \frac{m-1}{n-1} + \frac{n-m}{n-1} X_{m\beta}(n/m) \right)
 \,,\, \label{eq:Xn}  \\
X_b(k)& = \frac{1}{\overline{Z_b^{k}}} \int \prod_{a=1}^{k} \left[ \abs{\dif^d \xi_g}  \exp(-m \beta c(\xi_g) - (1 + m^2 \beta^2)l(\xi_g)) \right] \prod_{a\neq b} \exp(b^2 C(\xi_a, \xi_b)) (\xi_1.\xi_2) \,. \label{eq:EAintegral}
\end{align}
Here the two terms in the brackets represent contributions of two cases discussed above, and $X_\beta(k)$ is the ratio between the new continuum integral and the standard continuum integral of the logREM, $\overline{Z_b^{k}}$. In the case of the circular model we recall that the latter is given by \eqref{eq:dyson}.
Using the variable $t = - n \beta$, and defining free energy as $\mathcal{F} = -\beta^{-1}\ln\mathcal{Z}$, one can rewrite \eqref{eq:Xn} as:
\begin{equation}
 \left. \overline{e^{t\mathcal{F}} \abs{\langle \xi \rangle}^2} \right/ \overline{e^{t\mathcal{F}}} = \begin{dcases}  X_\beta(-t / \beta) & \beta < 1 \\ \frac{(1 + t) X_1(-t)
 	 + (\beta - 1)}{t + \beta}  &  \beta > 1\end{dcases}  \label{eq:etfEA}
\end{equation}
In particular, a pleasing general result can be obtained by setting $t = 0$ in \eqref{eq:etfEA}:
\begin{equation}
\overline{\abs{\langle \xi \rangle}^2}_{\beta > 1} = 1 - \beta^{-1} \left( 1 -\overline{\abs{\langle \xi \rangle}^2}_{\beta = 1}\right) \Leftrightarrow \beta \left(1 - \overline{\abs{\langle \xi \rangle}^2}_{\beta} \right) \text{ freezes when } \beta > 1\,. \label{eq:freezingofEA}
\end{equation}
In other words, we predict that for all spherical logREMs the EA order parameter is \textit{linear} in temperature $T = 1/\beta$ in the whole $\beta > 1$ phase. Since at $T = 0$ the EA order parameter is always $1$, only the slope is model-dependent.

Now we specify our consideration to the particular case of the circular model, to compare our result here with that of \cite{cao16maxmin}, Eq. (42). Its LHS is equal to that of \eqref{eq:etfEA} above. In \cite{cao16maxmin} an exact Coulomb gas integral is used to solve the high temperature phase. We quote the result for $\beta = 1$ (assuming continuity at that point): $X_1(n) = (2 - n)^{-1}$.
Plugging this into the $\beta > 1$ case of \eqref{eq:etfEA}, we obtain
$$ \left. \overline{e^{t\mathcal{F}} \abs{\langle \xi \rangle}^2} \right/ \overline{e^{t\mathcal{F}}} \stackrel{\beta > 1}=
\frac{\beta(2 + t) - 1}{(t + \beta)(t + 2)}\,, $$
in agreement with eq. (42) of \cite{cao16maxmin}. Given the $\beta < 1$-phase solution, obtained due to exact evaluation of the Coulomb gas integrals, a non-trivial application of the FDC was needed to obtain the correct $\beta > 1$ phase answer in \cite{cao16maxmin}. Here, the first-principle 1RSB calculation provides the correct non-analytical continuation, with the knowledge of the $\beta=1$ solution alone.


\section{Joint order statistics from randomly shifted decorated Gumbel Poisson point process}\label{sec:dppp}
Here we recall the notion of a randomly shifted decorated Gumbel Poisson point process (SDPPP) and calculate its minima value distribution. A reader recognising that \eqref{eq:pdfvalue} describes a SDPPP may safely skip this subsection.

A SDPPP is by definition generated by three steps. First, one generates a Gumbel Poisson point process (PPP) with the density $e^{x} \dif x$ . Let us recall a defining property of the PPP: the joint probability of the number of points
$N_{I_1}=n_1,..N_{I_k}=n_k$ in a set of disjoint intervals $I_1,..I_k$ is given by  $\prod_{j=1}^k \frac{1}{n_j!} (N_j)^{n_j} e^{- N_j}$ where $N_j= \int_{x \in I_j} e^x dx$.
From this, denoting {the sequence of ordered minima values} as $X_1 < X_2 < \dots $, it is easy to write down the following joint probability density, for $x_1< \dots < x_p < x$
\begin{equation}
\mathcal{P}(x_1, \dots, x_p, x) \stackrel{\text{def}}= \, \overline{\theta(X_{p+1} - x) \prod_{q=1}^{p} \delta(X_q - x_q)} = \exp \left(\sum_{q=1}^p x_q - \int_{-\infty}^{x} e^y \dif y \right)
= \exp \left(\sum_{q=1}^p x_q - \exp(x) \right) \,. \label{eq:ppp}
\end{equation}
a property which completely describes PPP and can equivalently be taken as its definition. In particular, $-X_1$ has a Gumbel distribution, hence the name.

To define the decorated Poisson point process (dPPP) \textit{without} random shift we need to specify a decoration process. It must be a point process that contains $0$ and contains only non-negative points, but can be otherwise arbitrary, so long as we can order its elements $0 = Y_0 < Y_1 < Y_2 < \dots$. Similarly as in \eqref{eq:decoration}, the decoration process $(Y_s)_{s=0}^{\infty}$ is completely determined by the following observables:
\begin{align}\label{eq:dppp}
d(\Delta_{0},\dots, \Delta_{\ell-1}) &\stackrel{\text{def}}= \,
\overline{\theta(Y_{\ell}- Y_{\ell-1} -\Delta_{\ell-1}) \prod_{s=0}^{\ell-2} \delta(Y_{s} + \Delta_s - Y_{\ell-1} - \Delta_{\ell-1} )} \\
&= \mathrm{Prob}(Y_{\ell}- Y_{0} -\Delta_{0} = Y_{\ell}- Y_{1} -\Delta_{1}= \dots =Y_{\ell}- Y_{\ell-1} -\Delta_{\ell-1} > 0) \,.
\end{align}
As is discussed in the main text (around \eqref{eq:Dlastzero}), this is equivalent to giving the joint pdfs of $Y_0, \dots, Y_{\ell-1}$ for any $\ell=1,2,\dots$. We now make an independent copy of $(Y_s)_{s=0}^{\infty}$ for each $q = 1, 2, \dots$, denoted $\left((Y_{q,s})_{s= 0}^{\infty}\right)_{q = 1}^{\infty}$, \textit{i.e.}, for any $q$, \eqref{eq:dppp} would hold if one replaces $Y_s \leadsto Y_{q,s}$. The copies are also jointly independent from the PPP. The dPPP is then obtained by replacing each point of the PPP $X_q$ by the points $X_q + Y_{q,0} ,X_q + Y_{q,1}, X_q + Y_{q,2}, \dots$, \textit{i.e.}, the $q$-th copy of the decoration process shifted by $X_q$.

Finally, the SDPPP is obtained by shifting all the points so far by one random variable $F$ independent of everything else.

Let us denote $X_{\min,0}  < X_{\min,1}  < \dots $ the minima of the dPPP, and thus $X_{\min,0} + F < X_{\min,1} + F < \dots$ the minima of the SDPPP. We first set the random shift aside and calculate the joint distribution $\overline{\theta(X_{\min,k}- y) \prod_{s=0}^{k-1} \delta(X_{\min,s} - y_{s})}$, where $y_0 < \dots < y_{k-1} < y$, in order to compare with \eqref{eq:pdfvalue}. By the above definition this is given by the sum over the partitions of ${y_0, \dots, y_{k-1}}$, indicating what points are in the same decoration. So let us fix a partition $\{y_{q,0}, \dots, y_{q,k_q -1}\}_{q=1}^p$ and calculate its contribution. First one has the factor arising due to PPP, which is equal to the probability (density) that $X_{q+1} > y$  and that $X_q = y_{q,0}, q=1, \dots, p$, which is given by \eqref{eq:ppp} with $x = y$ and $x_q = y_{q,0}, q = , \dots, p$. Then one has a factor coming from each decoration group, which is the probability that $x_q + Y_{q,k_q} > y$ and that $x_q + Y_{q, s} = y_{q, s}$ with $s = 0, 1, \dots, \ell = k_q - 1$. This is given by \eqref{eq:dppp} with $\Delta_s = y - y_{q,s}, s = 0, 1, \dots, \ell = k_q - 1$. In summary, we have:
\begin{align}
& \overline{\theta(X_{\min,k}- y)\prod_{s=0}^{k-1} \delta(X_{\min,s} - y_{s})} =
\sum_{\mathbb{P}(k)} \mathcal{P}(y_{1,0}, \dots, y_{p,0}, y) \prod_{q=1}^p
d(y - y_{q,0},\dots, y - y_{q,k_q-1}) \\
=& \sum_{\mathbb{P}(k)} \exp\left(-e^y + \sum_{q=1}^p y_{q,0}\right) \prod_{q=1}^p \left[d(y - y_{q,0},\dots, y - y_{q,k_q-1}) \right] \\
=&  \sum_{\mathbb{P}(k)} \exp\left(-e^y + p y\right) \prod_{q=1}^p \left[e^{y_{q,0} - y}  d(y - y_{q,0},\dots, y - y_{q,k_q-1}) \right] \\
=& \sum_{\mathbb{P}(k)} \mathcal{D}_p \mathsf{Gum}(y) \prod_{q=1}^p \left[e^{y_{q,0} - y}  d(y - y_{q,0},\dots, y - y_{q,k_q-1}) \right] \,,\quad \mathsf{Gum}(y) = \exp(-\exp(y))
\end{align}
using that $\mathcal{D}_p \exp(-\exp(y))= \exp(p y-\exp(y))$ where $\mathcal{D}_p$ is defined in \eqref{eq:Dp}. The above formula is identical to \eqref{eq:pdfvalue} and \eqref{eq:decoration}, with $G_{\infty}=\mathsf{Gum}$ (the REM-Gumbel $G_{\infty}$) and the decoration functions $D = d$. Now we incorporate the random shift, that is replace $y_{\dots} \rightarrow y_{\dots} - F$ in the above calculation and average over $F$. Namely, denoting the probability measure $\overline{\delta(F - f)} \dif f = \dif \mathbb{P}(f)$, we have
\begin{align}
& \overline{ \theta(X_{\min,k} + F - y) \prod_{s=0}^{k-1} \delta(X_{\min,s} + F - y_{s})} =
\int  \overline{ \theta(X_{\min,k} - (y - f)) \prod_{s=0}^{k-1} \delta(V_{\min,s} - (y_{s} - f))} \dif \mathbb{P}(f) \\
= &  \sum_{\mathbb{P}(k)} \int  \mathcal{D}_p \mathsf{Gum}(y-f) \dif \mathbb{P}(f) \prod_{q=1}^p \left[e^{y_{q,0} - y}  d(y - y_{q,0},\dots, y - y_{q,k_q-1}) \right]
\end{align}
Comparing with \eqref{eq:pdfvalue}, we identify $\int  \mathsf{Gum}(y-f) \dif \mathbb{P}(f) = G_{\infty}(y) \,.$
Since $\mathsf{Gum}(y) = \overline{\theta(- \mathrm{Gum} > y)}$ where $\mathrm{Gum}$ is a standard Gumbel distribution and
$ G_{\infty}(y) = \overline{\theta(V_{\min} > y)}$ ($V_{\min}$ here is the  minimum of the log-REM giving $G_{\infty}$), we get the convolution relation $V_{\min} = F - \mathrm{Gum}$. Comparing the latter with the freezing scenario (see \eqref{eq:Gfreezing} and Sect. \ref{sec:freezing}), we identify the random shift $F$ of the SDPPP with the critical-temperature free energy of the logREM.

Clearly, since the random shift $F$ has no effect on the gap distributions, the latter will be shared by the log-REM and the SDPPP with the same UV data/decoration process, and given by \eqref{eq:tipstat}.

\section{Marginal order and gap statistics of SDPPP}\label{sec:sdppp}
This section studies in more detail \eqref{eq:pdfvalue} copied here for convenience:
\begin{align}
&\overline{\theta(V_{\min,k} - y)\prod_{s=0}^{k-1}\delta(V_{\min,s} - y_s) } = \sum_{\mathbb{P}(k)} \mathcal{D}_p G_{\infty}(y) \prod_{q=1}^p \left[e^{y_{q,0}-y} D( y - y_{q,0},\dots,  y - y_{q,k_{q}-1})\right] \,. \label{eq:pdfvalueapp}
\end{align}
 By the previous Appendix the results of this section apply to extreme values statistics SDPPP in general.

\subsection{Gap statistics}\label{sec:gapstatistics}
 First we look at the gap statistics. For this we integrate \eqref{eq:pdfvalueapp} over $y \in \R$, while keeping all the differences $\Delta_s = y - y_s$, $s = 0, \dots, k-1$ fixed; the LHS then becomes the probability density of the event $V_{\min,k} - v_{\min,0} - \Delta_0 = V_{\min,k} - V_{\min,1} - \Delta_1 = \dots  = V_{\min,k} - V_{\min,k-1} - \Delta_{k-1} > 0$,  similarly to \eqref{eq:Dprobability}. In the RHS, all the differences being constant, we need to integrate only over $\mathcal{D}_p G_{\infty}(y)$, and it is not hard to show $\int \dif y \mathcal{D}_p G_{\infty}(y) = (p-1)!$ using  \eqref{eq:Dp} and $G_{\infty}(-\infty) = 1, G_{\infty}(+\infty) = 0$ (see \eqref{eq:Ginfty}).  Equating the two sides we obtain
\begin{align}
&\overline{ \theta(V_{\min,k} - V_{\min} - \Delta_0) \prod_{s=0}^{k-2} \delta(V_{\min,s} - V_{\min} + \Delta_{s} - \Delta_0)} = \sum_{\mathbb{P}(k)} (p-1)!  \prod_{q=1}^p \left[e^{-\Delta_{q,0}} D(\Delta_{q,0},\dots, \Delta_{q,k_q-1}) \right]  \label{eq:tipstat}
\end{align}
The above equation is the generalization of \eqref{eq:gapstat1} (which is the special case $k = 1$), and implies that the gap statistics, also known as the minimum process \textit{seen from the tip}, depends only on the UV data of the system.

In the rest of this subsection we shall fix some $\epsilon > 0$, and work with the domain $\mathbb{S}_k = \{\Delta_0 > \Delta_1 \dots > \Delta_{k-1} > \epsilon\}$. Integrating the LHS of \eqref{eq:tipstat} over $\mathbb{S}_k$ gives
\begin{equation} \int_\epsilon^{+\infty} \dif \Delta_{k-1} \overline{\theta(V_{\min,k} - V_{\min,k-1} - \Delta_{k-1})} = \overline{(V_{\min,k}-V_{\min,k-1} - \epsilon)_+} \,,\quad (x)_+ \defeq \max(x, 0) \,. \end{equation}
Note that the above identity holds for general non-negative random variables. Equating the above equation with the RHS of \eqref{eq:tipstat} gives
\begin{align}
&g_k \stackrel{\text{def}}{=} V_{\min,k} - V_{\min,k-1} \\
 \Rightarrow \,\, & \overline{(g_k - \epsilon)_+} = \int_{\mathbb{S}_k} \sum_{\mathbb{P}(k)} (p-1)!  \prod_{q=1}^p \left[e^{-\Delta_{q,0}} D(\Delta_{q,0},\dots, \Delta_{q,k_q-1}) \right]  \dif^{k} \Delta \,,\quad \dif^{k}\Delta = \dif \Delta_0 \dots \dif \Delta_{k-1}  \,.  \label{eq:kthgapD}
\end{align}

 We proceed to evaluate the RHS of \eqref{eq:kthgapD} in terms of $K_\infty(\Delta_0 = \epsilon,\dots,\Delta_{\ell-1} = \epsilon)$ (see \eqref{eq:defKgeneral} for definition and  \eqref{eq:partialKisD} for relation with $D(\Delta_0, \dots, \Delta_\ell)$), and in terms of \textit{integer} partitions. An integer partition (denoted $\lambda$) is that of identical (``bosonic'') elements, or equivalently, simply a decreasing sequence of positive integers giving the sizes of the parts $\lambda = (\lambda_1 \geq \dots \geq \lambda_{p} > 0)$, with $p$ being called the \textit{length}, $\abs{\lambda} = \sum_{q=1}^{l(\lambda)} \lambda_{q}$ the \textit{size}, and $m_j(\lambda) = \abs{\{q: \lambda_q = j\}}$ the \textit{multiplicities} of the partition $\lambda$. As we have seen in Sect. \ref{sec:derivationgeneral}, a single integer partition $\lambda$ of size $k$ can correspond to several partitions of $k$ distinct elements; the latter partitions are said to be of \textit{type} $\lambda$, and we denote by $\mathbb{P}(\lambda)$ all partitions of type $\lambda$.

With these notations we shall first show validity of a more general formula, which holds for any sequence $(a_p)_{p=1}^{\infty}$:
 \begin{align}
 & \int_{\mathbb{S}_k} \sum_{\mathbb{P}(k)} a_p  \prod_{q=1}^p \left[e^{-\Delta_{q,0}} D(\Delta_{q,0},\dots, \Delta_{q,k_q-1}) \right]  \dif^{k} \Delta  =   \sum_{\abs{\lambda} = k} \frac{a_p}{\prod_{j} m_j(\lambda)!} \prod_{q=1}^{p} \frac{\mathbf{K}(\lambda_q)}{\lambda_q!} \,, \label{eq:generalintegral} \\
 & \text{where }\mathbf{K}(\ell) \defeq (-1)^{\ell-1} K_\infty(\Delta_0 = \epsilon, \dots, \Delta_{\ell-1} = \epsilon)\,. \label{eq:kthgapK}
 \end{align}
\begin{itemize}
\item[] \textit{Proof.} Recall the relation \eqref{eq:partialKisD} implying the following Newton-Leibniz type identity:
  \begin{align}
  & \int_{\mathbb{S}_\ell} \dif^\ell \Delta e^{-\Delta_0} D(\Delta_0, \dots, \Delta_{\ell-1}) = - \int_{\mathbb{S}_\ell} \dif^\ell \Delta
   \partial_{0,\dots,\ell-1} K_\infty (\Delta_0, \dots, \Delta_{\ell-1}) \nonumber\\
   =&  - \frac{1}{\ell!} \int_{(\epsilon,\infty)^{\ell}} \dif^\ell \Delta
   \partial_{0,\dots,\ell-1} K_\infty (\Delta_0, \dots, \Delta_{\ell-1}) = \frac{\mathbf{K}(\ell)}{\ell!}  \,.
  \end{align}
  Here, in the first line $(\Delta_s)_{s=0}^{k-1}$ are still ordered so \eqref{eq:partialKisD} applies; only in the second line we do not use the ordered domain $\Delta_0 > \Delta_1 \dots $ to perform the integral, by exploiting the fact that $K_\beta$ is completely symmetric in its arguments and that $K_\beta \rightarrow 0$ when one of its arguments tends to $\infty$ (because then the Newton-Leibniz subtraction in \eqref{eq:defKgeneral} will have a pair of equal upper and lower bounds and thus vanishes). As a consequence,
  \begin{equation}
  \text{RHS of }\eqref{eq:generalintegral} =  \sum_{\abs{\lambda}} \frac{a_p}{\prod_{j} m_j(\lambda)!} \int_{S(\lambda)} \prod_{q=1}^{p} \prod_{s=0}^{\lambda_q - 1} \dif \Delta_{q,s}\prod_{q=1}^p \left[ e^{-\Delta_{q,0}} D(\Delta_{q,0}, \dots, \Delta_{q, \lambda_q-1}) \right] \,. \label{eq:rhskthgapK}
  \end{equation}
  where the integral is over the space $S(\lambda) = \{ \Delta_{q,s}: \epsilon < \Delta_{q,0} < \dots <\Delta_{q,\lambda_q -1}, q = 1, \dots, p \}$ in which the $\Delta$'s are ordered only inside same parts. Thus for each point in $S(\lambda)$ we can reorder its coordinates and obtain a point $\Delta_0 < \dots< \Delta_{k-1} \in \mathbb{S}_k$, and a partition of $\{\Delta_0, \dots, \Delta_{k-1}\}$ that is of type $\lambda$. This gives a mapping $S(\lambda) \rightarrow \mathbb{S}_k \times \mathbb{P}(\lambda)$. It is not hard to check that it covers all the target space, and it is $\prod_{j} m_j(\lambda)!$ to one: the degeneracy comes from permuting parts of \textit{same} sizes. This mapping equates \eqref{eq:rhskthgapK} to \eqref{eq:generalintegral}'s LHS and demonstrates the latter identity.
\end{itemize}
Using the standard exponential generating function technique \cite{stanleyEC2}, we can rewrite \eqref{eq:generalintegral} in a more elegant formal power series fashion:
\begin{align}
& \sum_{k=1}^{\infty} t^k \int_{\mathbb{S}_k} \sum_{\mathbb{P}(k)} a_p  \prod_{q=1}^p \left[e^{-\Delta_{q,0}} D(\Delta_{q,0},\dots, \Delta_{q,k_q-1}) \right]  \dif^{k} \Delta = F_a\left( \sum_{j=1}^{\infty} \mathbf{K}(j) \frac{t^j}{j!} \right) \,,\quad F_a(x) = \sum_{p=1}^{\infty} a_p \frac{x^p}{p!} \,, \label{eq:genfungen}
\end{align}
where we have used that for any $\mathbf{K}(j)$ and $a_p$ the following holds as an identity of formal power series (\textit{i.e.} term by term):
\begin{eqnarray}
\sum_{\lambda \neq 0} t^{|\lambda|} \frac{a_{\ell(\lambda)}}{\prod_{j} m_j(\lambda)!} \prod_{q=1}^{\ell(\lambda)} \frac{\mathbf{K}(\lambda_q)}{\lambda_q!} = \sum_{p=1}^{+\infty} \frac{a_p}{p!} \left[
\sum_{j=1}^{\infty} \mathbf{K}(j) \frac{t^j}{j!} \right]^p \,.
\end{eqnarray}
In particular, taking $a_p = (p-1)!$, which gives $F_a(x) = -\ln(1 - x)$, and recalling the equation \eqref{eq:kthgapD}, we have:
\begin{align}
\sum_{k=1}^{\infty}\overline{(g_k - \epsilon)_+} t^k = -\ln \left( 1 - \sum_{j=1}^{\infty} \mathbf{K}(j) \frac{t^j}{j!} \right) \,
\,, \label{eq:Kjgenfun}
\end{align}
which provides the marginal distribution of all global gaps, obtained by differentiation: $\overline{\delta(g_k - \epsilon)}= \partial_{\epsilon}^2 \overline{(g_k - \epsilon)_+}$ (recall $\mathbf{K}(j)$ depends on $\epsilon$ by \eqref{eq:kthgapK}).

We now show that \eqref{eq:Kjgenfun} is equivalent to \eqref{eq:gapgenerating0} in Sect. \ref{sec:synopsis}, by calculating explicitly $\mathbf{K}(j)$. For this we shall write down explicitly \eqref{eq:Kinftyumin}, by expressing the $2^j$ terms generated by equation \eqref{eq:NewtonLeib} as indexed by the subsets of $J = \{0, \dots, j-1\}$:
\begin{align}
\mathbf{K}(j) =  \sum_{J \supset S \neq \emptyset} (-1)^{1 + \abs{S}} \frac{\overline{\exp(-\min(\min_{s\in S} u_{\min,s} + \epsilon, u_{\min,j}))}}{\overline{\exp(-u_{\min})}} - \frac{\overline{\exp(-u_{\min,j})}}{\overline{\exp(-u_{\min})}}  \label{eq:Kj1}
\end{align}
Clearly, $\min_{s\in S} u_{\min,s} = u_{\min, \min S}$ because $u_{\min,s}$ is increasing. Therefore, all non-empty subsets with the same minimum element contribute the same magnitude, but with changing sign, and it is not hard to check that their sum is non-zero only if there is only one term, \textit{i.e.}, when $\min S = j-1$, $S = \{j-1\}$. This fact reduces \eqref{eq:Kj1} to
\begin{align}
\mathbf{K}(j) &= \frac{\overline{\exp(-\min(u_{\min,j-1} + \epsilon, u_{\min,j}))} - \overline{\exp(-u_{\min,j})}}{\overline{\exp(-u_{\min})}} \nonumber \\
 & = \frac{\overline{\exp(-u_{\min}) \left( \exp(u_{\min} - u_{\min,j-1} - \epsilon) - \exp(u_{\min} - u_{\min,j}) \right)_+}}{\overline{\exp(-u_{\min})}} \nonumber \\
& = \overline{\left( \exp(v_{\min} - v_{\min,j-1} - \epsilon) - \exp(v_{\min} - v_{\min,j}) \right)_+} \label{eq:Kj}
\end{align}
The last equality follows from the definition of the biased minima process \eqref{eq:defbiased}. Combining with \eqref{eq:Kjgenfun}, we have
\begin{align}
& \sum_{k=1}^{\infty} \overline{(g_k - \epsilon)_+} t^k = -\ln \left( 1 - \sum_{j=1}^{\infty} \frac{t^j}{j!}  \overline{\left( \exp(v_{\min} - v_{\min,j-1} - \epsilon) - \exp(v_{\min} - v_{\min,j}) \right)_+} \right)  \,, 
\label{eq:gapgenerating}
\end{align}
which is just \eqref{eq:gapgenerating0}. In particular, at order $t^1$, we have $\overline{(\exp(-\epsilon) - \exp(v_{\min} - v_{\min,1}))_+} = \overline{(g_1 - \epsilon)_+}$, which recovers \eqref{eq:gapstat}.
Therefore, \eqref{eq:gapgenerating} is a non-trivial generalisation of \eqref{eq:gapbiased}, which gives the distribution of all gaps in terms of the decoration process.

\subsection{Order statistics} \label{sec:kthmin}
Now we turn our attention to the marginal distribution of $(k+1)$-th minimum $V_{\min,k}$. For this, it is useful to observe that \eqref{eq:pdfvalueapp} provides two different expressions of joint pdf (a special case of this observation for first and second minima is in Sect. \ref{sec:lowTgen}):
\begin{itemize}
\item[-] The first is obtained by considering  \eqref{eq:pdfvalueapp} with $y = y_{k-1}$. Performing such a substitution makes the $\theta$-term in \eqref{eq:pdfvalueapp} always $1$, so we are left with the $k-1$ $\delta$'s. The result is:
\begin{align}
 \overline{\prod_{s=0}^{k-1} \delta(V_{\min,s} - y_s)} = \sum_{\mathbb{P}(k)} \mathcal{D}_p G_{\infty}(y_{k-1}) \prod_{q=1}^p  \left[ e^{y_{q,0}-y} D( y_{k-1} - y_{q,0},\dots,  y_{k-1} - y_{q,k_{q}-1}) \right] \,. \label{eq:pdfvalue1}
\end{align}
\item[-] The second is obtained by applying $-\partial_y$ to \eqref{eq:pdfvalueapp}:
\begin{align}
\overline{\delta(V_{\min,k} - y)\prod_{s=0}^{k-1} \delta(V_{\min,s} - y_s)} = &-\sum_{\mathbb{P}(k)} \mathcal{D}_{p} G'_{\infty}(y)
\prod_{q=1}^p \left[e^{y_{q,0}-y} D( y - y_{q,0},\dots,  y - y_{q,k_{q}-1}) \right] \nonumber \\  &-  \sum_{\mathbb{P}(k)} \mathcal{D}_{p} G_{\infty}(y)  \partial_{y} \prod_{q=1}^p \left[ e^{y_{q,0}-y} D( y - y_{q,0},\dots,  y - y_{q,k_{q}-1}) \right] \,.  \label{eq:pdfvalue3}
\end{align}
\end{itemize}

Now, to obtain the marginal distribution of the $(k+1)$-th minimum, we integrate out $\Delta_s = y - y_s, s = 0, \dots, k-1$ in the domain $\mathbb{S}_k = \{\Delta_0 > \dots >  \Delta_{k-1} > 0\}$ (in this subsection we set $\epsilon = 0$) in \eqref{eq:pdfvalue3}. The result can be rearranged into
\begin{align}
&\overline{\delta(V_{\min,k} - y)} + \sum_{\mathbb{P}(k)}  \mathcal{D}_p  G'_\infty(y) \int_{\mathbb{S}_k}  \dif^{k} \Delta
 \prod_{q=1}^p e^{-\Delta_{q,0}} D(\Delta_{q,0},\dots, \Delta_{q,k_q-1})   \nonumber \\
=& -\sum_{\mathbb{P}(k)} \mathcal{D}_{p} G_{\infty}(y)  \int_{\mathbb{S}_k}  \dif^{k} \Delta   \left(\sum_{s=0}^{k-1} \partial_{\Delta_{s}}\right) \left[\prod_{q=1}^{p} e^{-\Delta_{q,0}} D(\Delta_{q,0},\dots, \Delta_{q,k_q-1})\right] \nonumber \\
= & \sum_{\mathbb{P}(k)} \mathcal{D}_{p} G_{\infty}(y) \int_{\mathbb{S}_{k-1}}  \dif^{k-1} \Delta \left[\prod_{q=1}^{p} e^{-\Delta_{q,0}} D(\Delta_{q,0},\dots, \Delta_{q,k_q-1})\right]_{\Delta_{k-1} = 0} \,, \label{eq:Vminkminus1}
\end{align}
where the second to third line integrates out $\Delta_{k-1}$ while keeping $\Delta_s - \Delta_{k-1}, s = 0, \dots, k-2$ fixed (by applying Newton-Leibniz). On the other hand, it is not hard to see that if we integrate \eqref{eq:pdfvalue1} over $\Delta_s = y_{k-1} - y_{s}$, $s = 0,\dots,k-2$ while keeping $\Delta_{k-1} = 0$, we also obtain the last line of \eqref{eq:Vminkminus1}. This implies the following recursion relation:
\begin{align}
&\overline{\delta(V_{\min,k} - y)} - \overline{\delta(V_{\min,k-1} - y)} =  - \sum_{\mathbb{P}(k)}  \mathcal{D}_p  G'_\infty(y) \int_{\mathbb{S}_k}  \dif^{k} \Delta  \prod_{q=1}^p e^{-\Delta_{q,0}} D(\Delta_{q,0},\dots, \Delta_{q,k_q-1})  \,,
\end{align}
or, after integrating over $y$,
\begin{align}
\overline{\theta(V_{\min,k} - y)} - \overline{\theta(V_{\min,k-1} - y)} =  \sum_{\mathbb{P}(k)}  \mathcal{D}_p  G_\infty(y) \int_{\mathbb{S}_k}  \dif^{k} \Delta  \prod_{q=1}^p e^{-\Delta_{q,0}} D(\Delta_{q,0},\dots, \Delta_{q,k_q-1}) \,.
\end{align}
Writing the generating function $\sum_{k=1}^{\infty} t^k \dots$ of both sides, and then applying the generating function relation \eqref{eq:genfungen} with $a_p = \mathcal{D}_p  G_\infty(y)$, and after some rearrangement we obtain a power series generating all higher order statistics:
\begin{align}
&\sum_{k=0}^{\infty} \overline{\theta(V_{\min,k} - y)} t^k = \frac{1}{1 - t} \sum_{p=0}^{\infty} \frac{\mathcal{D}_p}{p!} G_\infty(y) \left( \sum_{j=1}^{\infty} \mathbf{K}(j) \frac{t^j}{j!} \right)^p  \,.
\end{align}
Now since $\mathcal{D}_p = -\partial_y (1 - \partial_y) \dots (p-1 - \partial_y)$ (see eq. \eqref{eq:Dp}) we can apply the binomial formula to the RHS of above, regarding $\partial_y$ formally as a variable:
\begin{align}
\sum_{k=0}^{\infty} \overline{\theta(V_{\min,k} - y)} t^k = \frac{1}{1-t} \left(1 - \sum_{j=1}^{\infty} \mathbf{K}(j) \frac{t^j}{j!}  \right)^{\partial_y}  G_\infty(y) \,,
\end{align}
and then apply \eqref{eq:Kjgenfun} (with $\epsilon = 0$) to arrive at
\begin{align}
&\sum_{k=0}^{\infty} \overline{\theta(V_{\min,k} - y)} t^k = \frac{1}{1-t} \exp\left(- \sum_{k=1}^{\infty} t^k \overline{g_k} \partial_y\right)G_\infty(y) =   \frac{1}{1-t} G_{\infty}\left(y -\sum_{k=1}^{\infty} t^k \overline{g_k} \right) \,, \label{eq:pdfkthmin}
\end{align}
which is \eqref{eq:marginalgenerating} in Sect. \ref{sec:fullsummary}. Expanding it out as a power series with respect to both $t$ and in $\partial_y$, we see that the cdf of any $V_{\min,k}$ is a linear combination of derivatives of $G_\infty(y)$:
\begin{align}
 \overline{\theta(V_{\min,k} - y)} = \sum_{s=0}^{k} c_{k,s} \partial_y^s G_\infty(y) \,,\label{eq:cks}
\end{align}
where the matrix of coefficients satisfies the following relations (we denote $\gamma_k \defeq \overline{g_k}$ in these formulae to avoid possible confusion caused by over-lines)
\begin{align}
& c_{k,0} = 1 \,,\, c_{k,1} = -\sum_{s=1}^k \gamma_k = \overline{V_{\min}} - \overline{V_{\min,k}} \,,\, c_{k,k} = \frac{(-1)^k}{k!} \gamma_1^k \,,  \label{eq:cksgeneral} \\
 &c_{3,2} =  \frac{\gamma_1^2}{2}+\gamma _2 \gamma _1 \,, c_{4,2} = \frac{\gamma _1^2}{2}+\gamma _2 \gamma _1+\gamma _3 \gamma _1+\frac{\gamma _2^2}{2} \,,\,  c_{4,3} = -\frac{\gamma _1^3}{6}-\frac{1}{2} \gamma _2 \gamma _1^2 \,,\, \\
& (c_{5,2}, c_{5,3}, c_{5,4})= \left( \frac{\gamma _1^2}{2}+\gamma _2 \gamma _1+\gamma _3 \gamma _1+\gamma _4 \gamma _1+\frac{\gamma _2^2}{2}+\gamma _2 \gamma _3 , - \frac{\gamma _1^3}{6}-\frac{1}{2} \gamma _2 \gamma _1^2-\frac{1}{2} \gamma _3 \gamma _1^2-\frac{1}{2} \gamma _2^2 \gamma _1 , \frac{\gamma _1^4}{24}+\frac{1}{6} \gamma _2 \gamma _1^3 \right) \,.
\end{align}
 Using computer algebra software this procedure can be easily continued to arbitrary order.   Note that \eqref{eq:cks} and \eqref{eq:cksgeneral} (for $k=1$) covers \eqref{eq:pdf2ndmin}. Although from \eqref{eq:pdfvalue1} it is not hard to see that \eqref{eq:cks} must hold for some set of coefficients $c_{k,s}$ that depend only on the biased minima process, it is not obvious that these coefficients are polynomial functions of the \textit{mean values} of the gaps.

As another check, let us apply $\int_\R \dif y (-y) \partial_y$ to both sides of \eqref{eq:pdfkthmin}:
The LHS gives
$$\int_\R \dif y (-y) \partial_y \sum_{k=0}^{\infty} \overline{\theta(V_{\min,k} - y)} t^k = \sum_{k=0}^\infty \overline{V_{\min,k}} t^k =\sum_{k=0}^\infty (\overline{V_{\min}} + \overline{g_1} + \dots + \overline{g_k})t^k \,.$$
For the RHS, recalling from \eqref{eq:Ginfty} that $-G'_\infty(y)$ is the pdf of $V_{\min}$, we have
$$\int_\R \dif y (-y) \partial_y  \frac{1}{1-t} G_{\infty}\left(y -\sum_{k=1}^{\infty} t^k \overline{g_k} \right) =   (1-t)^{-1}\left(\overline{V_{\min}} + \sum_{k=1}^\infty \overline{g_k} t^k\right) = \sum_{k=0}^\infty (\overline{V_{\min}} + \overline{g_1} + \dots + \overline{g_k})t^k\,.$$
The two sides give the same result as expected.

Although \eqref{eq:pdfkthmin} is a non-trivial general property of SDPPP, and can be used for numerical tests, it is highly unlikely to be a \textit{sufficient} characterization of SDPPP, since it concerns only marginal distributions.

\section{Minima positions}\label{sec:minposition}
This subsection extends the 1RSB approach to statistics of positions of the sequence of minima. We shall first rederive the result of \cite{biskup2016full} on the joint minima position-value process for general Euclidean logREM. Then we compare the 1RSB approach to the one based on FDC as used in  \cite{fyodorov2015moments,fyodorov2010freezing}.

\subsection{1RSB calculation}
As mentioned earlier, the general observable \eqref{eq:Hgeneral} contains also the minima position information if we do not sum over marker positions $j_0, \dots, j_{k-1}$. However, to take the $M \rightarrow \infty$ limit we need to specify how $(j_s)$ behave when performing the limit. We know now that the distance between extreme points are either of UV scale or of IR scale. This implies the \textit{clustering} of minima positions, namely, the joint distribution of continuous position of the minima will have delta singularities; for example, in the case of first and second minima we expect $Prob(\xi_0, \xi_1) = \delta(\xi_0 - \xi_1) P(\xi_0) + P'(\xi_0, \xi_1)$ where $P'(\xi_0, \xi_1)$ has no $\delta$-singularity at $\xi_0 = \xi_1$. In general, $Prob(\xi_0, \dots, \xi_{k-1})$ will be a sum over all the clusterings of the positions (\textit{i.e.}, over the partitions of $0, \dots, k-1$), with each term being the product of $\delta$'s imposing the clustering and a benign function of the cluster positions.

Therefore, to capture the individual terms we shall consider a limiting behaviour of $j_0, \dots, j_{k-1}$ in accordance with such a  picture, and also in agreement with the recent work \cite{biskup2016full}. For this let us fix a partition of the markers $\{(j_0,y_0), \dots, (j_{k-1}, y_{k-1})\}$ into $p$ parts (see Sect. \ref{sec:derivationgeneral} for notation of partition), and fix $p$ points $\xi_1, \dots, \xi_p \in X \subset \C$ (recall that $X$ is the geometric manifold on which the IR limit of the log REM is defined). Markers form blocks $\xi(j_{q,s}) = \eta_q$ for $q=1, \dots, p$, where $\eta_1, \dots ,\xi_p \in X$ are fixed continuum coordinates. We will resolve position information to the continuum coordinate precision, \textit{i.e.} the microscopic coordinates will be summed over. This leads to the following definition:
\begin{equation}
H_\beta( (\eta_q),(y_{q,s}), y) \stackrel{\text{def}} = \sum_{\xi(j_{q,s}) = \eta_q} H_\beta((j_s, y_{q,s}),y) \,. \label{eq:Hposition}
\end{equation}
Taking the $\beta\rightarrow\infty$ limit, evaluating the mixed partial derivative $\partial_{y_0} \dots \partial_{y_{k-1}}$, and setting $y$ to $y_{k-1}$ give rise to joint position-value distribution of the first $k$ minima:
\begin{equation}
\partial_{y_0 \dots y_{k-1}} H_\infty( (\eta_q),(y_{q,s}), y) = \overline{ \theta(V_{\min,k} - y) \prod_{s=0}^{k-1} \delta(V_{\min,s} - y_s)
 \prod_{q=1}^p \delta(\Pi_{q},\eta_{q})}  \label{eq:zeroTpos}
\end{equation}
In plain terms the above is the probability of the following event: the $k$ first minimal values are given by $y_0 < \dots < y_{k-1}$, and the $(k + 1)$-th minimum is larger than $y$; moreover,  the minima $\{y_0, \dots, y_{k-1}\}$ form $p$ blocks, described by a partition $\{ y_{q,s}: q = 1, \dots, p, s = 0, \dots, k-q - 1\}$, such that the $q$-th block is located at $\eta_{q}$. Let us emphasize that $\Pi_{1}, \dots, \Pi_{q} \in X$ are \textit{not} macroscopic positions of the first $q$ minima, but the first $q$ minima \textit{that are locally minimal}.  The normalisation of the Dirac-$\delta$'s on the positions will be clarified below, see \eqref{eq:deltapos}.

The method of sect. \ref{sec:fullminima} can be adapted here to calculate $H_{\beta}$ \eqref{eq:Hgeneral}. In fact, using \eqref{eq:HGen_gen} one easily rewrites \eqref{eq:Hposition} in terms of replica sums $\overline{\mathcal{W}^n}$ \eqref{eq:bigWdef}
\begin{equation}
H_\beta( (\eta_q),(y_{q,s}), y) = (-1)^{k} \sum_{\xi(j_{q,s}) = \eta_q} \sum_{n\geq 0} \frac{(-e^{\beta y})^n}{n!} \left. \overline{\mathcal{W}^n(j_s, \Delta_{s})}\right\vert^{y - y_{s}}_{\Delta_s = +\infty} \,,\, \label{eq:Hposgen}
\end{equation}
which in turn  can be treated using 1RSB; to simplify the presentation, we will focus on the $\beta > 1$ phase. When comparing to \eqref{eq:WnZn} the main new feature is
 that the marker positions are \textit{not} summed over the system, so replica groups that occupy the marker positions cannot move. On the other hand, the sums over microscopic coordinate remain the same as before and so are the UV factors of \eqref{eq:WnZn}; the combinatorics of matching replica groups and markers remain also intact. Therefore, \eqref{eq:WnZn} can be adapted as follows:
\begin{align}
 &\sum_{\xi(j_{q,s}) = \eta_q} \left. \overline{\mathcal{W}^n(j_s, \Delta_{s})}\right\vert^{y-y_{s}}_{\Delta_s = +\infty} \stackrel{\beta>1}=
 \frac{\Gamma(n\beta + 1)}{\Gamma(n\beta - p + 1)} \overline{ \mathcal{Z}^n} \times K \times \left. \frac{\overline{Y_b({n\beta - p}\vert\eta_1, \dots, \eta_{p})}}{\overline{Z_b^{n\beta}}}\right\vert_{b\rightarrow 1}  \,, \label{eq:Wposition}\\
 & \overline{Y_b(n\vert \eta_1,\dots, \eta_{p})} = \prod_{q<r}\exp(b^2 C(\eta_q, \eta_r)) \int_{X^n} \prod_{a=1}^{n} \left( \dif\mu_\beta(\xi_a) \prod_{s=1}^p \exp(b^2 C(\xi_a, \eta_s))  \right) \prod_{a<b} \exp(b^2 C(\xi_a, \xi_b)) \label{eq:defY}   \,,\, \\
 &\dif\mu_b(\xi_a)  = \exp(-b c(\xi_a) - (1 + b^2) l(\xi_a)) \abs{\dif^d \xi_a} \,,\,  K = \prod_{q=1}^p K_\beta(y - y_{q,0}, \dots, y-y_{q,k_q-1}) \label{eq:UV}
\end{align}

In \eqref{eq:defY} we defined new continuum Coulomb gas integrals $\overline{Y_b}$, which generalize \eqref{eq:contZ} by the insertion of $p$ charges at $\eta_0, \dots, \eta_{p}$, whereas $\overline{Z_b^n}$ can be seen as the $k = 0$ case of $\overline{Y_b}$. It is symmetric in its arguments $\eta_1, \dots, \eta_p$. Moreover, integrating $\overline{Y_b}$ over the charge positions gives back $\overline{Z^n_b}$
\begin{equation}
\int_{X^p} \overline{Y_b(n-p\vert \eta_1, \dots, \eta_{p})}  \prod_{q = 1}^p \dif\mu_b(\eta_q) = \overline{Z_b^{n}} \,. \label{eq:normalization}
\end{equation}
Since by \eqref{eq:partialZcont}, $\overline{Z^n_b} \sim (1 - b)^n \overline{\widetilde{Z}^n}$ when $b \rightarrow 1_-$, we expect a similar singular behaviour is therefore expected for $Y$:
\begin{align}
&\lim_{b \rightarrow 1_-}\overline{Y_b(n-p\vert \eta_1, \dots, \eta_{p})}(1-b)^{-n} \stackrel{\text{def}}=  \overline{\widetilde{Y}(n\vert\eta_1, \dots, \eta_{p})} \label{eq:Ytilde} \\
\Rightarrow& \int_{X^p} \overline{\widetilde{Y}(-t\vert \eta_1, \dots, \eta_{p})}  \prod_{q = 1}^p \dif\mu_1(\eta_q) = \overline{\widetilde{Z}^{-t}} \,, \label{eq:normalisation1}
\end{align}
so that the normalisation relation \eqref{eq:normalization} extends to the critical Coulomb-gas integrals. In terms of $\widetilde{Y}$, \eqref{eq:Wposition} becomes
\begin{equation}
\sum_{\xi(j_{q,s}) = \eta_q} \left. \overline{\mathcal{W}^n(j_s, \Delta_{s})}\right\vert^{y-y_{s}}_{\Delta_s = +\infty} \stackrel{\beta>1}=
 \frac{\Gamma(n\beta + 1)}{\Gamma(n\beta - p + 1)} \overline{ \mathcal{Z}^n}  K  \frac{\overline{\widetilde{Y}({n\beta}\vert\eta_1, \dots, \eta_{p})}}{\overline{\widetilde{Z}^{n\beta}}}  \label{eq:Wposition1}
 \end{equation} 

Now, by the freezing scenario \eqref{eq:Znfreezing}, the terms involving $Z$ and $\mathcal{Z}$ in \eqref{eq:Wposition1} essentially cancel out :
\begin{equation}
 \sum_{\xi(j_{q,s}) = \eta_q} \Gamma(1 - n) \left. \overline{\mathcal{W}^n(j_s, \Delta_{s})}\right\vert^{y-y_{k}}_{+\infty} \stackrel{\beta>1}=
 \frac{\Gamma(n\beta + 1)\Gamma(1 - n\beta)}{\Gamma(n\beta - p + 1)}   \overline{\widetilde{Y}(n\beta \vert \eta_1, \dots, \eta_{p})}
 \mathbf{Z}_0^n K \,.
\end{equation}
Applying (inverse) Laplace transform \eqref{eq:laplace}, \eqref{eq:laplaceder} (noting $n = -t/\beta$), and recalling \eqref{eq:Hposgen}, we obtain:
\begin{align}
&H_{\beta > 1}( (\eta_q),(y_{q,s}), y) = (-1)^{k+p} \mathcal{D}_p I_\beta(y\vert \eta_1, \dots, \eta_p ) \times K\,,\, \label{eq:Hposition0} \\
&\int_\R -\partial_y I_\beta(y\vert \eta_1, \dots, \eta_p ) e^{ty}\dif y = \Gamma(1 + t) \overline{\widetilde{Y}(-t \vert \eta_1, \dots, \eta_{p})}  \mathbf{Z}_0^{-t/\beta} \,,\, \beta > 1\,. \label{eq:defI}
\end{align}
Here $I_\beta(y\vert \eta_1, \dots, \eta_p) $ is determined by inverse Laplace transform, with freezing behaviour manifesting itself via its $\beta$ dependence being only a shift encoded in $e^{ - t \beta^{-1} \ln \mathbf{Z}_0}$. To obtain the final expression for the joint PDF, we combine \eqref{eq:zeroTpos} with \eqref{eq:Hposition0}, and plug in \eqref{eq:partialKisD} for the UV correction factor $K$ of \eqref{eq:UV}:
\begin{equation}
\overline{ \theta(V_{\min,k} - y)\prod_{s=0}^{k-1} \delta(V_{\min,s} - y_s) \prod_{q=1}^p \delta(\Pi_{s},\eta_{q})} =
  \mathcal{D}_p I_\infty(y\vert \eta_1, \dots, \eta_p ) \prod_{q=1}^p \left(e^{y_{q,0} - y} D(y - y_{q,0}, \dots, y - y_{q,k_q-1}) \right) \,. \label{eq:positionvalue}
\end{equation}

We now clarify the meaning of $\delta$'s involving position variables by comparing to \eqref{eq:pdfvalueapp}, which the above formula generalises. To do this we first combine \eqref{eq:normalisation1} and \eqref{eq:defI} to obtain
$$ \int_\R  \dif y \int_{X^p}  -\partial_y I_\beta(y\vert \eta_1, \dots, \eta_p ) e^{ty} \prod_{q = 1}^p \dif\mu_1(\eta_q) = \overline{\widetilde{Z}^{-t}}\Gamma(1 + t)  \mathbf{Z}_0^{-t / \beta} = \overline{\mathcal{Z}^{-t/\beta}}\Gamma(1 + t/\beta)$$
by \eqref{eq:Znfreezing}. Compared to \eqref{eq:laplace} the above implies
\begin{equation}  \int_{X^p} I_\beta(y\vert \eta_1, \dots, \eta_p )  \prod_{q = 1}^p \dif\mu_1(\eta_q) = G_\beta(y) \,. \label{eq:intIisG} \end{equation}
Using this formula, we integrate over the positions $\eta_{q}$ in both sides of \eqref{eq:positionvalue} and sum over all the partitions of $\{y_0, \dots, y_{k-1}\}$; its RHS gives then precisely that of \eqref{eq:pdfvalueapp}, so we obtain
$$ \sum_{\mathbb{P}(k)} \int_{X^p}  \overline{ \prod_{s=0}^{k-1} \delta(V_{\min,s} - y_s) \prod_{q=1}^{p}  \delta(\Pi_{q},\eta_{q}) } \prod_{q=1}^p \dif\mu_1(\eta_q)
 =   \overline{ \prod_{s=0}^{k-1} \delta(V_{\min,s} - y_s)} \,. $$
This tells us that Dirac-$\delta$ of the position is with respect to the measure $\dif\mu_1(\eta)$ in \eqref{eq:UV}; In terms of probability density, this means
\begin{equation}
\overline{ \dots  \delta(\Pi_{s}, \eta_{q}) \dots } =  \frac{\dif \textit{Probability}}{\dots \dif\mu_1(\eta_q) \dots } \,.   \label{eq:deltapos}
\end{equation}

\subsection{Random measure interpretation}
Let us connect the result \eqref{eq:positionvalue} with that of \cite{biskup2016full}. According to the main theorem (2.1) of the latter work, naïvely generalised, the probability at the LHS of \eqref{eq:positionvalue} is given by a decorated Poisson point process (DPPP) on the product space of $(y,\xi) \in \R \times X$ ($y\in\R$ standing for value and $\xi \in X$ for position), constructed as follows: one first generates a \textit{random} measure $\rho(\xi) \dif \mu_1 (\xi)$ on $X$, then generates a PPP $\{(X_q, \xi_q): q = 1, 2, \dots\}$ (ordered by $X_1 < X_2 < \dots$) on the product space $\R \times X$, with product density measure $e^{y} \dif y  \rho(\xi) \dif \mu_1 (\xi)$. Finally the decoration acts only on the value sector in the same way as described in \ref{sec:dppp}: one replaces each $(x_q, \eta_q)$ by a sequence $(X_{q,s} = X_q + Y_{q,s}, \xi_q), s = 0, ,1,\dots$ where $(Y_{q,s})_s$ are iid copies of the decoration process. We need no additional random shift, which can be seen as absorbed in the random measure. By a calculation similar to Appendix \ref{sec:dppp}, we can show that for such a process, its ordered minima (by value) $X_{\min,s}, s = 0, 1, \dots$ and the corresponding positions $\Pi_s$ have the following distribution:
\begin{align}
&\overline{\theta(X_{\min,k} - y)\prod_{s=0}^{k-1} \left[\delta(X_{\min,s} - y_s) \delta(\Pi_{s}, \eta_{q(s)})\right]}
\nonumber\\
 = & \prod_{q=1}^p \left[e^{y_{q,0} - y} D(y - y_{q,0}, \dots, y - y_{q,k_q-1}) \right] \mathcal{D}_p
 \overline{\exp\left(-e^{y} Z[\rho] \right) \prod_{q=1}^p \frac{\rho(\eta_q)}{Z[\rho]} }^{\rho}  \,,\, Z[\rho] =  \int_X \rho(\eta) \dif \mu_1 (\eta)
\end{align}
where $\overline{[\dots]}^{\rho}$ means averaging over the random measure $\rho \dif \mu_1$, and $Z[\rho]$ is the volume of $X$ under that measure. This formula is the same as our result \eqref{eq:positionvalue}, provided the identification
\begin{equation}
I_\infty(y\vert \eta_1, \dots, \eta_p ) = \overline{\exp\left(-e^{y} Z[\rho] \right) \prod_{q=1}^p  \hat{\rho}(\eta_q) }^{\rho} \,, \hat{\rho} = \frac{\rho}{Z[\rho]}
\label{eq:IisGZ}\end{equation}
which relates the random measure $\rho$ with $I_\infty$ which is in turn related to continuum Coulomb gas integrals \eqref{eq:Ytilde} via \eqref{eq:defI}. Let us briefly discuss some consequences of this correspondence. First, we look at the total mass $Z[\rho]$. Integrating the above over all $\eta_q$ and comparing with \eqref{eq:intIisG} we obtain
\begin{equation}
G_\infty(y) = \overline{\exp\left(-e^{y} Z[\rho] \right)}^{\rho} \,,
\end{equation}
Applying Laplace transform to this formula and comparing to \eqref{eq:Gfreezing} gives us
\begin{equation}
\overline{Z[\rho]^{-t}}\exp(-F(M, \beta = \infty, f_{ij}) t) = \overline{\widetilde{Z}^{-t}}  \,, \label{eq:Zrho}
\end{equation}
where $F(\dots)$,\eqref{eq:FM} is the non-fluctuating part of the minimum, and contributes a deterministic shift to $\ln Z[\rho]$. Up to translation, $-\ln Z[\rho]$ has the same distribution as the critical temperature free energy, as one can see from the above equation and \eqref{eq:freezing}. Note that up to now in the present paper the over-line in $\overline{\widetilde{Z}^t}$ \eqref{eq:partialZcont} was a formal symbol
and did not correspond to any genuine average, so \eqref{eq:Zrho} gives it a probability interpretation, in terms of the random measure $\rho$. Thus, from now on we shall identify $ \widetilde{Z}$ with $Z[\rho] \exp(F(M,\infty,f_{ij}))$. Next by applying the Laplace transform directly to \eqref{eq:IisGZ} and comparing with \eqref{eq:defI} gives after cancelling out  the extensive free energy factors by \eqref{eq:Zrho} and \eqref{eq:FM1}, we obtain
\begin{equation}
\overline{\widetilde{Y}(-t\vert\eta_1, \dots, \eta_{p})} = \overline{\widetilde{Z}^{-t} \prod_{q=1}^p  \hat{\rho}(\eta_q)}^{\rho} \label{eq:Ytrho}
\end{equation}
Eqs. \eqref{eq:Ytrho} and \eqref{eq:Zrho} express properties of the random measure $\rho \dif \mu_1$ in terms of analytical continuation of Coulomb gas integrals \eqref{eq:defY}, \eqref{eq:Ytilde}. Many general transformation properties of $\rho$ shown in \cite{biskup2014conformal} can be understood in terms of change of variables in \eqref{eq:defY}. In particular, if we set $-t = p$, and apply  \eqref{eq:defY}, \eqref{eq:Ytilde}, \eqref{eq:Ytrho} and $\hat{\rho} = \rho / Z[\rho]$ in \eqref{eq:IisGZ}, we obtain
\begin{equation}
 \overline{ \prod_{q=1}^p \rho(\eta_q)}^{\rho}  = \prod_{q<r}\exp(C(\eta_q, \eta_r))\,.
\end{equation}
This implies the formal identification of $\rho(\eta)$ with the regularized vertex operator $:e^{-V(\eta)}:$ where $V(\eta)$ satisfies $\overline{V(\eta) V(\xi)}^c = C(\eta,\xi)$.

\subsection{Relation with freezing approach}
\cite{fyodorov2010freezing} and \cite{fyodorov2015moments} studied also the position of the global minimum, using the replica-freezing approach, of which the derivation of \eqref{eq:positionvalue} above is a generalization to higher order statistics. To see this more clearly let us restrict \eqref{eq:positionvalue} to the particular case where $k = 1$ (so $p = 1$ and the partition is the trivial one) and $y = y_{k-1}$. Applying inverse Fourier transform to both sides and using \eqref{eq:defI} we have
\begin{equation}
\overline{\delta(V_{\min} - y)\delta(\Pi_1,\eta)} = - \partial_y I_\infty(y \vert \eta) \Leftrightarrow
\overline{\exp(V_{\min} t) \delta(\Pi_1,\eta) } = \Gamma(1 + t) \overline{\widetilde{Y}(-t \vert \eta)} \mathbf{Z}_0^{-t/\beta}  \,.
\end{equation}
Setting $t = 0$ we get the minimum position distribution
\begin{equation} \overline{\delta(\Pi_1, \eta)} = \overline{\widetilde{Y}(0\vert \eta)} = \lim_{\beta\rightarrow 1_-}
\overline{Y_\beta(-1\vert \eta)} = \overline{\hat{\rho}(\eta)}^{\rho} =  \overline{\rho(\eta) / Z[\rho]}^{\rho} \,, \label{eq:minpospdf} \end{equation}
see \eqref{eq:Ytilde} and \eqref{eq:Ytrho}. By \eqref{eq:defY}, the latter is an analytically continued Coulomb gas integral over $ n \rightarrow -1$ moving charges and one charge fixed at $\eta$, and the temperature fixed to the critical value: we obtained the same replica-trick recipe of minimum position in \cite{fyodorov2015moments}, sect. 2.3, which used the freezing-duality conjecture. Unfortunately, we know of no exactly solved integrals that give non-trivial position distribution. We note however that \cite{fyodorov2015moments} calculated integer moments of the minimum position of the interval model (see Sect \ref{sec:interval} for definition). Although the results are in principle equivalent to knowing $\overline{\delta(\Pi_0 - \eta)}$,  an explicit expression of the latter is still lacking.

\section{Asymptotic calculation of eq. \eqref{eq:defREMgamma}}  \label{sec:REMFyo}
 This appendix calculates the asymptotic behaviour of the following function, defined in \eqref{eq:defREMgamma}
 \begin{equation}\label{1}
 \gamma(x)=\int_{-\infty}^{\infty}\frac{dc}{\sqrt{4\pi\ln{M}}}e^{-\frac{c^2}{4\ln{M}}} e^{-e^{\beta(x-c)}} \,.
 \end{equation}
 First, using the Hubbard-Stratonovich transformation
 \begin{equation}\label{2}
 e^{-\frac{c^2}{4\ln{M}}}=\int_{-\infty}^{\infty}\frac{d\zeta}{\sqrt{2\pi}}e^{-\frac{\zeta^2}{2}-\im \frac{1}{\sqrt{2\ln{M}}}\zeta c}
 \end{equation}
 and changing the order of integrations we easily compute the integral of $c$ (which yields the Euler Gamma-function) and get:
 \begin{equation}\label{3}
 \gamma(x)=\int_{-\infty}^{\infty}\frac{d\zeta}{\sqrt{2\pi}} \Gamma(\im \zeta)e^{-\beta^2\ln{M}\zeta^2-\im \zeta\beta x}  =e^{-\frac{x^2}{4\ln{M}}}\int_{-\infty}^{\infty}\frac{d\zeta}{\sqrt{2\pi}} \Gamma(\im \zeta)e^{-\beta^2\ln{M}\left(\zeta+\frac{\im x}{2\beta \ln{M}}\right)^2}
 \end{equation}
 Obviously, for $\log{M}\gg 1$ this integral is dominated by the saddle-point at $\zeta=\im \left(-\frac{x}{2\beta \ln{M}}\right)$.
  At the same time, one needs to remember that  $\Gamma(\im\zeta)$ has poles at $\zeta=k\im, k = 0, 1, 2,\dots$. As is obvious, for $x>0$ we can deform the integration contour away from the real axis to pass through the saddle-point without crossing the pole at $\zeta=0$, and the large-$M$ asymptotic is given by the saddle-point contribution
  \begin{equation}\label{4}
 \gamma(x)\approx\gamma_{sp}(x)=\frac{1}{2\beta\sqrt{\pi \ln{M}}}\,e^{-\frac{x^2}{4\ln{M}}}\Gamma\left(\frac{x}{2\beta \ln{M}}\right), \quad x>0
 \end{equation}
Howewer, for $x<0$ (which is of interest to us here) when deforming the contour we will pick up the residue contribution from poles (the more negative is $x$, the more poles we will need). It turns out that for our goals two essentially different situations arise:
\begin{enumerate}
\item $-1<\frac{x}{2\beta \ln{M}}<0$. In this parameter range we need to cross only one pole at $\zeta=0$, close to which the singular part of the $\Gamma-$function is $\Gamma(\im \zeta)=\frac{\Gamma(1+\im \zeta)}{\im\zeta}$. Calculating  $2\pi \im \times $ residue of the integrand at the pole we find
\begin{equation}\label{5}
 \gamma(x)\approx 1+\gamma_{sp}(x)=1+\frac{1}{2\beta\sqrt{\pi \ln{M}}}\,e^{-\frac{x^2}{4\ln{M}}}\Gamma\left(\frac{x}{2\beta \ln{M}}\right), \quad -2\beta \ln{M}<x<0
 \end{equation}
 Let us now consider $x=f+\delta$ where $\delta$ is of the order of unity and $f<0$ is defined by the identity:
 \begin{equation}\label{6}
 \frac{1}{\sqrt{\ln{M}}}e^{-\frac{f^2}{4\ln{M}}}=\frac{1}{M}, \quad \Rightarrow \quad f=-2\ln{M}+\frac{1}{2}\ln{\ln{M}}+o(1)
 \end{equation}
 Note, that such $f$ together with $x>-2\beta\ln{M}$ immediately implies $\beta>1$. Finally, substituting $x=f+\delta$ to (\ref{5}) and using (\ref{6})
 yields:
 \begin{equation}\label{eq:gammalowT}
 \gamma(x)\approx 1-\frac{\Gamma\left(1-\frac{1}{\beta}\right)}{2\sqrt{\pi}}\frac{e^{\delta}}{M}, \quad \beta>1 \,.
 \end{equation}
 Now, if we absorb the $O(1)$-constant $C = \Gamma\left(1-1 /\beta\right) / (2\sqrt{\pi})$ into a redefinition of $f \to f - \ln C$, eq. \eqref{eq:gammalowT} is equivalent to the $\beta > 1$ case of \eqref{eq:gREM}. It is interesting to note that $-\ln C \rightarrow -\infty$ when $\beta \rightarrow 1_+$; this divergence is needed for matching with the $\beta > 1$ case which has a $\ln \ln M$ correction in $f$, whereas the $\beta < 1$ case does not have such a correction (see \eqref{9} and \eqref{eq:gammahighT} below).
 \item $\frac{x}{2\beta \ln{M}}<-1$. In this parameter range we need to cross not only the pole $\zeta=0$, but also $\zeta= \im$, close to which the singular part of the $\Gamma-$function is $\Gamma(\im\zeta)=\frac{\Gamma(2+\im\zeta)}{\im\zeta(1+\im\zeta)}$. Calculating $2\pi \im \times $residue of the integrand at the pole we find
   \begin{equation}\label{8}
  \gamma(x)\approx 1-e^{\beta^2\ln{M}+\beta x}+\ldots+\frac{1}{2\beta\sqrt{\pi \ln{M}}}\,e^{-\frac{x^2}{4\ln{M}}}\Gamma\left(\frac{x}{2\beta \ln{M}}\right), \quad x<-2\beta \ln{M}
  \end{equation}
  where $\dots$ stand for the omitted contributions from other poles (when they arise), as those can be easily written down and checked to be immaterial for our goals.
   Let us again consider $x=f+\delta$ where $\delta$ is of the order of unity but $f<0$ is defined by the identity:
  \begin{equation}\label{9}
  e^{\beta^2\ln{M}+\beta f}=\frac{1}{M}, \quad \Rightarrow \quad f=-\left(\beta+\frac{1}{\beta}\right)\ln{M}
  \end{equation}
  Note, that such $f$ together with $x<-2\beta\ln{M}$ immediately implies $0<\beta<1$. We also easily check that the last term in (\ref{8}) is subdominant.
  Finally we get
  \begin{equation}\label{eq:gammahighT}
  \gamma(x)\approx 1-\frac{e^{\beta\delta}}{M}, \quad \beta<1 \,,
  \end{equation}
  which is the $\beta < 1$ case of \eqref{eq:gREM}.
\end{enumerate}

\bibliography{rems.bib}

\end{document}